\documentclass[reprint,aps,showkeys]{revtex4-2}
\usepackage[utf8]{inputenc}
\usepackage{indentfirst}
\usepackage{amsmath}
\usepackage{amssymb}
\usepackage{bm}
\usepackage{hyperref}
\usepackage{braket}
\usepackage{mathptmx}
\usepackage{mathtools}
\usepackage{dcolumn}
\usepackage{ltablex}
\usepackage{multirow}
\usepackage{lipsum}
\usepackage{tabularray}
\usepackage{graphicx}

\newcommand{\od}[2]{\frac{d #1}{d #2}}
\newcommand{\pd}[2]{\frac{\partial #1}{\partial #2}}
\newcommand{\eq}[1]{Eq.\,(\ref{#1})}
\newcommand{\eqnm}[1]{(\ref{#1})}

\newcommand{\smbr}[1]{\left( #1 \right)}
\newcommand{\cubr}[1]{\left\{ #1 \right\}}

\newcommand{\sqbr}[1]{\left[#1\right]}
\newcommand{\abs}[1]{\left|#1\right|}

\newcommand{\mfk}[1]{\mathfrak{#1}}

\newcommand{\tqt}[1]{\textquoteleft #1\textquoteright}

\DeclareMathOperator\arctanh{arctanh}
\DeclareMathOperator\sech{sech}

\begin{document}
\title{Consecutive Supersymmetric Quantum Mechanical Algorithms for Resolving the Spectra of Legendre-type Differential Operators}
\author{Tianchun Zhou}
\email{tczhou@cdut.edu.cn}
\affiliation{College of Physics, Chengdu University of Technology, Chengdu 610059, P. R. China}


\begin{abstract}
  Quite comprehensive iterative algorithms based on consecutive supersymmetric quantum mechanics (SUSYQM) have been constructed heuristically, for resolving the eigenspectra of the Legendre operator and its associates parametrized in various coordinates, \emph{without receiving any input from traditional analytical methods}. On building these algorithms, the two symmetries \text{--} supersymmetry  and shape invariance symmetry \text{--} hidden in these operators are brought to the fore as the only two governing principles, with the aids of the newly devised algebraic techniques such as the \emph{active supersymmetrization transformations}, \emph{momentum operator maps} as well as plain and \emph{higher-order spin matrix maps}. In view of their algebraic simplicity and elegance, high efficiency, as well as the independency of traditional methods, some of these algorithms become the promising candidates to supplant the popular power series approach to the eigenspectra of Legendre-type operators.

 \end{abstract}
\maketitle

\section{Introduction}

``\emph{There is a pleasure in recognizing  old things from a new point of view}'', said by Richard Feynman.

Legendre function and its associate are special functions that are well-known for a long time. As for many special functions, traditionally these functions are obtained as finite power series solutions of Legendre and associated Legendre equations. However, as critically remarked by George Stokes, ``\emph{power series solutions have the advantage of being generally applicable, but are wholly devoid of elegance}''.  One question that arises naturally is: can these special functions be built via more elegant algebraic algorithms? Along the line of searching for answers to this question, the Infeld-Hull's algebraic factorization method \cite{Infeld_factorization_methods_1951} of resolving the spectra of a large class of Sturm-Liouville-type differential operators makes a great stride.  This factorization method, though having been around us for more than half a century,  is still not so popular as the power series method that prevails in many texts and monographs. Retrospectively, the reason for this situation may be attributed to the fact that, the Infeld-Hull's factorization method is still not heuristic enough in the sense that the symmetries behind this method are still not brought to the fore as the guiding principles for constructing the algorithms of tackling the issue of the spectra of those Sturm-Liouville-type differential operators. Nowadays, we know that the exact elegance that the analytical power series method falls short of is the innate elegance coming along with the symmetries hidden in the eigen differential operators.

Although it has long been well-known from analytical viewpoint that there are two distinct types of special functions called Legendre functions and the associated ones that are tied with the \emph{same} Legendre differential operator, one can still ask the innocent question: is there a more straightforward answer to this fact beyond the differential connection between the Legendre equation and its associated counterpart disclosed by the traditional analytical method?

At the advent of supersymmetric quantum mechanics (SUSYQM), people get to recognize that, there exist two types of symmetries \text{--} supersymmetry (SUSY) \cite{Witten_SUSY_breaking_1981, Witten_1982} and shape invariance symmetry (SISY) \cite{Gendenshtein_SUSY_Schr_Equ_1983} \text{--} behind the success of the Infeld and Hull's algebraic factorization method that is applicable to a large class of Sturm-Liouville operators, among which the Legendre operators are just such special examples, and bringing these two symmetries to the fore can make this factorization method more heuristic.

Many people have made efforts on elucidating SUSY and SISY owned by lots of linear eigen differential operators, among which the Legendre-type operators are just special examples. One key step is to find out the supersymmetric factorization of each of these operators \text{--} a product of a pair of ladder operators \text{--} and construct the two-term raising and lowering recurrence relations. In \cite{Lorente_2001, Cotfas_SI_raising_lowering_hypergeometric_equations_2002}, the two-term raising and lowering recurrence relations for hypergeometric-type operators are derived from the combination of the Rodriguez formula for hypergeometric-type functions and some algebraic three-term recurrence relations. However, these two inputs are not obvious and are usually the byproducts of traditional methods. In \cite{Cotfas_SI_raising_lowering_hypergeometric_equations_2002}, the same method is applied to constructing the ladder operators for associated hypergeometric-type operators. In \cite{Baziea_SUSY_Legendre_2012}, the ladder operators of the base Legendre operators are built out of the recurrence relations of the Legendre polynomials, which are supposed to be the outcomes of traditional methods. In \cite{Baziea_SUSY_Legendre_2012}, the eigenvalues of the base as well as the associated Legendre operators are also taken to be known beforehand from the traditional analytical methods. Probably because one main goal of these authors is just to exhibit the supersymmetry and the shape invariance symmetry behind these eigen differential operators, instead of building SUSYQM algorithms, that are independent  of any traditional method, for exhausting the spectra of these operators, they are free to receive any input from traditional methods. In the monograph \cite{Dong_Factorization_methods_in_QM_2007} and the numerous references cited therein, mainly because the authors' focus, among other things, is on revealing the hidden $SU(2)$ or $SU(1,1)$ dynamical symmetry realized by various  quantum mechanic models, instead of building the independent SUSYQM algorithms for resolving eigenspectra of these models, the ladder operators are obtained by differentiating the wave functions that have already been obtained through traditional analytical methods, and combining the recurrence relations of the classical special functions involved.

Judged by the heuristicity, efficiency, elegance, as well as the independency of traditional methods, it seems to be fair to say that none of these existing factorization algorithms, including those already explicitly incorporate the idea of SUSYQM, can work as the supplant of the traditional power series method to the classical special functions.  The SUSYQM algorithms that are more elegant and efficient are still welcome, in order to be one step close to the ultimate goal of supplanting the traditional analytical method. Such expected algorithms, if any, should receive inputs as few as possible, or even no input at all, from any traditional method.

Efforts have been made in \cite{Zhou_SUSY_hypergeometric_operator_2024} towards such end.  The algebraic SUSYQM algorithms for exhausting the spectra of hypergeometric-like operators and their associates, have been constructed. These algorithms are full of quantum mechanic flavor and do not receive any input from traditional methods, and the only input is the definition of hypergeometric-like operator itself. In particular,  the SUSYQM answer has been provided to the question on why there exist the base hypergeometric type functions as well as the associated ones for the same hypergeometric-like operators. That is, in these iterative SUSYQM algorithms, the same hypergeomeric-like operator admits two distinct types of initial supersymmetric factorizations, with two distinct sorts of quantum kinetic energy/momentum operators as well as two distinct sorts of active supersymmetrization transformations. These two supersymmetric factorizations then proliferate into two distinct branches of supersymmetric factorizations of higher and higher level, under the same two repetitive algebraic operations \text{--} \emph{intertwining action (leftward braiding)} plus  \emph{renormalization}, that are built based on the two common symmetry principles \text{--} the supersymmetry as well as the shape invariance symmetry hidden in hypergeometric-like operators. Along with such iterative procedure, the two distinct hierarchies of supersymmetric factorizations of eigen differential equations and the two distinct types of eigenfunctions \text{--} the base hypergeomeric-like functions as well as the associated counterparts \text{--} are built simultaneously. As Legendre operator is just a special hypergeometric-like operator, these systematic algorithms give the SUSYQWM answer to the question on why there exist the base Legendre function as well as the associated one that are related to the same Legendre operator.

In \cite{Zhou_SUSY_hypergeometric_operator_2024},  we have also proposed an efficient algebraic way of reaching the supersymmetric factorization of a Sturm-Liouville-type operator \text{--} combining the momentum operator map and the accompanying active supersymmtrization transformation, to replace the traditional analytical method (through solving another differential equation of second order for the required coordinate transformation) for passing from a Sturm-Liouville-type eigenequation to a Schr\"odinger equation. With this algebraic way of supersymmetric factorization, for constructing ladder operators, we do not need any more  inputs such as three-term algebraic or differential recurrence relations from traditional methods.

Despite of these attempts made in \cite{Zhou_SUSY_hypergeometric_operator_2024}, the Legendre operator has its own specialties. Firstly, as a special hypergeometric-type operator, its specialties will bring drastic simplifications to  its general SUSYQM algorithms. Secondly,  besides being parametrized as a special hypergeometric type operator in a certain coordinate, it admits other more flexible and convenient types of coordinate parametrization, and one can build its SUSYQM algorithms in each of these coordinates. In particular, because the important issue in which Legendre operator appears is the one with axi-symmetry and is described by Laplace equation, this is naturally parametrized in polar angle with respect to the symmetry axis. Thirdly, because the  associated Legendre operator is intimately connected to the Casimir operator of angular momentum algebra in 3-D Euclidean space \text{--} an $so(3)$ Lie algebra, and the eigenspectrum of the associated Legendre operator to the representation of this Lie algebra, the SUSYQM algorithm of this operator can be elegantly reduced from such Lie algebra and its representation, which is another natural starting route of building the SUSYQM algorithm of this operator.

As in \cite{Zhou_SUSY_hypergeometric_operator_2024}, in this paper, our purpose is not simply limited to demonstrating the SUSY and shape invariance symmetry behind the Legendre operators, but to promote these two symmetries to the height of dictating principles that guide us to build efficient algebraic SUSYQM algorithms for solving the eigenspectra of the Legendre operators, without reference to any input from traditional analytical methods, so that these independent algorithms raise the hope of becoming the candidates for supplanting the popular power series method. In these algorithms the ladder operators are built without any pre-knowledge of the eigenfunctions or without reference to any recurrence relation that is obtained through traditional methods. Also the eigenvalues of the Legendre operator and the associated Legendre operator are expected to come out of these algorithms themselves. This way we will demonstrate how SUSY and shape invariance symmetry owned by the Legendre differential operator are brought to the fore, and how they dictate the very existence of the two special functions \text{--} Legendre functions and the associated Legendre functions that are related to the same Legendre operator. In almost all sections except Section \ref{direct_factorization_of_associated_Legendre} of this paper, the associated Legendre differential equation is assumed not to be known beforehand, and it will be derived from symmetry algebra and the resultant algorithms.

The rest of this paper is mainly structured as the following.

Section \ref{SUSYQM_base_Legendre_operator} focuses on constructing the SUSYQM algorithms for resolving the eigenspectrum of the base Legendre operator in $x$ and $y$ coordinates. An explicit bottom-up iterative algorithm is also provided for simultaneously constructing the base Legendre equation of generic level and its eigen-solution, which serves to demonstrate from the angle of SUSYQM why there exists the Legendre polynomial that coincides with its traditional form.

In Section \ref{associated_SUSY_x_coordinate}, an inductive SUSYQM algorithm is designed for simultaneously building the associated Legendre equation as well as its eigen-spectrum level by level with both generic positive and negative association levels. This algorithm takes the base Legendre equation as its only input, assuming that the form of associated Legendre equation is unknown. Such algorithm reveals the SUSYQM ground for why there exist the base Legendre function as well as the associated Legendre function that are related to the same Legendre operator. One reason for working out this iterative SUSYQM algorithm for the associated Legendre operator that is similar to that developed in \cite{Zhou_SUSY_hypergeometric_operator_2024} lies in that, the positive and negative branches of association level in \cite{Zhou_SUSY_hypergeometric_operator_2024} has to be switched for Legendre operators in order to be consistent with the usual definition of associated Legendre functions.

Two SUSYQM algorithms for exhausting the spectra of the associated Legendre operators are developed in Section \ref{direct_factorization_of_associated_Legendre}, by \emph{directly} searching for the supersymmetric factorizations of these operators parameterized in $y$ and $\theta$ coordinates, assuming that the form of associated Legendre operator parameterized in $\theta$ coordinate is already known to us, e.g., which can be obtained by separating the two spherical coordinates $\theta$ and $\phi$ in the angular Laplacian or the square of angular momentum operator.

In section \ref{angular_momentum_algebra}, in terms of the supercharges, a two-parameter ${\cal N}=2$ $su(1,1)$ superalgebra  behind the angular momentum $so(3)$ algebra is firstly revealed, both through falling into the representation of the angular momentum algebra and directly through the \emph{plain spin matrix maps} of the ladder operators of $so(3)$ algebra. Such two-parameter superalgebra is then reduced into a one-parameter ${\cal N}=2$ $su(1,1)$ superalgebra in two approaches. In the first approach, the reduction is achieved \emph{component-wise}, by taking the azimuthal harmonic average of the two component eigenequations satisfied by the products of the 2-parameter ladder operators of $so(3)$ algebra. More concretely, a pair of 1-parameter ladder operators are firstly identified as the non-vanishing matrix elements of the pair of 2-parameter ladder operators of the $so(3)$ algebra between two appropriate azimuthal harmonics, and the two eigenequations of the product of this pair of 1-parameter ladder operators are then obtained by peeling off the azimuthal harmonic appearing in the corresponding two eigenequations of the products of pair of 2-parameter ladder operators of the $so(3)$ algebra.
In the second approach, without resort to the representation of the $so(3)$ algebra, at first the two-parameter ${\cal N}=2$ $su(1,1)$ superalgebra, that is equivalent to the $so(3)$ algebra, is enlarged to another two-parameter ${\cal N}=2$ $su(1,1)$ superalgebra satisfied by the enlarged super-Hamiltonian and two supercharges, all of which are $4\times4$ matrix operators, through \emph{higher-order spin matrix maps}. Such enlarged superalgebra is then reduced to the 1-parameter ${\cal N}=2$ $su(1,1)$ superalgebra satisfied by a certain super-Hamiltonian and the corresponding two supercharges, all of which are $2\times 2$ matrix operators, through averaging over a specific two-component superstate. In this second approach, reduction is achieved via manipulating the super-Hamiltonian (and the supercharges) as a whole. Eventually, the spectral representation of such one-parameter superalgebra in $\theta$ coordinate gives rise to the same SUSYQM algorithm of constructing the associated Legendre operator as one of those algorithms demonstrated already in Section \ref{direct_factorization_of_associated_Legendre},  yet in well-organized fashion from the very beginning, which saves the effort on searching for the supersymmetric factorization of the 'ill-organized' associated Legendre operator.

The conclusions and remarks are given in Section \ref{conclusion}.

For the purpose of comparing with our method of reaching a certain supersymmetric factorization of a second-order linear differential operator via active supersymmetrization transformation plus momentum operator map, the traditional method of passing to a non-supersymmetrically-factorized Schr\"odinger equation for further supersymmetric factorization, from a second-order linear differential equation that has first-order derivative, by focusing on eliminating the first-order derivative via analytical method, is provided In Appendix \ref{traditional_analytical_method}, taking the Legendre operator represented in both $x$ and $\theta$ coordinates as two examples. The supercharge formulation of the ${\cal N}=2$ $su(1,1)$ superalgebra behind the supersymetrically factorized Hamiltonian corresponding to the Legendre operator and its intertwining superpartner Hamiltonian is provided in Appendix \ref{super_charge_formulation}. The $\theta$ coordinate representation of the SUSYQM algorithm of constructing the eigenspectrum of the Legendre operator is given in Appendix \ref{theta_coord_J0} to demonstrate that the SUSQM algorithm can be established with the same amount of work in a certain non-standard coordinate, in which the factorization of an eigen differential operator is not symmetric at all.

\section{The Consecutive SUSYQM Algorithms of the Base Legendre Operator}\label{SUSYQM_base_Legendre_operator}

 Our first task is to reveal the SUSYQM structure that lurks behind the base Legendre operator $L(x)$. The knowledge of elementary quantum mechanics reminds us that we should first of all search for a standard $1$-D Hamiltonian operator consisting of the kinetic energy operator and the potential energy that is associated with $L(x)$. The question is how to quickly pin down these two ingredients \emph{in an elementarily algebraic way}.

Note that every non-relativistic Hamiltonian operator for a point particle with mass $M$ moving along $z$-axis  has the kinetic energy operator $-\smbr{\hbar^2/2M} d^2/dz^2$, where $z$ has the dimension of length. The characteristic length of this simple quantum system can be constructed as $l_0\equiv \smbr{\hbar^2/(2M {\cal E}_0)}^{1/2}$, where ${\cal E}_0$ is the characteristic energy. The dimensionless $y$ coordinate is then defined as $y\equiv z/l_0$,  and the kinetic energy operator becomes  $\hat{p}^2_y=-d^2/dy^2$, which is the square of the momentum operator $\hat{p}_y=-i d/d y$, with $i$ being the imaginary unit. Formally, this non-dimensionalization procedure  amounts to adopting the units in which $\hbar=2M={\cal E}_0=1$. This \emph{symmetric double-layer} $\hat{p}^2_y=-\smbr{d/dy}^2$ is the simplest form of the \emph{supersymmetric} factorization of a linear differential operator of second order.  This basic fact reminds us that, in order to prepare a Hamiltonian from a linear differential operator of second order like $L(x)$, we may firstly aim at preparing the negative of a \emph{square} of some simple first-order differential operator within the original coordinate $x$ and then conceptually identify such \emph{symmetrized} operator product with a quantum kinetic energy operator in some new coordinate.

\subsection{An Active Supersymmetrization Transformation of Base Hypergeometric-type Operators and The Momentum Operator Maps}\label{AST_hypergeometric_operator}

In classical mathematical physics, the base Legendre differential operator comes naturally in the lumped form
\begin{eqnarray}\label{J0_theta}
 J_0(\theta)=-\frac{1}{\sin\theta}\od{}{\theta}\sin\theta \od{}{\theta},
\end{eqnarray}
where $\theta$ is the polar angle in a cylindrical or spherical coordinate system. Upon introducing the variable $x=\pm\cos\theta$, this operator is rewritten as a special hypergeometric-like differential operator in the form
\begin{eqnarray}\label{operator_L_x_coord}
\begin{aligned}
 L(x) & \equiv- \od{}{x} p(x)\od{}{x} = - p(x) \frac{d^2 }{dx^2} +2x \od{}{x},
\end{aligned}
\end{eqnarray}
where $p(x)=1-x^2$.

In order to demonstrate how the specialties of $L(x)$ bring around the simplicities to its supersymmetric factorization, it is better to look at this issue in the wider context of the supersymmetric factorization of a generic hypergeometric-type of operator via the method termed active supersymmetrization transformation in \cite{Zhou_SUSY_hypergeometric_operator_2024}.

A hypergeometric-like differential operator $h(x)$ in $x$-coordinate reads
\begin{eqnarray}\label{hypergeometric-like_operator}
h(x) &\equiv& -p(x) \frac{d^2}{dx^2} - q(x) \frac{d}{dx},\nonumber \\
     &=&\frac{1}{p(x)}\sqbr{-p(x)\od{}{x}+2 W_0(x)}\sqbr{p(x)\od{}{x}},\nonumber \\
     &=&-w^{-1}(x)\od{}{x} w(x)p(x)\od{}{x},
\end{eqnarray}
where $p(x)$ is a binomial and $q(x)$ a monomial in $x$ in general,
\begin{eqnarray}\label{W_0}
 W_0(x)\equiv \frac{1}{2}\sqbr{p'(x)-q(x)}=-\frac{w'(x)}{2w(x)}p(x),
\end{eqnarray}
the prime denotes $x$-derivative and
\begin{eqnarray}\label{weight_w}
w(x)\equiv \exp\left\{-\int^x d{\tilde x}\frac{2W_0(\tilde x)}{p(\tilde x)}\right\}
\end{eqnarray}
is the weight function (the integration factor of the first factorial differential operator in the second line of \eq{hypergeometric-like_operator}) that allows us to write $h(x)$ in its self-adjoint form, i.e., the third line in \eq{hypergeometric-like_operator}.

Upon simply multiplying with $p(x)$, the lumped form of $h(x)$ in the second line of \eq{hypergeometric-like_operator} is rearranged as the following \emph{asymmetric} factorization
\begin{eqnarray}\label{asymmetric_form_ph}
 p(x)h(x)&=&\sqbr{-p(x) \od{}{x}+ 2W_0(x)}\sqbr{p(x)\od{}{x}},
\end{eqnarray}
which contains the expected square form $-\smbr{p(x)d/dx}^2$.

Now, the \emph{asymmetric} factorization \eqnm{asymmetric_form_ph} is expected to be turned into a certain \emph{supersymmetric} factorization by some algebraic transformation. There are several ways to achieve this. One simple way goes as the following.
Firstly, in terms of the lumped form of $h(x)$ (the third line) in \eq{hypergeometric-like_operator}, $p(x)h(x)$ is trivially factorized as
\begin{eqnarray}\label{asymmetric_form_ph1}
 p(x)h(x) &=&\frac{1}{\sqrt{w(x)}}\sqbr{\frac{1}{\sqrt{w(x)}}\smbr{-p(x) \od{}{x}}\sqrt{w(x)}}\nonumber \\
 &&\cdot\sqbr{\sqrt{w(x)}\smbr{p(x)\od{}{x}}\frac{1}{\sqrt{w(x)}}}\sqrt{w(x)}.
\end{eqnarray}
This particular structure suggests us to introduce the bilateral transformation ${\cal S}_b(\text{--})\equiv\sqrt{w}(\text{--})\sqrt{w}^{-1}$ and its inverse ${\cal S}^{-1}_b(\text{--})\equiv\sqrt{w}^{-1}(\text{--})\sqrt{w}$, and to rearrange the asymmetric factorization \eqnm{asymmetric_form_ph1} as
\begin{eqnarray}\label{supersymmetric_form_ph1}
 {\cal S}_b\smbr{p(x)h(x)} &=&{\cal S}^{-1}_b\smbr{-p(x) \od{}{x}}\cdot{\cal S}_b\smbr{p(x)\od{}{x}}.
\end{eqnarray}

It is easy to check that
\begin{eqnarray}\label{Sb_shift}
 {\cal S}_b\smbr{p(x)\od{}{x}}=p(x) \od{}{x} + W_0(x),
\end{eqnarray}
which says that ${\cal S}_b$ simply works to shift $p(x) d/dx$ by the amount $W_0(x)$. Consequently, Its inverse ${\cal S}^{-1}_b$ works to shift $p(x)d/dx$ by  $-W_0(x)$, and also to shift  $-p(x)d/dx$ by $W_0(x)$ again, i.e.,
\begin{eqnarray}\label{Sb_upward_shift}
 {\cal S}^{-1}_b\smbr{\pm p(x)\od{}{x}}=\pm p(x) \od{}{x} \mp W_0(x).
\end{eqnarray}

Therefore, \eq{supersymmetric_form_ph1} is rewritten as the supersymmetric factorization
\begin{eqnarray}\label{supersymmetric_form_ph2}
 {\cal S}_b\smbr{p(x)h(x)}=\sqbr{-p(x) \od{}{x}+ W_0(x)}\sqbr{p(x)\od{}{x}+W_0(x)}.\nonumber \\
\end{eqnarray}

Another slightly different way of reaching the supersymmetric factorization \eqnm{supersymmetric_form_ph2} directly starts with the asymmetric factorization \eqnm{asymmetric_form_ph}.
Looking at the upper branch of \eq{Sb_upward_shift} from the right of this equation to the left, it says that the right hand side is lumped into the form in the left hand side, with ${\cal S}^{-1}_b$ being made of the integration factor $w^{1/2}(x)$.  Also we note that  ${\cal S}^{-1}_b(\text{--})$ acts on an ordinary function such as $W_0(x)$ intactly. Therefore, by singling out the common function $W_0(x)$ as the superpotential from the two factors in the asymmetric factorization \eqnm{asymmetric_form_ph} and lumping the rest ordinary function term in each factor as the integration factors, these two factors are then tweaked into
\begin{eqnarray}\label{left_factor_partial_supersymmetrization}
 -p(x)\od{}{x}+ 2W_0(x)&=&\smbr{-p(x)\od{}{x}+ W_0(x)}+ W_0(x),\nonumber \\
                       &=&{\cal S}^{-1}_b\smbr{-p(x)\od{}{x}+W_0(x)},
\end{eqnarray}
\begin{eqnarray}\label{right_factor_partial_supersymmetrization}
 p(x)\od{}{x} &=&\smbr{p(x)\od{}{x} + W_0(x)} - W_0(x),\nonumber \\
              &=&{\cal S}^{-1}_b\smbr{p(x)\od{}{x}+W_0(x)}.
\end{eqnarray}
With these results in hand, applying ${\cal S}_b$ to the asymmetric factorization \eqnm{asymmetric_form_ph} and noticing ${\cal S}_b$ acts distributively on each factor, the action ${\cal S}^{-1}_b$ in Eqs. \eqnm{left_factor_partial_supersymmetrization} and \eqnm{right_factor_partial_supersymmetrization} is then nullified and the supersymmetric factorization \eqnm{supersymmetric_form_ph2} is reached.

Here we see that the transformation ${\cal S}_b(\text{--})$ just turns the asymmetric factorization \eqnm{asymmetric_form_ph} of $p(x)h(x)$ into its \emph{supersymmetric} factorization in $x$-coordinate, simply by partitioning \emph{evenly} the function $2 W_0(x)$ in the first factor of $p(x)h(x)$ over the two factorial operators, and the common function $W_0(x)$ reached in each factor in the supersymmetric factorization \eqnm{supersymmetric_form_ph2} is then identified as the superpoential.  In such supersymetric factorization, the two factors are almost the same except that the signs carried by two first-order differential operators are opposite to each other. Such structure automatically ensures that there is no net first-order differential term in its expanded form, a feature of any quantum Hamiltonian operator. It is for this very reason that the transformation ${\cal S}_b(\text{--})$ has been coined an \emph{active supersymmetrization} transformation in \cite{Zhou_SUSY_hypergeometric_operator_2024}.

Furthermore,  we can identify the \emph{common} operator $p(x) d/dx$ (or $-p(x)d/dx$) appearing in the two factors in \eq{supersymmetric_form_ph2} as the simplest differential operator $d/dy$ in some $y$ coordinate, i.e.,
\begin{eqnarray}\label{momentum_operator_map1}
 p(x)\od{}{x}\equiv \od{}{y}.
\end{eqnarray}
This implies that $ -i p(x) d/dx=-i d/dy\equiv \hat{p}_y$, where $i$ is the imaginary unit. Such operator $\hat{p}_y$ is the standard quantum momentum operator in $y$ coordinate in the units we adopted. For this reason, the identification in \eq{momentum_operator_map1} is termed the \emph{momentum operator map} \cite{Zhou_SUSY_hypergeometric_operator_2024}. Here we can also make the identification $-p(x) d/dx=d/dy$. Accordingly, there exists the quantum kinetic energy operator map
\begin{eqnarray}\label{KE_operator_map1}
 -\sqbr{p(x)\od{}{x}}^2 \equiv -\frac{d^2}{d y^2}.
\end{eqnarray}

\eq{momentum_operator_map1} is equivalent to the simple first-order differential equation $dy=dx/p(x)$ and its solution
\begin{eqnarray}\label{x_to_y_trans0}
 y(x)=\int^x \frac{d x}{p(x)}
\end{eqnarray}
 defines the coordinate transformation from $x$ to $y$ and vice versa.

Denoting the superpotential $W_0(x)$ in $y$ coordinate as
\begin{eqnarray}
 {\cal W}_0(y)= W_0(x),
\end{eqnarray}
the supersymmetric factorization \eqnm{supersymmetric_form_ph2} in $x$ coordinate is then mapped into that of the standard quantum Hamiltonian $\widehat{H}_0(y)$ in its supersymmetrically factorized form
\begin{eqnarray}\label{SUSY_factorization_H0}
 \widehat{H}_0(y)=\smbr{-\od{}{y}+ {\cal W}_0(y)}\smbr{\od{}{y}+ {\cal W}_0(y)}.
\end{eqnarray}

As a hypergeometric-like differential operator, Legendre operator $L(x)$ is special in that $q(x)=p'(x)$. This specialty leads to the trivial superpotential $W_0(x)=0$ and the trivial weight function $w(x)=1$, according to Eqs. \eqnm{W_0} and \eqnm{weight_w}, respectively. Consequently, we have the trivial active supersymmetrization ${\cal S}_b(\text{--})=1\cdot(\text{--})\cdot 1$. Because of such drastic simplification, the asymmetric factorization \eqnm{asymmetric_form_ph} or \eqnm{asymmetric_form_ph1} of $p(x)h(x)$ of the hypergeometric-type operator $h(x)$ degenerates into the square form
\begin{eqnarray}\label{base_L_squared_form}
 p(x)L(x)=-\sqbr{p(x)\od{}{x}}^2,
\end{eqnarray}
which is identified as the kinetic energy operator and is obtained simply by multiplying $L(x)$ in the form in \eq{operator_L_x_coord} with $p(x)$.  Later on we will see that, because of these specialities, the formalism of building the SUSYQM algorithms of the base Legendre operator $L(x)$ becomes much simpler than many other types of hypergeometric-like differential operators.

\subsection{The Supersymmetric Factorization and Shape Invariance Symmetry of Base Legendre Equation}\label{Hermitian_factorization}

\subsubsection{The Supersymmetric Factorizations of Legendre Equation in Both $x$ and $y$ Coordinates}\label{MO_map_SUSY_factorization}

The Legendre function $\Phi_j(x)$ of $j$-th level is defined as the eigenfunction of the Legendre differential equation
\begin{eqnarray}\label{Legendre_equ_x_coord}
 L(x) \Phi_j(x)=\lambda_j \Phi_j(x),
\end{eqnarray}
where $\lambda_j$ is the corresponding eigenvalue.

Along with the supersymmetrization of $L(x)$ into $p(x)L(x)$ by multiplying $p(x)$, as shown in \eq{base_L_squared_form}, \eq{Legendre_equ_x_coord} is then turned into
\begin{eqnarray}\label{Legendre_equ_x_coord1}
 &&\sqbr{-p(x)L(x)-\lambda_j p(x)+E_j}\Phi_j(x)\nonumber \\
 &\equiv& \sqbr{-\smbr{p(x)\od{}{x}}^2+V_j(x)} \Phi_j(x),\nonumber \\
 &=& E_j \Phi_j(x),
\end{eqnarray}
where an eigenvalue term has been added on both sides so that this equation is formally an eigenequation with eigenvalue $E_j$, and
\begin{eqnarray}\label{Vj_definition}
V_j(x)\equiv-\lambda_j p(x)+E_j
\end{eqnarray}
 works as the ‘potential'.

In view that $p(x)=1-x^2$ and the potential $V_j(x)$ is also of the same type of binomial as $p(x)$, both of which lack the term linear in $x$, the operator inside the square brackets in \eq{Legendre_equ_x_coord1} formally admits the following \emph{supersymmetric} factorization,
\begin{eqnarray}\label{pL_form_SUSY_factorization}
&&-\smbr{p(x)\od{}{x}}^2-\lambda_j p(x)+E_j \nonumber \\
&=&\sqbr{-p(x)\od{}{x}+ e_j x}\sqbr{p(x)\od{}{x}+ e_j x},
\end{eqnarray}
where $e_j$ is a constant.

Expanding the right hand side of \eq{pL_form_SUSY_factorization} and balancing the quadratic and constant terms in $x$ on both sides, we obtain
\begin{eqnarray}\label{lambdaj_Ej_ej_relation}
 \lambda_j=e_j(e_j+1),\quad E_j=e^2_j.
\end{eqnarray}

With the momentum operator map \eqnm{momentum_operator_map1}  and identifying the \emph{common} function $e_j x$ appearing in the two factors in \eq{pL_form_SUSY_factorization} as the superpotential
\begin{eqnarray}\label{calWjy1}
 {\cal W}_j(y)\equiv W_j(x)=e_j x,
\end{eqnarray}
the supersymmetric factorization \eqnm{pL_form_SUSY_factorization} in $x$ coordinate then becomes the supersymmetric factorization of the standard Hamiltonian $\widehat{H}_j(y)$ in $y$ coordinate in the form
\begin{eqnarray}\label{wdhtHjy_SUSY_factorization}
 \widehat{H}_j(y)\equiv {\widehat A}^\dagger_j(y) {\widehat A}_j(y),
\end{eqnarray}
where
\begin{eqnarray}\label{ladder_operators0}
{\widehat A}^\dagger_j(y) \equiv-\od{}{y} + {\cal W}_j(y), \quad {\widehat A}_j(y) \equiv \od{}{y}+ {\cal W}_j(y).
\end{eqnarray}
Here ${\widehat A}^\dagger_j(y)$ and ${\widehat A}_j(y)$ are Hermitian adjoint to each other with respect to the inner product $(f_1(y),f_2(y))\equiv \int^{\infty}_{-\infty}dy f_1(y)f_2(y)$ for two well-behaved functions $f_1(y)$ and $f_2(y)$. However, at this stage, it is not clear that which one of these two operators works as the lowering/raising operator.

\eq{x_to_y_trans0} gives the coordinate transformation from $x\in(-1,\,1)$  to $y$ for $y \in (-\infty, \infty)$
\begin{eqnarray}\label{x_to_y_trans1}
 y(x)=\frac{1}{2} \ln\frac{1+x}{1-x} \equiv \arctanh x,
\end{eqnarray}
whose inverse transformation is $x=\tanh y$.

Under the momentum operator map \eqnm{momentum_operator_map1}, the eigenequation \eqnm{Legendre_equ_x_coord1} is mapped into the supersymmetric factorization
\begin{eqnarray}\label{type1_S_Equ_standard}
 \widehat{H}_j(y) \widehat{\Phi}_j(y) \equiv {\widehat A}^\dagger_j(y) {\widehat A}_j(y)\widehat{\Phi}_j(y)= e^2_j \widehat{\Phi}_j(y)
\end{eqnarray}
of a standard Schr\"odinger equation, where $\widehat{\Phi}_j(y)\equiv \Phi_j(x(y))$ and the potential ${\cal V}_j(y)$ of $\widehat{H}_j(y)$ is related to the superpotential ${\cal W}_j(y)$ through the Ricatti equation
\begin{eqnarray}\label{potential_Vj}
 {\cal V}_j(y) =-\od{{\cal W}_j(y)}{y}+{\cal W}^2_j(y).
\end{eqnarray}

In the above procedure of searching for the standard supersymmetric factorization of the operator $-p(x)L(x)-\lambda_j p(x)+ E_j$, we firstly reached its supersymmetric factorization within $x$ coordinate and next standardized it by the momentum operator map. Of course, we can switch the order of these two steps.

With the kinetic energy operator identified in \eq{KE_operator_map1} and its accompanying coordinate transformation \eqnm{x_to_y_trans1} in hand,  \eq{Legendre_equ_x_coord1} is mapped to the standard (unfactorized) Schr\"odinger equation
\begin{eqnarray}\label{Sch_equ_y_coord}
 \widehat{H}_j(y) \widehat{\Phi}_j(y)\equiv \sqbr{-\frac{d^2}{d y^2}+ {\cal V}_j(y)}\widehat{\Phi}_j(y) = E_j\widehat{\Phi}_j(y)
\end{eqnarray}
where the potential
\begin{eqnarray}\label{Vj_definition1}
 {\cal V}_j(y)\equiv V_j\sqbr{x(y)} = -\lambda_j \sech^2(y) + E_j,
\end{eqnarray}
is of the P\"oschl-Teller-type, the $y$-coordinate version of $V_j(x)$ given in \eq{Vj_definition}.

In order to find out the supersymmetric factorization of \eq{Sch_equ_y_coord} directly in $y$ coordinate, the analytical form of the superpotential ${\cal W}_j(y)$ has to be fixed for the potential ${\cal V}_j(y)$ given in \eq{Vj_definition1}, trough the relation \eqnm{potential_Vj}. Note that ${\cal V}_j(y)$ is linear in $\tanh^2 y$ in view that $\sech^2 y =1-\tanh^2y$ and $d\tanh y/dy=1-\tanh^2y$. \eq{potential_Vj} then holds in such a special way that each term is linear in $\tanh^2 y$.  This consideration  restricts ${\cal W}_j(y)$ in the form
\begin{eqnarray}\label{cal_Wj(y)}
 {\cal W}_j(y)= e_j\tanh y
\end{eqnarray}
for some constant $e_j$. This form of ${\cal W}_j(y)$ is consistent with that given in \eq{calWjy1}, in view that $x=\tanh y$. Consequently, the combination of Eqs. \eqnm{potential_Vj}and \eqnm{Vj_definition1} leads to the expressions in \eq{lambdaj_Ej_ej_relation} for $\lambda_j$ and $E_j$ in terms of $e_j$.

\subsubsection{Comparison with the Traditional Method of Passing from a Linear Second-order Differential Equation to a Schr\"odinger Equation}

We have introduced the approach of reaching the supersymmetric factorization of a linear differential operator of second order (in particular, the hypergeometric-type operator) that consists of an active supersymmetrization transformation and a momentum operator map. At this point, it is worthy of comparing this approach with the traditional one of passing from a linear second-order differential equation to a standard Schr\"odinger equation. One example of the latter case for the Legendre equation is given in the first part of Appendix \ref{traditional_analytical_method}.

Traditionally, in order to turn a linear differential equation of second order into a Schr\"odinger type equation, people start with the \emph{expanded / broken} form of the linear differential operator, such as the form of $L(x)$ given after the second equal sign in \eq{operator_L_x_coord}, in which the first-order differentiation is singled out, and then aim at eliminating such first-order differentiation. Concretely, one conceptually treats the eigenfunction $\Phi(x)$ as a composite function $\widehat{\Phi}\smbr{y(x)}$, with $y$ being an intermediate coordinate as yet to be determined, then by the chain rule of differentiation, one obtains a differential equation of second order for $\widehat{\Phi}\smbr{y}$ with respect to the variable $y$. In general, this new differential equation consists of $y$-derivatives of first as well as second orders.  Forcing the coefficient of the first order $y$-derivative $\widehat{\Phi}'(y)$ to be vanishing then results in a new differential equation of second order for $y(x)$ with respect to $x$ coordinate. Upon solving this new differential equation, the coordinate transformation $y(x)$ is fixed. This traditional method is full of the flavor of \emph{analytical} mathematics.

Although the differential equation for $y(x)$ like \eq{ODE_2nd_order} is usually not hard to solve, this approach of obtaining the Shr\"odinger equation is not only mathematically march more involved, but is also not illuminating at all from the quantum mechanical perspective. This method does not pay attention to or ignores the \emph{defining differential structure} of a quantum Hamiltonian operator in non-relativistic quantum mechanics \text{---} the kinetic energy operator, which is the square of a momentum operator, a construct one could prepare from the scratch \emph{in an elementary algebraic way}, as we have shown in Subsection \ref{AST_hypergeometric_operator} for hypergeometric-type operator.

Our method that consists of supersymmetrization plus momentum operator map not only allows us to quickly build the standard SUSYQM for $L(x)$ shown in the previous subsubsection, we will see in Subsection \ref{second_momentum_map} that it also guides us to quickly pin down another type of supersymmetric factorization, hence another type of SUSYQM for the \emph{same} $L(x)$. Furthermore, this method is not only powerful in quickly transforming a large class of line differential equations of second order into certain Schr\"odinger equations that come out automatically in their partially or fully supersymmetrized forms, but also powerful in quickly finding out \emph{consecutive} SUSYQM algorithms of exhausting the spectra of these operators in pure algebraic ways, when an other symmetry \text{--} shape invariance \text{--} is also incorporated. An example of such symmetry we will come to in the next subsubsection. In view of the conceptual simplicity, the operational efficiency and the full flavor of \emph{algebraic} quantum mechanics of this method, we think that it is worthy of being advocated.

\subsubsection{Shape Invariance Symmetry of the Potential ${\cal V}_j(y)$}\label{SIS_Vjy}

Intertwiningly acting \eq{type1_S_Equ_standard} with ${\widehat A}_j(y)$ \text{--} the right factor of ${\widehat H}_j(y)$ in \eq{wdhtHjy_SUSY_factorization} \text{--} yields its superpartner Schr\"odinger equation
\begin{eqnarray}\label{sp_eigen_equ_before_rnmlz_y_coord}
 \widehat{H}^s_j(y) \widehat{\Phi}^s_j(y) = E_j \widehat{\Phi}^s_j(y),
\end{eqnarray}
where
\begin{eqnarray}\label{superpartner_eigenfunction_y_coord}
\widehat{\Phi}^s_j(y)\propto {\widehat A}_j(y) \widehat{\Phi}_j(y),
\end{eqnarray}
and
\begin{eqnarray}
\widehat{H}^s_j(y) \equiv {\widehat A}_j(y) {\widehat A}^\dagger_j(y)= -\frac{d^2}{d y^2} +{\cal V}^s_j(y)
\end{eqnarray}
is the superpartner Hamiltonian of $\widehat{H}_j(y) \equiv {{\widehat A}^\dagger_j(y) \widehat A}_j(y)$, and
\begin{eqnarray}\label{potential_Vsj}
 {\cal V}^s_j(y)\equiv \od{{\cal W}_j(y)}{y}+{\cal W}^2_j(y)=p\od{W_j(x)}{x}+W^2_j(x)
\end{eqnarray}
is the superpartner potential of ${\cal V}_j(y)$, according to the definitions of ${\widehat A}_j(y)$ and ${\widehat A}^\dagger_j(y)$  in \eq{ladder_operators0}.

One key idea of constructing a consecutive SUSYQM algorithm for exhausting the spectra of a hierarchy of the Hamiltonian pair $\widehat{H}_j(y)$ and $\widehat{H}^s_j(y)$ lies in that, formally both ${\cal V}^s_j(y)$ and ${\cal V}_{j\pm 1}(y)$ are \emph{functions of the  same type}, except that the superpartner potential ${\cal V}^s_j(y)$ differs from its nearest neighboring raw potential ${\cal V}_{j-1}(y)$ or ${\cal V}_{j+1}(y)$ simply by a certain constant. Such property owned by the potential series $\left\{{\cal V}_j(y)\right\}$ is termed \emph{shape invariance} symmetry \cite{Gendenshtein_SUSY_Schr_Equ_1983}. The exact role played by this symmetry will become clear soon. For instance, the following \emph{downward-connected} shape invariance relation between the potential pair ${\cal V}^s_j(y)$ and ${\cal V}_{j-1}(y)$ can be imposed,
\begin{eqnarray}\label{shape_invariance_base_level}
{\cal V}^s_j(y)-{\cal V}_{j-1}(y)=\Delta_j,
\end{eqnarray}
where $\Delta_j$ is a $j$-dependent constant as yet to be determined.

Upon adding and subtracting the kinetic energy operator $-d^2/d y^2$ on the left hand side, the shape invariance condition
\eqnm{shape_invariance_base_level} for the potential series is rewritten as the following \emph{quasi-commutator} relation
\begin{eqnarray}\label{SI_y_coord}
\widehat{H}^s_j(y)-\widehat{H}_{j-1}(y)&=&\left[{\widehat A}(y),\, {\widehat A}^\dagger(y)\right]^j_{j-1},\nonumber \\
&\equiv& {\widehat A}_j(y) {\widehat A}^\dagger_j(y)-{\widehat A}^\dagger_{j-1}(y) {\widehat A}_{j-1}(y),\nonumber \\
&=&\Delta_j,
\end{eqnarray}
where we have introduced the quasi-commutator notation $\left[{\widehat A}(y),\, {\widehat A}^\dagger(y)\right]^j_{j-1}$, which would reduce to the usual commutator if the superscript and the subscript were the same. We will see that in practice the quasi-commutator relation is more convenient than the shape invariance relation \eqnm{shape_invariance_base_level} of the potential series.  The quasi-commutator relation \eqnm{SI_y_coord} essentially says that the superpartner Hamiltonian of $j$-th level is different from the raw Hamiltonian of the nearest neighboring $(j-1)$-th level only by the constant $\Delta_j$. Such symmetry relation allows us to replace $\widehat{H}^s_j(y) = {\widehat A}_j(y) {\widehat A}^\dagger_j(y)$ in favor of $\widehat{H}_{j-1}(y)={\widehat A}^\dagger_{j-1}(y) {\widehat A}_{j-1}(y)$, so that the superpartner eigenequation \eqnm{sp_eigen_equ_before_rnmlz_y_coord} is \emph{renormalized} into
\begin{eqnarray}\label{superparnter_eigen_equ_after_rnmlz_y_coord}
 \widehat{H}_{j-1}(y)\Phi_{j-1}(y) = E_{j-1}\Phi_{j-1}(y),
\end{eqnarray}
where
\begin{eqnarray}
E_{j-1}\equiv E_j - \Delta_j
\end{eqnarray}
and
\begin{eqnarray}\label{A_lowering_relation}
\Phi_{j-1}(y)\equiv \Phi^s_j(y)\propto {\widehat A}_j(y) \widehat{\Phi}_j(y).
\end{eqnarray}
That is, the superpartner eigenequation \eqnm{sp_eigen_equ_before_rnmlz_y_coord} is nothing but the $(j-1)$-th version of the raw eigenequation. In other words, the shape invariance symmetry of the superpartner potential ${\cal V}^s_j(y)$ ensures the \emph{pattern invariance} (self-similarity) of the corresponding superpartner eigenequation with reference to the raw eigenequation. It is for this very reason that we will use shape invariance and pattern invariance interchangeably.

Such a process of turning the superpartner eigenequation of a raw eigenequation labeled by $j$ into the nearest neighboring one labeled by $j-1$ or $j+1$ is termed \emph{renormalization} in this paper, as the energy renormalization is involved and the form of the superpartner eigenequation is reshaped. We will see that this \emph{renormalization} plus the \emph{intertwining action} allow us to construct an iterative procedure in which we can keep going to smaller and smaller indices $j$  and eventually exhaust all the discrete eigen-spectra of the Legendre operator $L(x)$. According to the relation \eqnm{A_lowering_relation}, ${\widehat A}_j(y)$ works as the lowering operator that lowers the eigenfunction and eigenvalue by one level, and correspondingly ${\widehat A}^\dagger_j(y)$ works as the raising operator. This algorithm raises the hope to obtain all $\Phi_j(y)$ and $E_j$, by applying the raising operators repeatedly once the eigenfunction $\Phi_0(y)$ and eigenvalue $E_0$ of lowest level are easily identified, or by applying the lowering operators repeatedly once the eigenfunction $\Phi_{j_{max}}(y)$ and eigenvalue $E_{j_{max}}$ of highest level $j_{max}$ are easily identified.

It is worthy of pointing out that, identifying the ladder operators $\widehat{A}_j(y)$ and $\widehat{A}^\dagger_j(y)$ as a lowering operator and raising operator respectively is closely related to the particular choice of the \emph{downward-connected} shape invariance condition \eqnm{shape_invariance_base_level}. If one choose an \emph{upward-connected} shape invariance condition that relates ${\cal V}^s_j(y)$ to ${\cal V}_{j+1}(y)$, the roles played by $\widehat{A}_j(y)$ and $\widehat{A}^\dagger_j(y)$ will be switched.

One may notice that we do not care about the difference between ${\cal V}^s_j(y)$ and ${\cal V}_j(y)$ of the \emph{same} $j$-th level. In fact, in general these two functions may not be of the same shape. For instance, Eqs. \eqnm{potential_Vj} and \eqnm{potential_Vsj} together gives ${\cal V}^s_j(y)-{\cal V}_j(y)=2 d{\cal W}_j(y)d/dy =2p(x)d W_j(x)/dx$, which is not a constant for all $j$. In other words, we do not pay attention to the value of the usual commutator $\left[\widehat{A}_j(y),\,\widehat{A}^\dagger_j(y)\right]$ of a \emph{single} $j$-th level, instead, we do care about the value of the quasi-commutator $\left[\widehat{A}_j(y),\,\widehat{A}^\dagger_j(y)\right]^j_{j-1}$ between the two neighboring $j$-th and $(j-1)$-th levels, since this type relation help us to establish the \emph{cross-level} relation between the superpartner eigenequation of $j$-th level and the raw eigenequation of $(j-1)$-th level, which is what we need for constructing a SUSYQM algorithm.

In \eq{Vj_definition}, $\lambda_j$ and $E_j$ are legitimate to be vanishing for some $j$. Denoting such subscript as $j=0$, we have
\begin{eqnarray}\label{initial_lambda_and_E}
 \lambda_0=0,\quad E_0=0.
\end{eqnarray}
Consequently, $V_0(x)=0$, and because of the linearity of $W_j(x)\equiv e_j x$ (\eq{calWjy1}), we have to take $W_0(x)=0$. This implies that
\begin{eqnarray}\label{e0}
 e_0=0.
\end{eqnarray}
We have pointed out in the last paragraph of Subsection \ref{AST_hypergeometric_operator} that $W_0(x)=0$ is attributed to the specialty $q(x)=p'(x)$ for Legendre operator $L(x)$ as a special hypergeometric-like operator.

The eigenfunction $\Phi_0(x)$ is determined by the equation $pL(x)\Phi_0(x)=d\Phi_0(x)/dx=0$, which has the solution
\begin{eqnarray}
\Phi_0(x)=\text{const.}.
\end{eqnarray}
With this initial eigensolution $\Phi_0$ and the initial values in Eqs. \eqnm{initial_lambda_and_E} and \eqnm{e0} for the supersymmetric factorization in hand, we will see in the next subsubsection that, by running the iterative SUSYQM algorithm, the eigenspectra of all levels can be built.

\subsubsection{Constructing the Eigenspectrum of $L(x)$ in $y$ Coordinate}\label{detailed_factorization_scheme_y_coordinate}

In the above subsubsection, we have just obtained the initial data $E_0=\lambda_0=0$, $e_0=0$ and $\widehat{\Phi}_0(y)=\Phi_0(x)=\text{const.}$ for $j=0$. In the following, we will work out the data for all $j\ge1$.

In $y$ coordinate, making use of the identity $\sech^2y=1-\tanh^2y$, the potential ${\cal V}_j(y)$ in \eq{Vj_definition1} can be recast into
\begin{eqnarray}\label{Vj_original}
 {\cal V}_j(y)= \lambda_j \tanh^2y + E_j-\lambda_j.
\end{eqnarray}

On the other hand, upon substituting for ${\cal W}_j(y)$ via \eq{cal_Wj(y)}, \eq{potential_Vj} is then rewritten as
\begin{eqnarray}\label{Vj_Riccarti}
 {\cal V}_j(y) = e_j(e_j+1)\tanh^2y - e_j.
\end{eqnarray}

Comparing \eq{Vj_Riccarti} with \eq{Vj_original}, we have
\begin{eqnarray}\label{ej_lambdaj_relation}
 \begin{aligned}
  e_j(e_j+1)&=\lambda_j,\\
  e_j&=\lambda_j-E_j.
 \end{aligned}
\end{eqnarray}
These are consistent with the relations in \eq{lambdaj_Ej_ej_relation} that are obtained in $x$ coordinate.

Upon substituting for ${\cal W}_j(y)$ via \eq{cal_Wj(y)} in \eq{potential_Vsj}, the superpartner potential of ${\cal V}_j(y)$ is obtained as
\begin{eqnarray}\label{VsJ_reduced}
 {\cal V}^s_j(y)= e_j(e_j-1)\tanh^2y +e_j.
\end{eqnarray}

By making substitutions for ${\cal V}^s_j(y)$ and ${\cal V}_{j-1}(y)$ in \eq{shape_invariance_base_level}, which reflects the downward-connected shape invariance symmetry for $\{{\cal V}_j(y)\}$, via \eq{VsJ_reduced} and \eq{Vj_Riccarti}, respectively, we obtain
\begin{eqnarray}\label{shape_invariance_base_level1}
\sqbr{e_j(e_j-1)-e_{j-1}(e_{j-1}+1)}\tanh^2y + e_j+e_{j-1} =\Delta_j. \nonumber \\
\end{eqnarray}
This equation holds for arbitrary $y$, the coefficient of $\tanh^2y$ must be vanishing and the rest constant terms balance themselves. This yields the recurrence relations for series $\{e_i\}$ as
\begin{eqnarray}\label{ej_recurrence}
\begin{aligned}
 e_j&=e_{j-1}+1,\\
 e_j+e_{j-1} &= \Delta_j.
 \end{aligned}
\end{eqnarray}

With $e_0$ given in \eq{e0}, \eq{ej_recurrence} is then solved and we obtain
\begin{eqnarray}\label{e_j_value}
\begin{aligned}
 e_j&=j,\\
 \Delta_j&=2j-1
 \end{aligned}
\end{eqnarray}
for $j$ being non-negative integer.

Substituting $e_j=j$ in \eq{ej_lambdaj_relation} further, we obtain the eigenvalues as
\begin{eqnarray}\label{eigenvalues_lambda_and_E}
\begin{aligned}
 \lambda_j&=j(j+1),\\
 E_j&=j^2.
 \end{aligned}
\end{eqnarray}

In this algorithm, we see that it is the combination of shape invariance symmetry of the potential and the formal supersymmetric factorization (in particular, the Ricatti equation) that forces the eigenvalues $\lambda_j$ and $E_j$ to be quantized in pure algebraic ways, which is in sharp contrast to how the eigenvalue $\lambda_j$ is quantized in the traditional power series method.

With ${\cal W}_j(y)$ given in \eq{cal_Wj(y)} and $e_j$ in \eq{e_j_value}, ${\widehat A}_j(y)$ and ${\widehat A}^\dagger_j(y) $ in \eqnm{ladder_operators0} become
\begin{eqnarray}\label{ladder_operators_hat_Aj_and_hat_Aj}
\begin{aligned}
{\widehat A}_j(y) &= \od{}{y}+ j \tanh{y},\\
                  &=\cosh^{-j} y \od{}{y} \cosh^j y, \\
                  &={\cal S}^{2j}_y\smbr{\od{}{y}},\\
{\widehat A}^\dagger_j(y) &=-\od{}{y} + j \tanh{y},\\
                          &=- \cosh^j y \od{}{y} \cosh^{-j} y,\\
                          &={\cal S}^{-2j}_y\smbr{-\od{}{y}},
\end{aligned}
\end{eqnarray}
where ${\cal S}^{2j}_y(-)$ is the $2j$-th power of the bilateral transformation ${\cal S}_y(-)\equiv\cosh^{-1/2}y(-)\cosh^{1/2}y$ and ${\cal S}^{-2j}_y(-)$ is the $2j$-th power of the corresponding inverse transformation ${\cal S}^{-1}_y(-)\equiv\cosh^{1/2}y(-)\cosh^{-1/2}y$.

Choosing a particular normalization, the proportional relation \eqnm{superpartner_eigenfunction_y_coord} then becomes the lowering recurrence relation
\begin{eqnarray}\label{lowering_recurrence_relation_y_coord}
 \widehat{\Phi}_{j-1}(y) =\frac{1}{j}{\widehat A}_j(y)\;\widehat{\Phi}_j(y)
\end{eqnarray}
for $j\ge 1$.

The combination of Eqs. \eqnm{type1_S_Equ_standard}, \eqnm{eigenvalues_lambda_and_E} and \eqnm{lowering_recurrence_relation_y_coord} then yields the raising recurrence relation
\begin{eqnarray}\label{base_raising_recurrence_y_coord}
 \widehat{\Phi}_j(y) = \frac{1}{j}{\widehat A}^\dagger_j(y) \;\widehat{\Phi}_{j-1}(y)
\end{eqnarray}
for $j\ge 1$.

The self-adjoint of $L(x)$ / $J_0(\theta)$ in $x$ / $\theta$ coordinate is defined with respect to the measure $dx$ / $\sin\theta d\theta$. Under the momentum operator map \eqnm{momentum_operator_map1} or the equivalent coordinate transformation $x=\tanh y$, there is the measure map
\begin{eqnarray}
dx =\sech^2 y\, dy
\end{eqnarray}
from $x$ coordinate to $y$ coordinate. Later on, we will introduce $\Theta_{00}(\theta)=1/\sqrt{2}$ and $\Phi_{00}(x)=1/\sqrt{2}$, which are respectively the $\theta$ and $x$-coordinate incarnations of $\widehat{\Phi}_0(y)$. This $\Theta_{00}(\theta)$ has normalized to unity with respect to the  measure  $\sin\theta d\theta$, and this $\Phi_{00}(x)$ has been normalized to unity with respect to the measure $dx$. In order that $\int^{\infty}_{-\infty} d y \sech^2y \widehat{\Phi}^2_0(y)=1$ with respect to the measure $\sech^2y\, dy$, we choose $\widehat{\Phi}_0(y)=1/\sqrt{2}$.

By consecutively applying the raising recurrence relation \eqnm{base_raising_recurrence_y_coord} for $j\rightarrow j,\; j-1,\;\cdots,\;2,\;1$, we arrive at the general expression for Legendre polynomial of $j$-th order
\begin{eqnarray}\label{L_polynomial_lth_y_coord}
 \widehat{\Phi}_j(y)&=&\frac{1}{j!}\widehat{A}^\dagger_j\cdot \widehat{A}^\dagger_{j-1}\cdots \widehat{A}^\dagger_2 \cdot \widehat{A}^\dagger_1 \cdot \widehat{\Phi}_0,\nonumber \\
                    &=&\frac{(-1)^j}{\sqrt{2}j!} \cosh^j y \smbr{\od{}{y} \cosh^{-1}y}^j \cdot 1,
\end{eqnarray}
where to shorten the long operator product, the lumped form of $\widehat{A}^\dagger_j$ in \eq{ladder_operators_hat_Aj_and_hat_Aj} has been made use of.

Finally, we just mention the very deep reason why the symmetry carried by the Hamiltonian $\widehat{H}_j(y)$ and its superpartner $\widehat{H}^s_j(y)$ or by the pair of eigenequations \eqnm{type1_S_Equ_standard} and \eqnm{sp_eigen_equ_before_rnmlz_y_coord} is called supersymmetry. These two equations are just the two component eigenequations of the eigenequation of a super-Hamiltonian $\widehat{H}^{ss}_j$ that is a matrix operator with $\widehat{H}_j(y)$ and $\widehat{H}^s_j(y)$ diagonally distributed, and such super-Hamiltonian and the \emph{two} (${\cal N}=2$) supercharges $\widehat{Q}^+_j$ and $\widehat{Q}^+_j$ together form an $su(1,1)$ superalgebra, hence the corresponding symmetry is called ${\cal N}=2$ $su(1,1)$ supersymmetry. See the detailed supercharge formulation of this symmetry algebra in Appendix \ref{super_charge_formulation}.

\subsection{The $x$-coordinate Representation of the SUSYQM Algorithm of the Base Legendre Operator}\label{nonhermitian_factorization}

In Subsection \ref{Hermitian_factorization}, we have successfully built the SUSYQM algorithm for the spectrum of the base Legendre operator in the $y$ coordinate. The advantage of this $y$-coordinate representation lies in that, at each level, both the raw eigenequation and its superpartner eigen-equation assume the familiar standard form of Schr\"odinger equation, and the supersymmetric quantum mechanic structure is explicitly demonstrated. However, for the convenience of application, one has to go back to $x$ coordinate, by the inverse momentum map or the inverse coordinate transformation from $y$ coordinate to $x$ coordinate. One may wonder if the SUSYQM algorithm of the base Legendre operator $L(x)$ can be established \emph{within} the $x$ coordinate. As pointed out in Subsection \ref{AST_hypergeometric_operator} and in  \cite{Zhou_SUSY_hypergeometric_operator_2024}, for generic hypergeometric-like operator, there is a $1$-to-$1$ correspondence between the SUSYQM algorithm represented in the $x$ coordinate and the standard one represented in the $y$ coordinate. In general, one can transform back from $y$-coordinate representation to $x$-coordinate representation directly by the combination of the inverse momentum operator map and the inverse of supersymmetrization transformation. However, since $L(x)$ is such a simple hypergeometric-type operator that its own specialties bring around several extreme simplifications, e.g., the active supersymmetrization transformation is trivial, as pointed out at the end of Subsection \ref{AST_hypergeometric_operator}, it is quite straightforward to build the SUSYQM algorithm \emph{directly} in $x$-coordinate, \emph{without reference to} the standard $y$-coordinate representation of this algorithm.

In $x$ coordinate, starting with \eq{Legendre_equ_x_coord}, by the same reasoning for preparing the kinetic energy operator $-\smbr{p(x)d/dx}^2$, we have reached the supersymmetric factorization \eqnm{pL_form_SUSY_factorization} of $p(x)L(x)$
and the relations in \eq{lambdaj_Ej_ej_relation}.  That is, \eq{Legendre_equ_x_coord1} is supersymmetrically factorized into
\begin{eqnarray}\label{Legendre_equ_non_Hermitian_factorized0}
B_jA_j \Phi_j &= E_j\Phi_j,
\end{eqnarray}
where $A_j$ and $B_j$ are designed as
\begin{eqnarray}\label{Aj-Bj}
\begin{aligned}
 B_j &=-A_0+ W_j(x),\\
A_j &=A_0+W_j(x)
 \end{aligned}
\end{eqnarray}
for $A_0(x)\equiv p(x) d/dx$ and
\begin{eqnarray}
 W_j(x)\equiv e_j x.
\end{eqnarray}
Here we see that one advantage of $x$-coordinate representation lies in that the superpotential $W_j(x)$ is as simple as being proportional to $x$, which is a special case of the linear superpotential for hypergeometric-like operator $h(x)$. That is, this fits into the general feature of the superpotential  for the supersymetric factorization of $p(x)h(x)$, a linear function of $x$ \cite{Zhou_SUSY_hypergeometric_operator_2024}. The other advantage is that, since we stay within $x$ coordinate, we actually do not perform the momentum operator map \eqnm{momentum_operator_map1}, though we still conceptually view $- \smbr{p(x)d/dx}^2$ as the kinetic energy operator.

From this point on, instead of going to $y$ coordinate, we stay \emph{within} $x$ coordinate and complete the rest of the SUSYQM algorithm.

Intertwiningly acting on the eigenequation in \eqnm{Legendre_equ_non_Hermitian_factorized0} with $A_j$, we obtain the superparter eigenequation for $A_j B_j$ in the form
\begin{eqnarray}\label{A_B_A_product}
A_j B_j \Phi_{j-1} =E_j \Phi_{j-1},
\end{eqnarray}
where $\Phi_{j-1}\propto A_j \Phi_j$.  We can choose a normalization so that this proportionality relation is promoted to the lowering recurrence relation
\begin{eqnarray}\label{Phij_minus1_lowering_recurrence}
\Phi_{j-1} =\frac{1}{\sqrt{E_j}} A_j \Phi_j
\end{eqnarray}
for all $j\ge1$, where $A_j$ is the lowering ladder operator.

The combination of \eq{Phij_minus1_lowering_recurrence} and \eq{Legendre_equ_non_Hermitian_factorized0} then yields the raising recurrence relation
\begin{eqnarray}\label{Phij_raising_recurrence}
\Phi_j =\frac{1}{\sqrt{E_j}} B_j \Phi_{j-1}
\end{eqnarray}
for $l\ge 1$, where $B_j$ is the raising ladder operator.

In order to renormalize the superpartner eigenequation \eqnm{A_B_A_product} into
\begin{eqnarray}\label{BA_j_minus_1_eign_equ}
 B_{j-1} A_{j-1}\Phi_{j-1} =E_{j-1}\Phi_{j-1},
\end{eqnarray}
which assumes the same pattern as the original eigenequation \eqnm{Legendre_equ_non_Hermitian_factorized0} (in fact, which is just the $(j-1)$-th version of the latter), the downward-connected pattern invariance condition
\begin{eqnarray}\label{shape_invariance_condition_x_coord}
  A_j B_j - B_{j-1}A_{j-1}= \Delta_j,
\end{eqnarray}
is needed  for some constant $\Delta_j$. Substituting for $A_jB_j$ via \eq{shape_invariance_condition_x_coord}, \eq{A_B_A_product} is renormalized into \eq{BA_j_minus_1_eign_equ} and $E_{j-1}=E_j-\Delta_j$.

Substitute for $A_j$, $B_j$,  $A_{j-1}$ and $B_{j-1}$ in \eq{shape_invariance_condition_x_coord} via \eq{Aj-Bj} and then balance the coefficients of $x^2$ and $x^0$, we obtain
\begin{eqnarray}\label{e_recurrenc_relation}
 \begin{aligned}
 e_j+e_{j-1} &=\Delta_j,\\
 e_j&=e_{j-1} +1.
 \end{aligned}
\end{eqnarray}

For $j=0$, \eq{Legendre_equ_non_Hermitian_factorized0} reduces to $A_0 \Phi_0=0$ when taking $\lambda_0=0$, which has constant solution $\Phi_0$. With $\lambda_0=0$, \eq{lambdaj_Ej_ej_relation} implies that $e_0=0$ and $E_0=0$. With these initial values,  \eq{e_recurrenc_relation} is then solved as
\begin{eqnarray}\label{ej}
 \begin{aligned}
  e_j&=j,\\
  \Delta_j&=2j-1.
 \end{aligned}
\end{eqnarray}
It then follows from \eq{lambdaj_Ej_ej_relation} that
\begin{eqnarray}\label{base_eigen_values}
 \begin{aligned}
 \lambda_j&=j(j+1),\\
       E_j&=j^2.
 \end{aligned}
\end{eqnarray}

 Now the two ladder operators in \eq{Aj-Bj} can be rewritten as
\begin{eqnarray}\label{Bj_Aj_ladder_operators}
\begin{aligned}
 B_j&=-p\od{}{x}+ j\; x=-\sqrt{p}^{-j}\smbr{p\od{}{x}}\sqrt{p}^j,\\
 A_j&=p\od{}{x} + j\; x = \sqrt{p}^j\smbr{p\od{}{x}}\sqrt{p}^{-j}.
\end{aligned}
\end{eqnarray}

Consecutively using the raising recurrence relation \eqnm{Phij_raising_recurrence}  and the lumped form of $B_j$ in \eq{Bj_Aj_ladder_operators} for $j\rightarrow j,\; j-1,\;\cdots,\;2\;\text{and}\;1$, we arrive at
\begin{eqnarray}\label{unity_normed_Legendre_polynomial}
 \Phi_j(x) &=&\frac{1}{j!}B_jB_{j-1}\cdots B_2 B_1 \Phi_0, \nonumber \\
        &=&\frac{\smbr{-1}^j}{j!} \sqrt{p}^{-j}\smbr{p \od{}{x}\sqrt{p}}^j\Phi_0.
\end{eqnarray}

In \eq{unity_normed_Legendre_polynomial}, if we choose $\Phi_0=1/\sqrt{2}$, we have $\Phi_j(x)=\widehat{\Phi}_j(y)$ with $\widehat{\Phi}_j(y)$ given by \eq{L_polynomial_lth_y_coord} and $x=\tanh y$. Such $\Phi_j(x)$ is the Legendre polynomial with unity norm. In contrast, the usual Legendre function $P_j(x)$ is not normalized to unity.

If, instead of the normalization adopted in \eq{Phij_minus1_lowering_recurrence}, we choose a normalization such that the lowering recurrence relation is
\begin{eqnarray}\label{Phij_minus1_lowering_recurrence1}
\Phi_{j-1} =\frac{1}{E_j} A_j \Phi_j,
\end{eqnarray}
then it follows from \eq{Legendre_equ_non_Hermitian_factorized0} the raising recurrence relation
\begin{eqnarray}\label{Phij_raising_recurrence1}
 \Phi_j =B_j \Phi_{j-1}.
\end{eqnarray}

Consecutively applying \eq{Phij_raising_recurrence1} for $j\rightarrow j,\; j-1,\;\cdots,\;2\;\text{and}\;1$, we obtain
\begin{eqnarray}\label{non_normalized_Legendre_polynomial}
 \Phi_j(x) &=&B_jB_{j-1}\cdots B_2 B_1 \Phi_0, \nonumber \\
        &=&\smbr{-1}^j \sqrt{p}^{-j}\smbr{p \od{}{x}\sqrt{p}}^j\Phi_0,
\end{eqnarray}
which is an analog of \eq{unity_normed_Legendre_polynomial}. Choosing $\Phi_0=1$, \eq{non_normalized_Legendre_polynomial} then defines the usual Legendre polynomial $\Phi_j(x)=P_j(x)$.

Combining \eq{Phij_minus1_lowering_recurrence}, the $(j+1)$-th version of \eq{Phij_raising_recurrence}, and \eq{Bj_Aj_ladder_operators}, we obtain the well-known three-term recurrence relations for Legendre polynomials as
\begin{eqnarray}\label{three-term_relations}
\begin{aligned}
 &\smbr{2 p(x) \od{}{x}+ x}\Phi_j=\smbr{j+1}\Phi_{j+1} - j\Phi_{j-1},\\
 &\smbr{j+1}\Phi_{j+1} -\smbr{2j+1}\Phi_j+j \Phi_{j-1} =0.
\end{aligned}
\end{eqnarray}

We know that $L(x) = \sqbr{\smbr{-d/dx}\sqrt{p}}\sqbr{\sqrt{p}d/dx}$ is a self-adjoint factorization with trivial weight function $w(x)=1$, i.e., $L^\dagger(x)=L(x)$ , with respect to the inner product $(f_1, f_2)\equiv\int^1_{-1} dx f_1(x) f_2(x)$, which is unweighted or weighted by $w(x)=1$. However, with respect to this inner product, the splitting $p(x)L(x)=B_0 A_0 $ with $-B_0=A_0\equiv p(x)d/dx$ is not a mutually adjoint factorization. In fact, the conjugate of $A_0$ is $A^\dagger_0=-A_0+2x=B_2\ne B_0$. Consequently, according to \eq{Aj-Bj}, $A^\dagger_j=B_{j+2}\ne B_j$ for all nonnegative $j$, and the factorization $B_j A_j$ in \eqnm{Legendre_equ_non_Hermitian_factorized0} is \emph{not} a product of mutually adjoint pair. However, this non-adjointness of the factorization \eqnm{Legendre_equ_non_Hermitian_factorized0} does not mean that it does have the SUSYQM structure in the $x$ coordinate representation. Indeed, there are \emph{one-to-one} correspondences between the SUSYQM structures in the $x$ and $y$ coordinate representations, established via the momentum operator map \eqnm{momentum_operator_map1} (and the trivial active supersymmetrization transformation). For instance, the non-adjoint pair of ladder operators $B_j$ and $A_j$ in \eq{Aj-Bj} correspond respectively to the mutually adjoint ladder operator pair $\widehat{A}^\dagger_j$ and $\widehat{A}_j$ in \eq{ladder_operators0}, the shape invariance condition \eqnm{shape_invariance_condition_x_coord} to \eqnm{SI_y_coord}, the ladder recurrence relations \eqnm{Phij_minus1_lowering_recurrence} and \eqnm{Phij_raising_recurrence}  respectively to \eqnm{lowering_recurrence_relation_y_coord}  and \eqnm{base_raising_recurrence_y_coord}, etc. Because of such correspondence, the SUSYQM algorithms can be built in either coordinates with almost the same amount of work. We will call the $x$-coordinate representation the \emph{implicit/non-standard} representation of the SUSYQM algorithm, and the $y$-coordinate \emph{explicit/standard} representation of it.

Bazeia and Das \cite{Baziea_SUSY_Legendre_2012} have achieved a factorization amounting to $B^+_{j+2}B_j\Phi_{j-1}=j^2\Phi_{j-1}$ in terms of our notations, which is just \eq{A_B_A_product}, in view that $A^\dagger_j=B^+_{j+2}$ and $E_j=j^2$. However, their method rests on the \emph{three-term} recursion relations of Legendre polynomials in \eq{three-term_relations}  as the inputs for deriving the \emph{two-term} raising and lowering recurrence relations. We know that in the past the three-term recursion relations come out as the results of traditional approaches, such as via the generating function or via the generic solution coming out of power series method. In this sense, Bazeia and Das's method essentially depends on these traditional approaches. In contrast, in our approach these three-term recurrence relations given in \eq{three-term_relations} come out as the byproduct of the SUSYQM algorithm itself, instead of as the inputs for constructing the ladder operators and the SUSYQM algorithm.

It is worthy of noticing that, parallel to the $x$-coordinate representation of the SUSYQM algorithm of the base Legendre operator, there is the $\theta$-coordinate counterpart. The latter can be translated from the former by the coordinate transformation $x=-\cos\theta$ or can be directly built in $\theta$ coordinate (see Appendix \ref{theta_coord_J0} for the direct construction of this algorithm in $\theta$-coordinate representation).

\subsection{The Explicit Bottom-up Iterative SUSYQM Algorithm Represented in $x$ Coordinate for Simultaneously Building Legendre Equation as Well as Its Eigensolution}\label{inductive_iterative_base_Legendre_function}

In Sububsection \ref{MO_map_SUSY_factorization} and Subsection \ref{nonhermitian_factorization}, as the Legendre equation \eq{Legendre_equ_x_coord} of generic level $j$ is given as an input, only the \emph{one-go} supersymmetric factorization is conducted for constructing the whole SUSYQM algorithm. Such algorithm can be equivalently interpreted as the following bottom-up \emph{iterative} SUSYQM algorithm for simultaneously building the Legendre equation as well as its eigensolution \emph{in multiple steps}.

In this algorithm, only the initial level eigenequation
\begin{eqnarray}\label{0th_eigenequ}
 p(x)L(x)\Phi_0(x)=B_0A_0\Phi_0(x)=0
\end{eqnarray}
is assumed to be known, where
\begin{eqnarray}
B_0= -A_0=-p\od{}{x},
\end{eqnarray}
while the eigenequation of generic level for $L(x)$ itself is assumed to be unknown to us. Instead, such eigenequation and its eigensolution are needed to be constructed simultaneously level by level. In the following, we will illustrate how such iterative algorithm is constructed, by taking the $x$-coordinate representation of this algorithm as an example.

We note that \eq{0th_eigenequ} is equivalent to $A_0\Phi_0(x)=0$. This says that $A_0$ is a lowering operator that annihilates the ground state $\Phi_0$, which is a constant. Because the operator product $B_0 A_0=-A^2_0$ in \eq{0th_eigenequ} is a symmetric form, the intertwining action with $A_0$ goes nowhere, i.e., left acting $A_0$ on \eq{0th_eigenequ} does not generate the renewed eigenequation for keeping the iteration going. In order to break the symmetry of the square form $B_0 A_0$, we can substitute for $B_0 A_0$ via the \emph{asymmetric}  product $A_1 B_1$, which is subject to the shape invariance relation of first level
\begin{eqnarray}
 A_1 B_1-B_0A_0=\Delta_1
\end{eqnarray}
for operators $A_1$ and $ B_1$ to be appropriately designed and for some constant $\Delta_1$ to be determined. \eq{0th_eigenequ} is then rearranged into the asymmetric factorization
\begin{eqnarray}\label{0th_asymmeric_AB_eigenequ}
 A_1 B_1\Phi_0=E_1\Phi_0,
\end{eqnarray}
where $B_1$ is supposed to play the role of a raising operator and $E_1=\Delta_1$.

In view that $A_0\equiv p(x) d/dx$, $p(x)=1-x^2$ and $[x,\, A_0]=p(x)$, in order to allow condition \eqnm{0th_asymmeric_AB_eigenequ} to hold, $B_1$ and $A_1$ have to be designed as the following specific linear shifts of $B_0$ and $A_0$, respectively,
\begin{eqnarray}
\begin{aligned}
 B_1&=B_0+ x=-A_0+x,\\
 A_1&=A_0+x.
\end{aligned}
\end{eqnarray}
As a result, $\Delta_1=1$.

Now we can perform the intertwining action on \eq{0th_asymmeric_AB_eigenequ} by acting with $B_1$ and generate the superpartner eigenequation
\begin{eqnarray}\label{1st_asymmeric_BA_eigenequ}
 B_1A_1 \Phi_1=E_1\Phi_1,
\end{eqnarray}
where
\begin{eqnarray}
\Phi_1=B_1\Phi_0.
\end{eqnarray}

Furthermore, if $B_2$ and $A_2$ can be designed respectively as the shifts of $B_1$ and $A_1$ by the same amount $x$, i.e.,
\begin{eqnarray}
\begin{aligned}
 B_2&=B_1+x=B_0+ 2x,\\
 A_2&=A_1+x=A_0+2x,
\end{aligned}
\end{eqnarray}
then there exists the shape invariance condition of second level
\begin{eqnarray}
 A_2 B_2-B_1A_1 =\Delta_2=3.
\end{eqnarray}
Substitute for $B_1A_1$ via the above condition, \eq{1st_asymmeric_BA_eigenequ} is rearranged as
\begin{eqnarray}\label{2nd_asymmeric_AB_eigenequ}
 A_2 B_2\Phi_1=E_2\Phi_1,
\end{eqnarray}
with $E_2\equiv \Delta_1+\Delta_2$.

Similarly, intertwiningly acting \eq{2nd_asymmeric_AB_eigenequ} with $B_2$ leads
\begin{eqnarray}\label{2nd_asymmeric_BA_eigenequ}
 B_2 A_2 \Phi_2 =E_2 \Phi_2,
\end{eqnarray}
where
\begin{eqnarray}
 \Phi_2=B_2\Phi_1.
\end{eqnarray}

Eqs. \eqnm{2nd_asymmeric_BA_eigenequ}, \eqnm{1st_asymmeric_BA_eigenequ} and \eqnm{0th_eigenequ} are of the same pattern. Keeping such iteration going \emph{inductively} to the generic $j$-th level ($j\ge 3$), we \emph{simultaneously} obtain the supersymmetrically factorized Legendre equation which is the same as that given by \eq{Legendre_equ_non_Hermitian_factorized0}, and its eigenfunction which is the same as that given by \eq{unity_normed_Legendre_polynomial}, and eigenvalue $E_j=j^2$.

\section{Simultaneously Constructing a Hierarchy of Associated Legendre Differential Equations and Their Solutions in an Iterative Approach in $x$ coordinate}\label{associated_SUSY_x_coordinate}

An innocent question to ask is, why do there exist the base Legendre functions as well the associated Legendre functions for the same Legendre operator $L(x)$? A standard answer to this question is that, mathematically one can derive the associated Legendre differential equation from the base Legendre differential equation and the differential connection between the associated Legendre function and the base Legendre function is constructed along with this derivation. This sort of answer pervades predominantly in literature and textbooks. In the framework of SUSYQM, we could provide another elegant answer to this question. This story begins with the basic fact that, from the same Legendre operator $L(x)$, one could construct two types of squared forms of first-order differential operators, one is $-(p(x)d/dx)^2$ that is contained in $p(x)L(x)$ and has been discussed in Subsection \ref{Hermitian_factorization}, the other one is $-\smbr{\sqrt{p(x)}d/dx}^2$ that can be constructed from $L(x)$ alone and will be addressed in this section. Both types of squared forms can be viewed as the kinetic energy operators, hence there are two type of Hamiltonians and momentum operators that are associated with the same Legendre operator $L(x)$. What is more is that, both Hamiltonians admit distinct supersymmetric factorizations, and these two initial supersymmetric factorizations can iteratively proliferate into two hierarchies of supersymmetric factorizations in a self-similar fashion. One leads to the generic base Legendre equation and its eigenfunction \text{--} the Legendre polynomial, which have been addressed extensively in Section \ref{SUSYQM_base_Legendre_operator}, while the other leads to the generic associated Legendre equation and the associated Legendre function.  In fact, from the perspective of SUSYQM, this is the very reason why there exist the associated Legendre equation and the associated Legendre function besides the base Legendre equation and the base Legendre function for the \emph{same} operator $L(x)$. However, the situation facing us here is different from building the SUSYQM algorithm of the base Legendre operator $L(x)$, once we assume that the associated Legendre differential operator of generic level is unknown to us, and we have to simultaneously construct it as well as its eigenspectrum \emph{inductively} in an iterative way, which is the main task of this section.

\subsection{Another Momentum Operator Hidden in $L(x)$ and the Related Active Supersymmetrization Transformation of $L(x)$}\label{second_momentum_map}

We now start to demonstrate that there exists another momentum operator and another Hamiltonian that can be derived from $L(x)$ alone and such Hamiltonian admits a certain supersymmetric factorization.

The first term in \eq{operator_L_x_coord} for $L(x)$ can be rewritten as
\begin{eqnarray}
 -p\frac{d^2}{ d x^2}=\smbr{-\sqrt{p} \od{}{x} + \sqrt{p}'}\smbr{\sqrt{p}\od{}{x}}
\end{eqnarray}
and the second term as $-p'd/dx=-\smbr{2\sqrt{p}'}\smbr{\sqrt{p}d/dx}$. $L(x)$ then admits the natural asymmetric factorization
\begin{eqnarray}\label{LA_asymmetric_factorization}
 L(x)= \mfk{L}^\dagger_0\mfk{L}_0,
\end{eqnarray}
where
\begin{eqnarray}
W^a_0(x) &\equiv& -\frac{1}{2}\sqrt{p}^{(1)},\label{Wa0_x}\\
\mfk{L}^\dagger_0(x)&\equiv& -\sqrt{p}\od{}{x}+2 W^a_0(x) = - \od{}{x}\sqrt{p},\label{mfkLd_0}\\
\mfk{L}_0(x) &\equiv& \sqrt{p} \od{}{x}\label{mfkL_0}
\end{eqnarray}
and the superscript $(1)$ denotes the first-order derivative with respect to $x$.

It is evident that
\begin{eqnarray}\label{self_adjoint_form1}
\mfk{L}^\dagger_0\mfk{L}_0 =-\od{}{x} p\od{}{x},
\end{eqnarray}
and the right hand side of this equation is nothing but the \emph{self-adjoint form} of $L(x)$, where, for two well-behaved functions $f_1(x)$ and $f_2(x)$, the adjoint is defined with respect to the inner product $(f_1,\, f_2)\equiv \int^{1}_{-1}dx f_1(x)f_2(x)$ with trivial weight function $w(x)=1$, recalling that the weight function is $1$ in the self-adjoint form of $L(x)$. The factorization \eqnm{LA_asymmetric_factorization} is thus an \emph{adjoint factorization} of $L(x)$, obtained simply by partitioning the middle factor $p$ inside the self-adjoint form \eqnm{self_adjoint_form1} of $L(x)$ over the left and right factor differential operators by equal amount $\sqrt{p}$, and the dagger denotes adjoint operation. Here we have chosen a sign convention by assigning the minus sign to the first factor $\mfk{L}^\dagger_0(x)$ for the purpose that will become clear as we proceed.

The factorization \eqnm{LA_asymmetric_factorization} is not explicitly supersymmetric because beside the two differential operators $\pm \sqrt{p} d/dx$, the rest are not the same functions. Soon we will see that this asymmetric adjoint factorization is related to a certain supersymetric factorization by some unitary transformation. It is in this sense that we call such asymmetric adjoint factorization of $L(x)$  an implicit supersymmetric factorization, that is represented in $x$ coordinate, or in this loose sense, still a supersymmetric factorization.

Like the square form $-\smbr{p(x) d/dx}^2$ contained in $p(x)L(x)$ is regarded as a kinetic energy operator $p^2_y$ of Hamiltonian $\widehat{H}_j(y)$, with $-i p d/dx\equiv \hat{p}_y$ the momentum operator (see \eq{momentum_operator_map1}), the square form
\begin{eqnarray}
 -\smbr{\sqrt{p}\od{}{x}}^2 =\smbr{\pm i\sqrt{p}\od{}{x}}^2
\end{eqnarray}
in the factorization \eqnm{LA_asymmetric_factorization} can be regarded as another kinetic energy operator $\hat{p}^2_\theta$ of a certain Hamiltonian, once the base operator $\pm i\sqrt{p}d/dx$ is identified as \emph{another} type of momentum operator $\hat{p}_\theta$, which is defined as
\begin{eqnarray}\label{momentum_operator_map2}
 \pm i\sqrt{p}\od{}{x} \equiv -i \od{}{\theta} =\hat{p}_\theta
\end{eqnarray}
in some coordinate $\theta$, which turns out to be the usual polar angle coordinate (see this in Subsection \ref{Legendre_operator_theta_coordinate}). According to our experience obtained in Subsection \ref{Hermitian_factorization}, this implies that we can construct another Hamiltonian and its corresponding Schr\"odinger equation that is related to the Legendre equation \eqnm{Legendre_equ_x_coord}, for which $-\smbr{\sqrt{p}d/dx}^2$ works as the kinetic energy operator.

However, $L(x)$ itself can not be directly identified with a standard Hamiltonian, in view that the factorization \eqnm{LA_asymmetric_factorization} is asymmetric in that $2W^a_0(x)$ is \emph{unevenly} located in the first factor $\mfk{L}^\dagger_0$. We can seek for a bilateral transform on $L(x)$ that is distributive over each factor of $L(x)$  and evenly partitions $2W^a_0(x)$ over its two factors $\mfk{L}^\dagger_0$ and $\mfk{L}_0$, in the sense that it \emph{lowers} the first factor $\mfk{L}^\dagger_0$ by $W^a_0(x)$ and \emph{raises} the second one $\mfk{L}_0$ by the \emph{same} amount. When such a transformation is applied to $L(x)$, a supersymmetric factorization of $L(x)$ is then reached, with $W^a_0(x)$ being identified as the superpotential of such factorization. For this reason, such transformation is dubbed as \emph{active supersymmetrization} transformation.

To pin down such a supersymmetrization transformation, we simply single out the expected $-\sqrt{p}d/dx+ W_0^a$ from the first factor  in the factorization \eqnm{LA_asymmetric_factorization}. This part can be lumped
simply by looking for its integration factor, i.e.,
\begin{eqnarray}\label{int_factor}
-\sqrt{p} \od{}{x} + W^a_0 &=&{\cal S}^{-1}_x\smbr{-\sqrt{p} \od{}{x}},
\end{eqnarray}
where the bilateral transformation ${\cal S}_x(\text{--})$ and its inverse ${\cal S}^{-1}_x(\text{--})$ are defined as
\begin{eqnarray}
 {\cal S}_x\smbr{\text{--}}&\equiv& e^{-\int dx \frac{W^a_0}{\sqrt{p}}}\smbr{\text{--}} \,e^{\int dx \frac{W^a_0}{\sqrt{p}}}= p^{\frac{1}{4}}\smbr{\text{--}}p^{-\frac{1}{4}}, \label{Sa_transformation}\\
 {\cal S}^{-1}_x\smbr{-}&\equiv& e^{\int dx \frac{W^a_0}{\sqrt{p}}}\smbr{\text{--}} \,e^{-\int dx \frac{W^a_0}{\sqrt{p}}}= p^{-\frac{1}{4}}\smbr{\text{--}}p^{\frac{1}{4}}, \label{Sa_inverse_transformation}
\end{eqnarray}
where to reach the second equality in \eq{Sa_transformation} or in \eq{Sa_inverse_transformation}, $W^a_0$ has been substituted for via \eq{Wa0_x}. When acting on an ordinary function such as $W^a_0(x)$, ${\cal S}^{\pm1}_a(-)$ work as identity transformations, i.e., they keep that function intact.

It follows from \eqnm{int_factor} that
\begin{eqnarray}\label{shift_effects_of_Sas}
\begin{aligned}
{\cal S}_x\smbr{\pm \sqrt{p}\od{}{x} } &=\pm \sqrt{p}\od{}{x} \pm W^a_0,\\
{\cal S}^{-1}_x\smbr{\pm\sqrt{p}\od{}{x} } &=\pm \sqrt{p}\od{}{x} \mp W^a_0,
\end{aligned}
\end{eqnarray}
That is, ${\cal S}_x$ works to shift $\pm\sqrt{p}d/dx$ up/down-wards by $W^a_0$, while the inverse ${\cal S}^{-1}_x$ works to shift $\pm\sqrt{p}d/dx$ in the opposite directions by the same amount. These shifting roles of ${\cal S}^{-\pm1}_x$ can also be directly checked. Later on, we will see that these roles of ${\cal S}^{\pm1}_x$ on the basic differential operators $\pm \sqrt{p}d/dx$ are also very crucial for directly building the associated Legendre operator of generic level in its supersymmetrically factorized form, when they are applied \emph{multiple times} to $\pm\sqrt{p}d/dx$ (e.g., see this in Eqs. \eqnm{tldLm_symmetric_factorization}, \eqnm{tldmfkLmplus1} and \eqnm{tldmfkLdaggermplus1}).

Adding $2W_0^a$ to the minus branch of the upper equation in \eqnm{shift_effects_of_Sas} yields
\begin{eqnarray}\label{shift_effects_by_S0}
{\cal S}_x\smbr{-\sqrt{p}\od{}{x} + 2 W^a_0} = -\sqrt{p}\od{}{x} + W^a_0.
\end{eqnarray}
The minus branch of the upper equation in \eqnm{shift_effects_of_Sas} reads
\begin{eqnarray}
 {\cal S}_x\smbr{\sqrt{p}\od{}{x}} = \sqrt{p}\od{}{x} + W^a_0.
\end{eqnarray}
Taking the product of the above two equations, and making use of the distributivity of ${\cal S}_x$ over each factor of an operator product, we then turn the asymmetric factorization \eqnm{LA_asymmetric_factorization} of $L(x)$ into the \emph{supersymetric} factorization
\begin{eqnarray}\label{SUSY_associated_factorization}
 {\cal S}_x\smbr{L(x)}=\smbr{-\sqrt{p} \od{}{x} +W^a_0}\smbr{\sqrt{p} \od{}{x} + W^a_0},
\end{eqnarray}
where the \emph{common} function $W^a_0$ in the two factors is identified as the superpotential of this supersymmetric factorization. For this very reason, ${\cal S}_x$ is termed as an \emph{active supersymmetrization} transformation, and its inverse ${\cal S}^{-1}_s$ as the de-supersymmetrization transformation.

In the above, we just followed a rather general while relatively lengthy algebraic procedure for pinning down the SUSY factorization of a hypergeometric-like differential operator \cite{Zhou_SUSY_hypergeometric_operator_2024}, assuming that we first unfortunately confront with the \emph{expanded} hypergeometric form \eqnm{operator_L_x_coord} of the Legendre differential operator. There is also another more straightforward and almost effortless algebraic procedure \cite{Zhou_SUSY_hypergeometric_operator_2024} of supersymmetrizing a hypergeometric-type differential operator like $L(x)$. In this latter procedure, one starts with the \emph{self-adjoint} form of Legendre operator in \eq{self_adjoint_form1}, one then attaches one copy of $\sqrt{p}$ \text{--} the square root of the middle factor $p$ to its right neighboring differential operator $d/dx$ for preparing $\sqrt{p}d/dx$, and at this time  prepare $-\sqrt{p}d/dx$ by dividing and multiplying the factor $\sqrt{p}$ in the front. Next, one \emph{evenly} partition the two copies of $p^{1/4}$ \text{--} the square root of the middle factor $\sqrt{p}$ that is left from the last step, to its immediate left and right blocks and wraps $\mp\sqrt{p}d/dx$ respectively with the bilateral transformations ${\cal S}^{\mp 1}_x\equiv p^{\mp 1/4}(-)p^{\pm 1/4}$, which are the same as those introduced in Eq. \eqnm{shift_effects_of_Sas}.  In so doing, $L(x)$ is rewritten as
\begin{eqnarray}\label{supersymmetrizing_Lx}
L(x) &=&\frac{1}{\sqrt{p}}\smbr{-\sqrt{p}\od{}{x} }\sqrt{p}\smbr{\sqrt{p}\od{}{x}},\nonumber \\
     &=& \frac{1}{p^{1/4}}\left[\frac{1}{p^{1/4}} \smbr{-\sqrt{p}\od{}{x} } p^{1/4}\right] \nonumber \\
     &&\quad \cdot \left[ p^{1/4}\smbr{\sqrt{p}\od{}{x}} \frac{1}{p^{1/4}} \right]p^{1/4}.
\end{eqnarray}

\eq{supersymmetrizing_Lx} is rearranged  as
\begin{eqnarray}\label{supersymmetrizing_Lx1}
{\cal S}_x\smbr{L(x)} &=& \left[{\cal S}^{-1}_x\smbr{-\sqrt{p}\od{}{x} }\right] \cdot \left[{\cal S}_a\smbr{\sqrt{p}\od{}{x}}\right].
\end{eqnarray}
On the other hand, the combination of the lower branch of the lower equation in \eqnm{shift_effects_of_Sas} and \eq{shift_effects_by_S0} gives us
\begin{eqnarray}
 {\cal S}^{-1}_x\smbr{-\sqrt{p}\od{}{x} }& =-\sqrt{p}\od{}{x} + W^a_0={\cal S}_x\smbr{\mfk{L}^\dagger_0}.
\end{eqnarray}
 Therefore, the supersymmetric factorization \eqnm{supersymmetrizing_Lx1} is indeed the same as that given in \eq{SUSY_associated_factorization}.

With the notations
\begin{eqnarray}\label{L_tilde_symmetric_factorization}
\begin{aligned}
 \tilde{\mfk{L}}_0 \smbr{x} &\equiv {\cal S}_x\smbr{{\mfk{L} }_0}= \sqrt{p}\od{}{x}+ W^a_0(x),\\
 \tilde{\mfk{L}}^\dagger_0 \smbr{x}\equiv& {\cal S}_x\smbr{\mfk{L}^\dagger_0}=-\sqrt{p}\od{}{x}+ W^a_0(x),
\end{aligned}
\end{eqnarray}
we have
\begin{eqnarray}\label{L_tilde_symmetric_factorization1}
 \tilde{L}(x)\equiv {\cal S}_x\smbr{L(x)}=\tilde{\mfk{L}}^\dagger_0\smbr{x} \tilde{\mfk{L}}_0 \smbr{x}.
\end{eqnarray}

In view of the supersymmetric factorization \eqnm{L_tilde_symmetric_factorization1} for $L(x)$, by simply multiplying with $p^{1/4}$ to prepare ${\cal S}_a$, \eq{Legendre_equ_x_coord} is then supersymmetrically factorized as
\begin{eqnarray}\label{SUSY_factorized_Legendre_equ_base_x_coord}
 \tilde{L}(x)\Psi_j(x)=\tilde{\mfk{L}}^\dagger_0\smbr{x} \tilde{\mfk{L}}_0 \smbr{x}\Psi_j(x)=\lambda_j\Psi_j(x),
\end{eqnarray}
where $\Psi_j(x)\equiv p^{1/4}\Phi_j(x)$.

We could also assign the minus sign in $L(x)$ to the second factor in the adjoint form \eqnm{self_adjoint_form1} of $L(x)$ and introduce the notations
\begin{eqnarray}\label{mfkL_1_def}
 \begin{aligned}
  \mfk{L}_1(x)\equiv \od{}{x}\sqrt{p}=-\mfk{L}^\dagger_0(x),\\
  \mfk{L}^\dagger_1(x)\equiv -\sqrt{p}\od{}{x}=-\mfk{L}_0(x),\\
 \end{aligned}
\end{eqnarray}
so that $L(x)$ has another form of natural factorization
\begin{eqnarray}\label{L_factorization_minus_right}
 L(x)=\mfk{L}_1(x)\mfk{L}^\dagger_1(x).
\end{eqnarray}

The definitions of $\mfk{L}_1(x)$ and $\mfk{L}^\dagger_1(x)$ here together with those in \eq{L_tilde_symmetric_factorization} imply that
\begin{eqnarray}\label{mfkL_1_series}
\begin{aligned}
 \mfk{L}_1(x)&=\mfk{L}_0(x) + \sqrt{p}^{(1)}={\cal S}^{-2}_x\smbr{\mfk{L}_0(x)}, \\
 \mfk{L}^\dagger_1(x)&=\mfk{L}^\dagger_0(x) + \sqrt{p}^{(1)}={\cal S}^{2}_x\smbr{\mfk{L}^\dagger_0(x)},
\end{aligned}
\end{eqnarray}
where ${\cal S}^{\pm2}_x$ denote the square of ${\cal S}^{\pm1}_x$.

Although $\mfk{L}_1(x)$ and $\mfk{L}^\dagger_1(x)$ has initially been introduced simply as notations, soon we will see in Subsections \ref{Positive_Association_Levels} and \ref{Negative_Association_Levels}  that $\mfk{L}_0(x)$/$\mfk{L}^\dagger_0(x)$ can proliferate systematically into the operator series $\{\mfk{L}_m(x)\}$/ $\{\mfk{L}^\dagger_m(x)\}$ for $m$ being integers, and $\mfk{L}_1(x)$ and $\mfk{L}^\dagger_1(x)$ are just the $m=1$ elements inside these two series. Thus, \eq{mfkL_1_def} actually reflect symmetry relations between $\mfk{L}_1(x)$/$\mfk{L}^\dagger_1(x)$ and $\mfk{L}^\dagger_0(x)$/$\mfk{L}_0(x)$.

\subsection{The $\theta$-coordinate Parametrization of the SUSYQM of Legendre Operator}\label{Legendre_operator_theta_coordinate}

In Subsection \ref{second_momentum_map}, we are unfortunately dwelling in the $x$-coordinate representation of Legendre operator from the very beginning and we then search for its supersymmetric factorization in $x$ coordinate. Of course, working in this $x$-coordinate representation has its own advantages, one of such advantages is that the Legendre operator $L(x)$ is just a very special type of the hypergeometric-like differential operators, and we are ensured that there is a generic SUSYQM algorithm  developed in \cite{Zhou_SUSY_hypergeometric_operator_2024} for simultaneously building the corresponding associated eigenequation and its solution. However, in many physical problems, the Legendre differential operator emerges as the reduction of the angular Laplacian (see \eq{angular_Laplacian}) when the physics systems own axial symmetry, i.e., they are independent of the azimuthal angle $\phi$. In terms of the polar $\theta$, this operator appears naturally in the lumped (factorized) form given in \eq{J0_theta}. Such $J_0(\theta)$ comes with the natural factorization
\begin{eqnarray}\label{lumped_L_thet0}
 J_0(\theta) =\mfk{J}^+_0(\theta) \mfk{J}^-_0(\theta),
\end{eqnarray}
where
\begin{eqnarray}\label{mfkJpm_0}
\begin{aligned}
 \mfk{J}^+_0(\theta) &\equiv \frac{1}{\sin\theta}\od{}{\theta} \sin\theta = \od{}{\theta} +\cot\theta=\od{}{\theta}+2 {\cal W}^a_0(\theta), \\
 \mfk{J}^-_0(\theta) &\equiv -\od{}{\theta}
\end{aligned}
\end{eqnarray}
and
\begin{eqnarray}\label{Wa0_theta}
{\cal W}^a_0(\theta)\equiv \frac{1}{2}\cot\theta.
\end{eqnarray}

The factorization \eqnm{lumped_L_thet0}, though asymmetric, is an adjoint factorization, as $\mfk{J}^+_0(\theta)$ and $\mfk{J}^-_0(\theta)$ are an adjoint /conjugate pair with respect to the inner product with the measure $\sin\theta d\theta$.

The Legendre equation in its factorized form in $\theta$ coordinate reads
\begin{eqnarray}\label{Legendre_equ_theta_coord0}
 J_0(\theta)\Theta_j(\theta)=\mfk{J}^+_0(\theta) \mfk{J}^-_0(\theta)\Theta_j(\theta)=\lambda_j\Theta_j(\theta).
\end{eqnarray}

Evidently the factorization \eqnm{lumped_L_thet0} of $J_0$ is not supersymmetric, because the two $\theta$-functions (one is $2{\cal W}^a_0(\theta)$ and the other is zero) in $\mfk{J}^\pm_0$ are obviously not the same, hence neither of them can work as the superpotential, although the two differential operators $\pm d/d\theta$ are the expected conjugate pair. We can single out a \emph{common} function ${\cal W}^a_0(\theta)=\cot\theta/2$ in $\mfk{J}^\pm_0(\theta)$ as the superpotential and lump the remaining pieces $\pm\cot\theta/2$ into the integration factors $\sin^{\pm1/2}\theta$ that wrap $\pm d/d\theta$. That is, because
\begin{eqnarray}
\od{}{\theta}+\frac{1}{2}\cot\theta=\sin^{-1/2}\theta\smbr{\od{}{\theta}}\sin^{1/2}\theta\equiv{\cal S}^{-1}_\theta\smbr{\od{}{\theta}}\nonumber \\
\end{eqnarray}
for the bilateral transformation
\begin{eqnarray}\label{S_theta_inverse}
{\cal S}^{-1}_\theta\smbr{-}\equiv \sin^{-1/2}\theta\smbr{-}\sin^{1/2}\theta
\end{eqnarray}
and  its inverse
\begin{eqnarray}\label{S_theta}
{\cal S}_\theta\smbr{-}\equiv \sin^{1/2}\theta\smbr{-}\sin^{-1/2}\theta,
\end{eqnarray}

$\mfk{J}^\pm_0(\theta)$ in \eq{mfkJpm_0} are then rewritten as
\begin{eqnarray}\label{mfkJpm_theta}
\begin{aligned}
 \mfk{J}^+_0(\theta) &= {\cal S}^{-1}_\theta\smbr{\od{}{\theta}} +\frac{1}{2}\cot\theta={\cal S}^{-1}_\theta\smbr{\od{}{\theta} +\frac{1}{2}\cot\theta}, \\
 \mfk{J}^-_0(\theta) &= -{\cal S}^{-1}_\theta\smbr{\od{}{\theta}} +\frac{1}{2}\cot\theta={\cal S}^{-1}_\theta\smbr{-\od{}{\theta} +\frac{1}{2}\cot\theta}.
\end{aligned}
\end{eqnarray}

Therefore, the transformation ${\cal S}_\theta\smbr{-}$ works as the \emph{active supersymmtrization} transformation that supersymmetrizes the asymmetric factorization \eqnm{lumped_L_thet0} of $J_0(\theta)$, i.e.,
\begin{eqnarray}\label{J0_theta_SUSY_factorization}
 \tilde{J}_0(\theta)&\equiv& {\cal S}_\theta\smbr{J_0(\theta)}={\cal S}_\theta\smbr{\mfk{J}^+_0(\theta)}{\cal S}_\theta\smbr{\mfk{J}^-_0(\theta)},\nonumber \\
                    &=& \tilde{\mfk{J}}^+_0(\theta) \tilde{\mfk{J}}^-_0(\theta),
\end{eqnarray}
where
\begin{eqnarray}\label{tldmfkJpm0}
\begin{aligned}
 \tilde{\mfk{J}}^+_0(\theta) &\equiv {\cal S}_\theta\smbr{\mfk{J}^+_0(\theta)}=\od{}{\theta} +\frac{1}{2}\cot\theta, \\
 \tilde{\mfk{J}}^-_0(\theta) &\equiv {\cal S}_\theta\smbr{\mfk{J}^-_0(\theta)}=-\od{}{\theta} +\frac{1}{2}\cot\theta,\\
\end{aligned}
\end{eqnarray}
which are the rearrangements of the two equations in \eq{mfkJpm_theta} by applying ${\cal S}_\theta$ to both sides of each equation. Formally ${\cal S}_\theta$ lowers $d/d\theta$ and $\mfk{J}^+_0(\theta)$, and raises $\mfk{J}^-_0(\theta)$, by the same amount $\cot\theta/2$, while
${\cal S}^{-1}_\theta$ does the opposite thing --- raises $d/d\theta$ and lowers $-d/d\theta$, by the same amount $\cot\theta/2$.

The supersymmetric factorization of $\tilde{J}_0(\theta)$ in \eq{J0_theta_SUSY_factorization} is just the $m=0$ case of the supersymmetric factorization $\tilde{J}_m(\theta)=\tilde{\mfk J}^+_m(\theta)\tilde{\mfk J}^-_m(\theta)$ with $\tilde{\mfk J}^\pm_m(\theta)$ defined in \eq{tildeJ_mpm_theta}, for the generic associated Legender differential operator $J_m(\theta)\equiv{\mfk J}^+_m(\theta){\mfk J}^-_m(\theta)$, with ${\mfk J}^+_m(\theta)$ and ${\mfk J}^-_m(\theta)$ defined in \eq{mfkJpm_def0}. ${\cal W}^a_0(\theta)$ in \eq{Wa0_theta} turns out to be the superpotential of the supersymmetric factorization in \eq{J0_theta_SUSY_factorization} and to be just the $m=0$ version of the superpotential ${\cal W}^a_m(\theta)=(1/2-m)\cot\theta$ in \eq{Wam_superpotential}, which emerges when supersymmetrizing the factorization $J_m(\theta)\equiv \mfk{J}^+_m(\theta)\mfk{J}^+_m(\theta)$ of the associated Legendre differential operator $J_m(\theta)$ in $\theta$-coordinate (see \eq{Jm_theta_supersymmetric_factorization} for this).

We see that one of the advantages of the $\theta$-coordinate representation of the Legendre operator is that ${\cal S}_\theta\smbr{J_0(\theta)}$ is the supersymmetric factorization of a \emph{standard} Hamiltonian, as the momentum operator $-i d/d \theta$ naturally assumes the standard form in this $\theta$ coordinate itself, i.e., the momentum operator map accompanying this supersymmetrization transformation is trivial.

There exists an even more efficient algebraic procedure of pinning down the supersymmetrization transformation ${\cal S}_\theta$ and reaching the supersymmetric factorization \eqnm{J0_theta_SUSY_factorization}, which is highly similar to the procedure of supersymmetrizing the $x$-coordinate representation of the asymmetric factorization of $L(x)$, i.e., the passage from \eq{supersymmetrizing_Lx} to \eq{L_tilde_symmetric_factorization1}. Concretely,  one starts  with the \emph{lumped} form of $J_0(\theta)$ in \eq{J0_theta}
and \emph{evenly} partitions the two copies of the factor $\sin^{1/2}\theta$ \text{--} the square root of the middle factor $\sin\theta$ in $J_0(\theta)$ to the immediate left and right factorial operators, one then wraps $d/d\theta$ by the supersymmetrization transformation ${\cal S}_\theta$ given in \eq{S_theta} and $-d/d\theta$ by the inverse transformation ${\cal S}_\theta\smbr{\cdots}$ given \eq{S_theta_inverse}. That is,
\begin{eqnarray}\label{lumped_L_theta1}
 J_0(\theta) &=&\frac{1}{\sin^{1/2}\theta} \left[\frac{1}{\sin^{1/2}\theta} \smbr{\od{}{\theta}} \sin^{1/2}\theta \right] \nonumber \\
 &&\cdot \left[ \sin^{1/2}\theta \smbr{-\od{}{\theta}}\sin^{-1/2}\theta\right] \sin^{1/2}\theta, \nonumber \\
             &=&{\cal S}^{-1}_\theta\sqbr{{\cal S}^{-1}_\theta\smbr{\od{}{\theta}} \cdot {\cal S}_\theta\smbr{-\od{}{\theta}}}.
\end{eqnarray}
Therefore,
\begin{eqnarray}\label{supersymmetrized_J0_theta}
 {\cal S}_\theta \smbr{J_0(\theta)} &=&{\cal S}^{-1}_\theta\smbr{\od{}{\theta}} \cdot {\cal S}_\theta\smbr{-\od{}{\theta}}.
\end{eqnarray}

It is easy to check that
\begin{eqnarray}\label{essential_roles_Spm}
 \begin{aligned}
  {\cal S}^{-1}_\theta\smbr{\od{}{\theta}} &=\od{}{\theta} + \frac{1}{2}\cot\theta={\cal S}_\theta\smbr{{\mfk J}^+_0(\theta)}, \\
  {\cal S}_\theta\smbr{-\od{}{\theta}} &= -\od{}{\theta} + \frac{1}{2}\cot\theta = {\cal S}_\theta\smbr{{\mfk J}^-_0(\theta)}.
 \end{aligned}
\end{eqnarray}
So the supersymmetric factorization \eqnm{supersymmetrized_J0_theta} is the same as that given in \eq{J0_theta_SUSY_factorization}.

Multiplying with $\sin^{1/2}(\theta)$, which is the left side of the double-sided transformation ${\cal S}_\theta$, and making use of the supersymmetric factorization of $\tilde{J}_0(\theta)$ in \eq{J0_theta_SUSY_factorization}, the asymmetrically factorized Legendre equation \eqnm{Legendre_equ_theta_coord0} is then transformed into the supersymmetrically factorized form of the standard Schr\"odinger equation with Hamiltonian $\tilde{J}_0(\theta)$,
\begin{eqnarray}\label{SUSY_factorized_Legendre_equ_base_theta_coord}
 \tilde{J}_0(\theta)\widetilde{\Theta}_j(\theta)= \tilde{\mfk J}^+_0(\theta) \tilde{\mfk J}^-_0(\theta) \widetilde{\Theta}_j(\theta) = \lambda_j \widetilde{\Theta}_j(\theta),
\end{eqnarray}
where $\widetilde{\Theta}_j(\theta)\equiv \sin^{1/2}\theta \Theta_j(\theta)$. This equation is just  the $\theta$-coordinate representation of \eq{SUSY_factorized_Legendre_equ_base_x_coord}.

The momentum operator map \eqnm{momentum_operator_map2} are equivalent to the differential equation
\begin{eqnarray}
 d\theta =\mp \frac{d x}{\sqrt{1-x^2}},
\end{eqnarray}
which has the solution $x=\pm \cos\theta$ for $\theta \subset [0,\pi)$. This $\theta$ is in fact the polar angle of a spherical coordinate system, in view that $J_0(\theta)=L(x)$.

Given the definitions for $\mfk{L}^\dagger_0(x)$ in \eq{mfkLd_0} and $\mfk{L}_0(x)$ in \eq{mfkL_0} and those for $\mfk{J}^\pm_0(\theta)$ in \eq{mfkJpm_0}, we choose the coordinate transformation $x=\cos\theta$ and then obtain
\begin{eqnarray}
\begin{aligned}
 \mathfrak{L}^\dagger_0(x)&=\mfk{J}^+_0(\theta),\\
\mathfrak{L}_0(x)&=\mfk{J}^-_0(\theta),
 \end{aligned}
\end{eqnarray}
\begin{eqnarray}
 W^a_0(x)&={\cal W}^a_0(\theta),
\end{eqnarray}
\begin{eqnarray}
 \tilde{L}(x)=\tilde{J}_0(\theta)
\end{eqnarray}
and
\begin{eqnarray}
 {\cal S}_x={\cal S}_\theta.
\end{eqnarray}
That is, the momentum map \eqnm{momentum_operator_map2} (a coordinate transformation) maps the $x$-coordinate representation of the supersymmetric factorization of $L(x)$ to its $\theta$-coordinate representation. In other words, these two coordinate representations are equivalent. Later on, we will see that the momentum map and the supersymmetrization transformation also preserve the shape invariance symmetry. For these reasons, one can choose either one of these two coordinate representations to proceed to construct the SUSYQM algorithm for building the associated Legendre differential equation and its eigensolutions.

The initial asymmetric factorization \eqnm{LA_asymmetric_factorization} of $L(x)$ is related to the \emph{supersymmetric} factorization \eqnm{L_tilde_symmetric_factorization1} simply by the active supersymmetrization transformation ${\cal S}_x$, and the initial asymmetric factorization \eqnm{lumped_L_thet0}  of $J_0(\theta)$ is related to  the \emph{supersymmetric} factorization \eqnm{J0_theta_SUSY_factorization} simply by the active supersymmetrization transformation ${\cal S}_\theta$. In the next subsection, we will see that, this type of simple relations between the asymmetric factorization and the explicit supersymmetric factorization still hold for higher association levels, not only for the factorization of associated Legendre operator of each association level but also for the shape invariance symmetry between two neighboring association levels. In this sense, the asymmetric factorizations just mentioned are called \emph{implicit} supersymmetric factorizations.  For this very reason, in building the SUSYQM algorithms for simultaneously generating the associated Legendre equation of generic association level and its eigensolution, we can just start with the asymmetric factorization \eqnm{LA_asymmetric_factorization} and do not need to care about the explicit supersymmetry at each forward step, since such supersymmetry is guaranteed by the active supersymmetrization transformation level-wise. In fact, this is what we will do in the next subsection.

Traditionally, one approaches this sort of supersymmetric factorization \emph{in two steps}. Firstly, one aims at eliminating the first-order derivative in the expanded form of the raw eigenequation by solving another differential equation of second order for the coordinate transformation function and reaches a certain Schr\"odinger-like equation from the raw eigenequation. One such example on eliminating the first-order differentiation in $J_0(\theta)$ and turning the associated Legendre equation parametrized in $\theta$ coordinate into a Schr\"odinger equation is shown in the second part of Appendix \ref{traditional_analytical_method}. Secondly, one passively searches for the supersymmetric factorization of the resultant Schr\"odinger-like equation. In contrast, we have adopted the more efficient \emph{algebraic method} by combining the active supersymmetrization transformation and momentum operator map and reached actively the supersymmetric factorization of the raw eigenequation \emph{in one go}.

\subsection{Positive Association Levels}\label{Positive_Association_Levels}

Choosing the factorization \eqnm{LA_asymmetric_factorization} of $L(x)$ and introducing the notations $\Phi_{j0}(x)\equiv \Phi_j(x)$ and $\lambda_{j0}\equiv \lambda_j$, the Legendre equation \eqnm{Legendre_equ_x_coord} takes the following factorized form.
\begin{eqnarray}\label{Legendre_equ_factorization_0}
\mfk{L}^\dagger_0(x) \mfk{L}_0(x)\Phi_{j0}(x)=\lambda_{j0}\Phi_{j0}(x).
\end{eqnarray}
 However, this eigenequation can not work as the start-up eigenequation for proliferating all factorized associated Legendre equations of higher \emph{positive} association levels via the SUSYQM algorithm, since $\mfk{L}_0\equiv \sqrt{p}d/dx$ works as the lowering operator that lowers the association level by one unit each step  and $\mfk{L}^\dagger_0$ the raising one, and these two operators are situated in \tqt{wrong} positions, one can not obtain the eignequation and the eigenfunction $\Phi_{j1}(x)$ for the positive association level $m=1$, by intertwiningly acting the lowering operator $\mfk{L}_0(x)$ on this equation. In view of the relations in \eq{mfkL_1_def}, $\mfk{L}^\dagger_1(x)/\mfk{L}_1(x)$ plays the role of raising/lowering the association level by one unit, we should adopt the other factorization $L_0(x)=\mfk{L}_1(x) \mfk{L}^\dagger_1(x)$ (see \eq{L_factorization_minus_right}) start with the following factorized Legendre equation
\begin{eqnarray}\label{Legendre_equ_right_dagger}
L_0(x) \Phi_{j0}(x) \equiv  \mfk{L}_1(x) \mfk{L}^\dagger_1(x) \Phi_{j0}(x) = \lambda_{j0}\Phi_{j0}(x).
\end{eqnarray}
This amounts to making use of the shape invariance relation of zeroth level
\begin{eqnarray}
\mfk{L}^\dagger_0(x) \mfk{L}_0(x) -\mfk{L}_1(x) \mfk{L}^\dagger_1(x)=\delta_0
\end{eqnarray}
for $\delta_0=0$ to renomalize \eq{Legendre_equ_factorization_0} into \eq{Legendre_equ_right_dagger}.

Intertwiningly acting \eq{Legendre_equ_right_dagger} with $\mfk{L}^\dagger_1$ yields its superpartner eigen-equation
\begin{eqnarray}\label{eigen_equ_1st_level_before_rmnlz}
\mfk{L}^\dagger_1  \mfk{L}_1 \Phi_{j1} = \lambda_{j0} \Phi_{j1},
\end{eqnarray}
where
\begin{eqnarray}\label{Phi_j1_proportinal_to}
 \Phi_{j1}\propto \mfk{L}^\dagger_1 \Phi_{j0}= - \sqrt{p} \Phi^{(1)}_{j0}.
\end{eqnarray}
We can normalize $\Phi_{j1}$ to unity, though $\Phi_{j0}$ is not necessarily normalized to unity. Note that $\mfk{L}_1(x)$ and $\mfk{L}^\dagger_1(x)$ are Hermitian conjugate to each other. \eq{eigen_equ_1st_level_before_rmnlz} then implies that $\int dx \abs{\mfk{L}^\dagger_1 \Phi_{j0}}^2=\lambda_{j0}$. Thus, the proportionality relation \eqnm{Phi_j1_proportinal_to} can be refined to the equation
\begin{eqnarray}\label{raising_ladder_1}
 \Phi_{j1}\equiv \frac{\mfk{L}^\dagger_1}{\sqrt{\lambda_{j0}}} \Phi_{j0}= - \frac{\sqrt{p}}{\sqrt{\lambda_{j0}}} \Phi^{(1)}_{j0}.
\end{eqnarray}
Such $\Phi_{j1}$ has a unity norm, and this $\Phi_{j1}$ happens to be the same type of function as the traditional associated Legendre polynomial of first positive association level $m=1$.

Since $\mfk{L}^\dagger_1=-\mfk{L}_0$ is different from the raising operator $B_{j+1}$ in \eq{Bj_Aj_ladder_operators}, when they act on the same function $\Phi_{j0}(x)\equiv\Phi_j(x)$, it starts to branch into two different types of functions. According to \eq{unity_normed_Legendre_polynomial} or \eq{non_normalized_Legendre_polynomial},  by the action of $B_{j+1}$, $\Phi_{j0}$ is raised into $\Phi_{(j+1)0}$, which still belongs to the polynomial family $\{\Phi_j\}$ in $x$. In contrast, according to \eq{raising_ladder_1}, $\Phi_{j1}$ is not even a polynomial-type function of $x$, simple become the factor $\sqrt{p}\equiv \smbr{1-x^2}^{1/2}$ is not a polynomial in $x$. Here we see how two \emph{distinct} factorizations of the \emph{same} operator $L(x)$ start to generate different types of eigenfunctions at the immediate next level of the initial level of factorization, via the same standard intertwining action. We will see that, by incorporating another symmetry -- pattern invariance -- and repeating the intertwining action, eventually two distinct families of eigenfunctions will naturally come out, one is the family of base Legendre functions, which has already been obtained in Section \ref{SUSYQM_base_Legendre_operator}, the other is is the family of associated Legendre functions, which we will continue to work on.

Obviously, the definition $\Phi_{j1}$ in \eq{raising_ladder_1} can be regarded as a raising recurrence relation, and $\mfk{L}^\dagger_1$ works to raise the association level from $m=0$ to $m=1$. 
This relation is instructive in that acting with $\mfk{L}^\dagger_1$ one more time would possibly generate information on $\Phi_{j2}$. In fact, acting $\mfk{L}^\dagger_1$ in the form given in \eq{mfkL_1_def} on \eq{raising_ladder_1} yields
\begin{eqnarray}\label{2nd_level_introduction}
 \mfk{L}^\dagger_1 \Phi_{j1} = - \sqrt{p}^{(1)} \Phi_{j1} + (-1)^2 \frac{\sqrt{p}^2}{\sqrt{\lambda_{j0}}} \Phi^{(2)}_{j0}.
\end{eqnarray}
Collect the two terms involving $\Phi_{j1}$ together and introducing $\mfk{L}^\dagger_2$ and its Hermitian conjugate $\mfk{L}_2$ as
\begin{eqnarray}\label{mfkL2}
\begin{aligned}
 \mfk{L}^\dagger_2 &\equiv \mfk{L}^\dagger_1 + \sqrt{p}^{(1)}=\mfk{L}^\dagger_0 + 2 \sqrt{p}^{(1)}={\cal S}^4_x\smbr{\mfk{L}^\dagger_0}, \\
 \mfk{L}_2 &\equiv \mfk{L}_1 + \sqrt{p}^{(1)}=\mfk{L}_0 + 2 \sqrt{p}^{(1)}={\cal S}^{-4}_x\smbr{\mfk{L}_0},
\end{aligned}
\end{eqnarray}
where ${\cal S}_x^{4}$ and ${\cal S}_x^{-4}$ are respectively the $4$-th power of ${\cal S}_x$ and ${\cal S}^{-1}_x$.
\eq{2nd_level_introduction} is then reorganized as
\begin{eqnarray}\label{1to2_raising_relation}
\mfk{L}^\dagger_2 \Phi_{j1}=(-1)^2 \frac{\sqrt{p}^2}{\sqrt{\lambda_{j0}}} \Phi^{(2)}_{j0}.
\end{eqnarray}
Now we see that the relation between $\mfk{L}^\dagger_2/\mfk{L}_2$ and  $\mfk{L}^\dagger_1/\mfk{L}_1$ and the relation between$\mfk{L}^\dagger_1/\mfk{L}_1$ and $\mfk{L}^\dagger_0/\mfk{L}_0$ shown in \eq{mfkL_1_series} just follow the same pattern.
We also see that there is a \emph{positive} amount of $\sqrt{p}^{(1)}$ increment from $\mfk{L}_0/\mfk{L}^\dagger_0$ to $\mfk{L}_1/\mfk{L}^\dagger_1$ and then from $\mfk{L}_1/\mfk{L}^\dagger_1$ to $\mfk{L}_2/\mfk{L}^\dagger_2$, as a consequence of attributing the minus sign to the definition of $\mfk{L}^\dagger_0$ in our convention ( for this, see \eq{mfkL_1_def}). This is why in this convention the higher association levels are labeled by \emph{positive} integers.

According to the definitions \eqnm{mfkL_1_def}, $\mfk{L}_1$ and $\mfk{L}^\dagger_1$ are related to $\mfk{L}_0$ through
\begin{eqnarray}
 \mfk{L}_1 = \mfk{L}_0 +\sqrt{p}^{(1)},\quad \mfk{L}^\dagger_1 =-\mfk{L}_0.
\end{eqnarray}
As a consequence, we have
\begin{eqnarray}\label{commutator_mfkL1}
 \mfk{L}^\dagger_1\mfk{L}_1 =-\mfk{L}^2_0-\mfk{L}_0\sqrt{p}^{(1)}.
\end{eqnarray}

 On the other hand, $\mfk{L}_2$ and $\mfk{L}^\dagger_2$ are related to $\mfk{L}_0$ through Eqs. \eqnm{mfkL2}, we then obtain
\begin{eqnarray}
 \mfk{L}_2 \mfk{L}^\dagger_2&=&-\mfk{L}^2_0-\mfk{L}_0\sqrt{p}^{(1)}+2\left[\mfk{L}_0,\,\sqrt{p}^{(1)}\right]+2\smbr{\sqrt{p}^{(1)}}^2.\nonumber \\
 &=&-\mfk{L}^2_0-\mfk{L}_0\sqrt{p}^{(1)}+p^{(2)}.
\end{eqnarray}

Consequently, the following shape invariance relation of the first level is reached,
\begin{eqnarray}\label{SI_plus1_level}
 \mfk{L}^\dagger_1 \mfk{L}_1 - \mfk{L}_2 \mfk{L}^\dagger_2 = \delta_1,
\end{eqnarray}
where
\begin{eqnarray}\label{delta1}
 \delta_1 \equiv -p^{(2)}=2.
\end{eqnarray}

Substituting for $\mfk{L}^\dagger_1 \mfk{L}_1$ via the relation \eqnm{SI_plus1_level}, \eq{eigen_equ_1st_level_before_rmnlz} can be renormalized into the following one,
\begin{eqnarray}\label{eigen_equ_1st_level__rmnlzd}
L_1 \Phi_{j1} &\equiv \mfk{L}_2 \mfk{L}_2^\dagger   \Phi_{j1} = \lambda_{j1} \Phi_{j1},
\end{eqnarray}
where
\begin{eqnarray}
 \lambda_{j1} \equiv \lambda_{j0}- \delta_1.
\end{eqnarray}
\eq{eigen_equ_1st_level__rmnlzd} assumes the same pattern as \eq{Legendre_equ_right_dagger} in terms of the super/sub-scripts. This self-similar feature further justifies the definitions of $\mfk{L}_1$ and $\mfk{L}_2$ just introduced and is why the relation \eqnm{SI_plus1_level} is also called \emph{pattern/shape invariance} condition. \eq{eigen_equ_1st_level__rmnlzd} will work as the renewed-level eigenequation for generating the eigenequation of next neighboring higher association level.

It is amazing that in this procedure the eigenequation and its eigensolution $\Phi_{j1}$ and $\lambda_{j1}$ are obtained \emph{simultaneously}, simply by the two \emph{routine} manipulations in a consecutive SUSYQM algorithm \text{--} intertwining action plus renormalization, assuming that there exist the base-level eigensolution $\Phi_{j0}$ and eigenvalue $\lambda_{j0}$, without knowing their concrete forms. We will repeat this type of manipulations and construct the eigensolutions for all higher association levels.

Intertwiningly acting \eq{eigen_equ_1st_level__rmnlzd} with $\mfk{L}^\dagger_2$ yields
\begin{eqnarray}\label{eigen_equ_minus_2nd_level_before_rmnlz}
\begin{aligned}
 \mfk{L}^\dagger_2 \mfk{L}_2  \Phi_{j2} &= \lambda_{j1} \Phi_{j2},
\end{aligned}
\end{eqnarray}
where $\Phi_{j2} \propto \mfk{L}^\dagger_2 \Phi_{j1}$. On the other hand,  \eq{eigen_equ_1st_level__rmnlzd} implies that $\mfk{L}^\dagger_2 \Phi_{j1}$  has the norm $\sqrt{\lambda_{j1}}$. These two facts together with \eq{1to2_raising_relation} allow us to define the unity-normed $\Phi_{j2}$ as
\begin{eqnarray}\label{Phi_j2}
\begin{aligned}
 \Phi_{j2} &\equiv\frac{{\cal L}^\dagger_2}{\sqrt{\lambda_{j1}}} \Phi_{j1}=\frac{{\cal L}^\dagger_2}{\sqrt{\lambda_{j1}}} \frac{{\cal L}^\dagger_1}{\sqrt{\lambda_{j0}}} \Phi_{j0}, \\
           &=(-1)^2 \frac{\sqrt{p}^2}{\sqrt{\lambda_{j1}}\sqrt{\lambda_{j0}}} \Phi^{(2)}_{j0},
\end{aligned}
\end{eqnarray}
which exhibits the same pattern as $\Phi_{j1}$ given in \eq{raising_ladder_1}.

In order to renormalize \eq{eigen_equ_minus_2nd_level_before_rmnlz} into the same pattern of  \eq{eigen_equ_1st_level__rmnlzd} or \eq{Legendre_equ_right_dagger}, certain operators $\mfk{L}_3$ and $\mfk{L}^\dagger_3$ are needed to be appropriately devised. To this end, similar to the operation that leads to  \eq{2nd_level_introduction}, we act on $\Phi_{j2}$ in the form given in \eq{Phi_j2} with $\mfk{L}^\dagger_2=-\sqrt{p}^2 (d/dx) \sqrt{p}^{-1}$ one more time, and obtain
\begin{eqnarray}\label{introduction_3rd_level}
 \mfk{L}^\dagger_2 \Phi_{j2} &=&- \sqrt{p}^{(1)} \Phi_{j2} + (-1)^3 \frac{\sqrt{p}^3}{\sqrt{\lambda_{j1}}\sqrt{\lambda_{j0}}} \Phi^{(3)}_{j0}.
\end{eqnarray}
Collecting the two terms involving $\Phi_{j2}$ together, the above equation can be reorganized as
\begin{eqnarray}\label{lowering_recursion_minus_3rd_level}
\mfk{L}^\dagger_3 \Phi_{j2}=(-1)^3\frac{\sqrt{p}^3}{{\sqrt{\lambda_{j1}}\sqrt{\lambda_{j0}}}} \Phi^{(3)}_{j0},
\end{eqnarray}
where $\mfk{L}^\dagger_3$ and its Hermitian conjugate $\mfk{L}_3$ are defined as
\begin{eqnarray}\label{mfkL3}
\begin{aligned}
 \mfk{L}^\dagger_3 &\equiv \mfk{L}^\dagger_2 + \sqrt{p}^{(1)}=\mfk{L}^\dagger_0 + 3 \sqrt{p}^{(1)}={\cal S}^{-6}_a\smbr{\mfk{L}^\dagger_0},\\
 \mfk{L}_3 &\equiv \mfk{L}_2 + \sqrt{p}^{(1)}=\mfk{L}_0 + 3 \sqrt{p}^{(1)}={\cal S}^6_a\smbr{\mfk{L}_0}.
\end{aligned}
\end{eqnarray}

Note that $\mfk{L}_2$ and $\mfk{L}^\dagger_2$ are related to $\mfk{L}_0$ through Eqs. \eqnm{mfkL2}. We then obtain
\begin{eqnarray}\label{mfkLd2_mfkL2_product}
 \mfk{L}^\dagger_2 \mfk{L}_2 =-\mfk{L}^2_0-\mfk{L}_0\sqrt{p}^{(1)}-\left[\mfk{L}_0,\,\sqrt{p}^{(1)}\right]+2\smbr{\sqrt{p}^{(1)}}^2.\nonumber \\
\end{eqnarray}

On the other hand, $\mfk{L}_3$ and $\mfk{L}^\dagger_3$ are related to $\mfk{L}_0$ through Eqs. \eq{mfkL3}, and we have
\begin{eqnarray}\label{mfkL3_mfkLd3_product}
 \mfk{L}_3 \mfk{L}^\dagger_3=-\mfk{L}^2_0-\mfk{L}_0\sqrt{p}^{(1)}+3\left[\mfk{L}_0,\,\sqrt{p}^{(1)}\right]+6\smbr{\sqrt{p}^{(1)}}^2.\nonumber \\
\end{eqnarray}

Subtracting \eq{mfkL3_mfkLd3_product} from \eq{mfkLd2_mfkL2_product}, we reach the shape invariance relation of the second level as
\begin{eqnarray}\label{SI_plus2_level}
\mfk{L}^\dagger_2 \mfk{L}_2 - \mfk{L}_3 \mfk{L}^\dagger_3 = \delta_2,
 \end{eqnarray}
where
\begin{eqnarray}
 \delta_2 \equiv -4\left[\mfk{L}_0,\,\sqrt{p}^{(1)}\right]-4\smbr{\sqrt{p}^{(1)}}^2=-2 p^{(2)}=4.
\end{eqnarray}

Substituting for $\mfk{L}^\dagger_2 \mfk{L}_2$ via the relation \eqnm{SI_plus2_level}, \eq{eigen_equ_minus_2nd_level_before_rmnlz} is renormalized to
\begin{eqnarray}\label{eigen_equ_minus_2nd_level_after_rmnlz}
L_2\Phi_{j2} &\equiv \mfk{L}_3 \mfk{L}^\dagger_3 \Phi_{j2} = \lambda_{j2} \Phi_{j2},
\end{eqnarray}
where
\begin{eqnarray}
 \lambda_{j2} &\equiv \lambda_{j1}-\delta_2=\lambda_{j0}-(\delta_1+\delta_2).
\end{eqnarray}

Intertwiningly acting \eq{eigen_equ_minus_2nd_level_after_rmnlz} with $\mfk{L}^\dagger_3$ yields
\begin{eqnarray}\label{Superpartner_eigen_equ_2nd_level}
\mfk{L}^\dagger_3 \mfk{L}_3 \Phi_{j3} = \lambda_{j2} \Phi_{j3},
\end{eqnarray}
where $\Phi_{j3}\propto \mfk{L}^\dagger_3 \Phi_{j2}$.

\eq{eigen_equ_minus_2nd_level_after_rmnlz} implies that $\mfk{L}^\dagger_3 \Phi_{j2}$ has the norm $\sqrt{\lambda_{j2}}$. This fact together with \eq{Phi_j2} allow us to define $\Phi_{j3}$ with unity norm as
\begin{eqnarray}\label{Phi_j3}
\begin{aligned}
\Phi_{j3} &\equiv \frac{\mfk{L}^\dagger_3}{\sqrt{\lambda_{j2}}} \Phi_{j2}=\frac{\mfk{L}^\dagger_3}{\sqrt{\lambda_{j2}}} \frac{\mfk{L}^\dagger_2}{\sqrt{\lambda_{j1}}}\frac{\mfk{L}^\dagger_1}{\sqrt{\lambda_{j0}}}\Phi_{j0},\\
          &=(-1)^3\frac{\sqrt{p}^3}{\sqrt{\lambda_{j2}}{\sqrt{\lambda_{j1}}\sqrt{\lambda_{j0}}}} \Phi^{(3)}_{j0},
\end{aligned}
\end{eqnarray}
which exhibits the same pattern as $\Phi_{j2}$ given in \eq{Phi_j2} and $\Phi_{j1}$ given in \eq{raising_ladder_1}.

In order to renormalize \eq{Superpartner_eigen_equ_2nd_level} into
\begin{eqnarray}\label{eigen_equ_3nd_level}
L_3 \Phi_{j3} \equiv \mfk{L}_4 \mfk{L}^\dagger_4  \Phi_{j3} = \lambda_{j3} \Phi_{j3}
\end{eqnarray}
which assumes the same pattern as Eqs. \eqnm{eigen_equ_minus_2nd_level_after_rmnlz}, \eqnm{eigen_equ_1st_level__rmnlzd} and \eqnm{Legendre_equ_right_dagger}, we need the pattern invariance relation of the third level
\begin{eqnarray}\label{SI_plus3_level}
\mfk{L}^\dagger_3 \mfk{L}_3 - \mfk{L}_4 \mfk{L}^\dagger_4= \delta_3
\end{eqnarray}
for the operator $\mfk{L}_3$ and its Hermitian conjugate $\mfk{L}^\dagger_3$ to be appropriately devised and a certain constant $\delta_4$ as yet to be determined. In \eq{eigen_equ_3nd_level},
\begin{eqnarray}
 \lambda_{j3}\equiv \lambda_{j2}+\delta_3=\lambda_{j0}+\smbr{\delta_1+\delta_2+\delta_3}.
\end{eqnarray}

To such end, we compute $\mfk{L}^\dagger_3\Phi_{j3}$ with $\mfk{L}^\dagger_3=-\sqrt{p}d/dx+ 2\sqrt{p}^{(1)}$ and $\Phi_{j3}$ in the form given in the second line of \eq{Phi_j3}  and obtain
\begin{eqnarray}\label{introduction_4th_level}
 \mfk{L}^\dagger_3 \Phi_{j3} &=&- \sqrt{p}^{(1)} \Phi_{j3} + (-1)^4 \frac{\sqrt{p}^4}{\sqrt{\lambda_{j2}}\sqrt{\lambda_{j1}}\sqrt{\lambda_{j0}}} \Phi^{(4)}_{j0}.\nonumber \\
\end{eqnarray}
Merging the two terms involving $\Phi_{j3}$, this equation can be reorganized as
\begin{eqnarray}\label{lowering_recursion_minus_3rd_level1}
\mfk{L}^\dagger_4 \Phi_{j3}=(-1)^4\frac{\sqrt{p}^4}{\sqrt{\lambda_{j2}}\sqrt{\lambda_{j1}}\sqrt{\lambda_{j0}}} \Phi^{(4)}_{j0},
\end{eqnarray}
where $\mfk{L}^\dagger_3$ and its Hermitian conjugate $\mfk{L}_3$ are defined as
\begin{eqnarray}\label{mfkL4}
\begin{aligned}
 \mfk{L}^\dagger_4 &\equiv \mfk{L}^\dagger_3 + \sqrt{p}^{(1)}=\mfk{L}^\dagger_0 + 4 \sqrt{p}^{(1)}={\cal S}^{-8}_x\smbr{\mfk{L}^\dagger_0},\\
 \mfk{L}_4 &\equiv \mfk{L}_3 + \sqrt{p}^{(1)}=\mfk{L}_0 + 4 \sqrt{p}^{(1)}={\cal S}^8_x\smbr{\mfk{L}_0}.
\end{aligned}
\end{eqnarray}

With these $\mfk{L}_4$ and $\mfk{L}^\dagger_4$ in hand, we have
\begin{eqnarray}\label{mfkL4_mfkLd4_product}
\mfk{L}_4\mfk{L}^\dagger_4= -\mfk{L}^2_0-\mfk{L}_0\sqrt{p}^{(1)} + 4\left[\mfk{L}_0,\,\sqrt{p}^{(1)}\right] + 12\smbr{\sqrt{p}^{(1)}}^2.\nonumber \\
\end{eqnarray}
On the other hand, with $\mfk{L}_3$ and $\mfk{L}^\dagger_3$ given in \eq{mfkL3}, we have
\begin{eqnarray}\label{mfkLd4_mfkL4_product}
 \mfk{L}^\dagger_3 \mfk{L}_3 = -\mfk{L}^2_0-\mfk{L}_0\sqrt{p}^{(1)}-2\left[\mfk{L}_0,\,\sqrt{p}^{(1)}\right]+6\smbr{\sqrt{p}^{(1)}}^2.\nonumber \\
\end{eqnarray}

Subtracting \eq{mfkL4_mfkLd4_product} from \eq{mfkLd4_mfkL4_product}, we reach the relation \eqnm{SI_plus3_level} with the constant $\delta_3$ given by
\begin{eqnarray}\label{delta_3}
 \delta_3 \equiv-6\left[\mfk{L}_0,\,\sqrt{p}^{(1)}\right]-6\smbr{\sqrt{p}^{(1)}}^2=-3 p^{(2)}=6.
\end{eqnarray}

Keeping this iteration going to the $m$-th positive association level ($m>3$), the following generic patterns emerge.
The implicitly supersymmetrically factorized associated Legendre equation of $m$-th association level reads
\begin{eqnarray}\label{eigen_equ_normalized_ascending_generic}
 L_m \Phi_{jm}&\equiv \mfk{L}_{m+1}  \mfk{L}^\dagger_{m+1} \Phi_{jm} = \lambda_{jm} \Phi_{jm},
\end{eqnarray}
where the two ladder operators are defined as
\begin{eqnarray}\label{mfkL_mplus1}
\begin{aligned}
\mfk{L}_{m+1} &\equiv \mfk{L}_0 + \smbr{m+1} \sqrt{p}^{(1)},\\
               &=\sqrt{p}^{-m}\smbr{\od{}{x}}\sqrt{p}^{m+1},\\
               &={\cal S}^{-2(m+1)}_x\smbr{\mfk{L}_0},
\end{aligned}
\end{eqnarray}
\begin{eqnarray}\label{mfkLd_mplus1}
\begin{aligned}
  \mfk{L}^\dagger_{m+1} &\equiv \mfk{L}^\dagger_0 + (m+1) \sqrt{p}^{(1)},\\
                      &=-\sqrt{p}^{m+1}\smbr{\od{}{x}}\sqrt{p}^{-m}, \\
                      &={\cal S}^{2(m+1)}_x\smbr{\mfk{L}^\dagger_0},
\end{aligned}
\end{eqnarray}
where ${\cal S}^{\pm 2(m+1)}_x$ are the $2\smbr{m+1}$-th power of ${\cal S}^{\pm1}_x$,
the eigenvalue is given by
\begin{eqnarray}\label{lambda_jm}
\lambda_{jm} &\equiv \lambda_{j0} -\sum^m_{n=1}\delta_n=\lambda_{j0}-m\smbr{m+1},
\end{eqnarray}
the pattern invariance relation is
\begin{eqnarray}\label{shape_inviance_condition_ascending_generic}
 \mfk{L}^\dagger_n \mfk{L}_n-\mfk{L}_{n+1}\mfk{L}^\dagger_{n+1} = \delta_n= 2n.
\end{eqnarray}

The generic raising recurrence relation reads
\begin{eqnarray}\label{Phi_jm_raising_ladder}
 \Phi_{jm} &=& \frac{\mfk{L}^\dagger_m}{\lambda^{1/2}_{jm-1}} \Phi_{j,m-1}.
\end{eqnarray}

Combining \eq{Phi_jm_raising_ladder} and  \eq{eigen_equ_normalized_ascending_generic} with $m+1 \rightarrow m$, we obtain the lowering recurrence relation
\begin{eqnarray}\label{Phi_jmminus1_lowering_ladder}
  \Phi_{jm-1} = \frac{{\mfk L}_m}{\lambda^{1/2}_{jm-1}} \Phi_{jm}.
\end{eqnarray}

The consecutive accumulation of the raising relation \eqnm{Phi_jm_raising_ladder} for $m\rightarrow m-1,\, m-2,\, ,\cdots,\,2,\,1$ then gives us the relation between $\Phi_{jm}$ and $\Phi_{j0}$ as
\begin{eqnarray}
 \Phi_{jm}(x) &=&\frac{\mfk{L}^\dagger_m}{\lambda^{1/2}_{jm-1}}\frac{\mfk{L}^\dagger_{m-1}}{\lambda^{1/2}_{jm-2}}\cdots\frac{\mfk{L}^\dagger_2}{\lambda^{1/2}_{j1}} \frac{\mfk{L}^\dagger_1}{\lambda^{1/2}_{j0}} \Phi_{j0}(x),\nonumber\\
           &=&\frac{(-1)^m \sqrt{p}^m}{\lambda^{1/2}_{jm-1}\lambda^{1/2}_{jm-2}\cdots \lambda^{1/2}_{j1}\lambda^{1/2}_{j0}} \Phi^{(m)}_{j0}(x),\label{Phi_jm_bottom_up}
\end{eqnarray}
where we have made use of the lumped form of $\mfk{L}^\dagger_m$ in \eq{mfkLd_mplus1} to shorten the product of $m$ raising operators.

Up to this point, we have iteratively built the associated Legendre equation of generic association level as well as it eigensolution \emph{simultaneously} in a \emph{bottom-up} approach. In this approach, the concrete forms (details) of the initial solutions $\Phi_{j0}=\Phi_j$ and $\lambda_{j0}=\lambda_j$ are assumed to exist but not invoked during the whole derivations. If these concrete forms are inputted from other approaches such as from those demonstrated in Section \ref{SUSYQM_base_Legendre_operator}, we immediately obtain the concrete forms of the generic eigensolutions $\Phi_{jm}$ and $\lambda_{jm}$. However, in what follows we will demonstrate that the concrete forms of the initial solutions $\Phi_{j0}=\Phi_j$ and $\lambda_{j0}=\lambda_j$ from outside of this approach are not indispensable for obtaining these concrete forms of the generic eigensolutions. Instead, they (including those of $\Phi_j$ and $\lambda_j$ as the special $m=0$ case) can be obtained in a certain \emph{top-down} approach \emph{within} the present algorithm, once the finiteness of the association levels is taken into account.

If we assume that there exists a maximum positive association level $m_{max}$, denoted by $m_{max}\equiv j$, such that $\lambda_{jj}=0$, $\lambda_j=\lambda_{j0}$ is then fixed from \eq{lambda_jm} as
\begin{eqnarray}\label{lambda_j0}
\lambda_{j0}=j(j+1).
\end{eqnarray}
This is the same as the eigenvalue $\lambda_j$ of the base Legendre operator $L(x)$, obtained by other approaches in  \eq{eigenvalues_lambda_and_E} and in \eq{base_eigen_values}.

Recall that  in Subsections \ref{detailed_factorization_scheme_y_coordinate} and \ref{nonhermitian_factorization}, the egienvalue $\lambda_j$ is fixed by imposing the shape invariance conditions there. Here $\lambda_{jm}$ or $\lambda_{j0}$ cannot be fixed by the shape invariance condition \eqnm{shape_inviance_condition_ascending_generic} itself. This is because now the relevant ladder operators $\mfk{L}_m$ defined in \eq{mfkL_mplus1} and $\mfk{L}^\dagger_m$ in \eq{mfkLd_mplus1} do not involve $\lambda_{jm}$ or $\lambda_{j0}$ at all.

As a result of $\lambda_{jj}=0$ for $m=j$, \eq{eigen_equ_normalized_ascending_generic} reduces to the \emph{annihilation} condition
\begin{eqnarray}\label{annihilation_condition_top_m}
 \mfk{L}^\dagger_{j+1} \Phi_{jj}=\sqbr{-\sqrt{p}\od{}{x} + j \sqrt{p}^{(1)}}\Phi_{jj}=0.
\end{eqnarray}
This means that the iterative raising procedure, following the recurrence relation \eqnm{Phi_jm_raising_ladder}, terminates beyond the highest positive association level $m_{max}=j$ and $\Phi_{jj+1}=0$.

\eq{annihilation_condition_top_m} is easily integrated and we obtain $\Phi_{jj}\propto \sqrt{p}^j$.  Denote $I^j \equiv \int^1_{-1} d x \sqbr{\sqrt{p}^j}^2$, we then have $I^0=2$ and the recurrence relation $I^j/I^{j-1}= 2j/(2j+1)$. It then follows that $I^j= 2^{j+1}j!/(2j+1)!!$ for integer $j\ge 0$. Thus, $\Phi_{jj}(x)$ that is normalized to unity can be defined as
\begin{eqnarray}\label{Phi_jj_from_annihilation_condition}
 \Phi_{jj}(x)=(-1)^j\frac{\sqrt{p}^j}{\sqrt{I^j}}=(-1)^j \frac{\sqrt{(2j+1)!!}}{\sqrt{2^{j+1}j!} } \sqrt{p}^j,
\end{eqnarray}
where the phase factor $(-1)^j$ is attached so that the resulting $\Phi_{jm}(x)$ is the same as $\Theta_{jm}(\theta)$ in \eq{Theta_jm_top_down} with the substitution $x=\cos\theta$ ($\Theta_{jm}(\theta)$ will be obtained in Subsubsection \ref{eigen_solutions_associated_equ_theta_coord} via directly factorizing the associated Legendre equation in polar angle $\theta$ coordinate). On the other hand, the $m_{max}=j$ case of \eq{Phi_jm_bottom_up} reads
\begin{eqnarray}\label{Phi_jj_from_bottom_up_iteration}
\Phi_{jj}(x) = \frac{(-1)^j\Phi^{(j)}_{j0}}{\lambda^{1/2}_{jj-1}\lambda^{1/2}_{jj-2}\cdots \lambda^{1/2}_{j1} \lambda^{1/2}_{j0}} \sqrt{p}^j= \frac{(-1)^j \Phi^{(j)}_{j0}}{\sqrt{(2j)!}}\sqrt{p}^j.\nonumber \\
\end{eqnarray}
Comparing \eq{Phi_jj_from_bottom_up_iteration} with \eq{Phi_jj_from_annihilation_condition}, $\Phi^{(j)}_{j0}$ is fixed as
\begin{eqnarray}
 \Phi^{(j)}_{j0} = \sqrt{\frac{2j+1}{2}} (2j-1)!!.
\end{eqnarray}

With $\lambda_{j0}$ in \eq{lambda_j0} and $\Phi_{jj}$ in \eq{Phi_jj_from_annihilation_condition} as inputs, we can obtain a generic formula of $\Phi_{jm}$ for $0\le m< j$, by consecutively applying the lowering recurrence relation \eqnm{Phi_jmminus1_lowering_ladder}. Soon we will see that, through such a top-down approach, even the base Legendre function $\Phi_{j}=\Phi_{j0}$ (in a nice compact form) can be obtain, without resort to any results out of the SUSYQM algorithms developed in Section \ref{SUSYQM_base_Legendre_operator} for solving the base Legendre equation. In such an algorithm, the only input is the base Legendre equation itself.

The consecutive accumulation of the lowering relation \eqnm{Phi_jmminus1_lowering_ladder} for $m-1 \rightarrow m,\, m+1,\,\cdots,\, j-1 \,\text{and}\, j$ yields
\begin{eqnarray}\label{Phi_jm_top_down}
 \Phi_{jm}(x)=\frac{{\mfk L}_{m+1}}{\lambda^{1/2}_{jm}}\frac{{\mfk L}_{m+2}}{\lambda^{1/2}_{jm+1}}\cdots \frac{{\mfk L}_{j-1}}{\lambda^{1/2}_{jj-2}}\frac{{\mfk L}_j}{\lambda^{1/2}_{jj-1}}\Phi_{jj}(x).
\end{eqnarray}

Making use of the lumped form of $\mfk{L}_m$ in \eq{mfkL_mplus1} for $m \rightarrow m+1,\,m+2,\,\cdots,\,j-1\,\text{and}\, j$, the operator product in \eq{Phi_jm_top_down} can be shortened as
\begin{eqnarray}\label{product_mfkL_mplus1_to_j}
 {\mfk L}_{m+1}{\mfk L}_{m+2}\cdots {\mfk L}_{j-1}{\mfk L}_j= \sqrt{p}^{-m}\smbr{\od{}{x}}^{j-m}\sqrt{p}^j.
\end{eqnarray}
On the other hand, substituting for $\lambda_{jm}=(j-m)(j+m+1)$ for $m \rightarrow m,\,m+1,\,m+2,\,\cdots,\,j-2 \,\text{and}\, j-1$, we have
\begin{eqnarray}\label{product_lambda_jm_tojj-1}
 \lambda_{jm}\lambda_{jm+1}\cdots \lambda_{jj-2} \lambda_{jj-1}=(2j)!\frac{(j-m)!}{(j+m)!}.
\end{eqnarray}

Inserting Eqs. \eqnm{Phi_jj_from_annihilation_condition}, \eqnm{product_mfkL_mplus1_to_j} and \eqnm{product_lambda_jm_tojj-1}, \eq{Phi_jm_top_down} is shortened into
\begin{eqnarray}\label{Phi_jm_top_down_final}
\begin{aligned}
 \Phi_{jm}(x) &= \frac{(-1)^j}{2^j j!} \sqrt{\frac{2j+1}{2}}\sqrt{\frac{(j+m)!}{(j-m)!}}\sqrt{p}^{-m}\smbr{\od{}{x}}^{j-m} p^j, \\
 &= (-1)^m \sqrt{\frac{2j+1}{2}} \sqrt{\frac{(j-m)!}{(j+m)!}} P^m_j(x),
 \end{aligned}
\end{eqnarray}
where
\begin{eqnarray}\label{Pmj_1st_form}
 \hspace{-10pt} P^m_j(x)\equiv \frac{(-1)^{j+m}}{2^jj!}\frac{(j+m)!}{(j-m)!} \sqrt{p}^{-m }(x) \smbr{\od{}{x}}^{j-m} p^j(x)
\end{eqnarray}
is the usual associated Legendre function of $m$-th association level that is not normalized to unity, see, e.g., \cite{Merzbacher_QM_1998}.

For $m=0$, \eq{Phi_jm_top_down_final} reduces to the base Legendre polynomial in its Rodriguez's formula
\begin{eqnarray}\label{Phi_j0_top_down}
 \Phi_{j0}(x)= \frac{(-1)^j}{2^j j!} \sqrt{\frac{2j+1}{2}} \smbr{\od{}{x}}^j p^j,
\end{eqnarray}
which has a unity norm.

Making use of $\lambda_{jm}=(j-m)(j+m+1)$ for $m \rightarrow m-1,\, m-2,\,\cdots,\, 1 \,\text{and}\, 0$, we have
\begin{eqnarray}\label{lambda_jm_product_mminus1_to_1}
 \lambda_{jm-1}\lambda_{jm-2}\cdots \lambda_{j1}\lambda_{j0} =\frac{(j+m)!}{(j-m)!}.
\end{eqnarray}

Substituting $\Phi_{j0}$ via \eq{Phi_j0_top_down} and $\lambda_{jm-1}\lambda_{jm-2}\cdots \lambda_{j1}\lambda_{j0}$ via \eq{lambda_jm_product_mminus1_to_1}, the bottom-up form of $\Phi_{jm}$ in \eq{Phi_jm_bottom_up} is recast into
\begin{eqnarray}\label{Phijm_plus_alternative}
\begin{aligned}
\Phi_{jm}&=\frac{(-1)^{j+m}}{2^j j!}\sqrt{\frac{2j+1}{2}}\sqrt{\frac{(j-m)!}{(j+m)!}}\sqrt{p}^m \smbr{\od{}{x}}^{j+m} p^j,\\
         &= (-1)^m \sqrt{\frac{2j+1}{2}}\sqrt{\frac{(j-m)!}{(j+m)!}} P^m_j,
\end{aligned}
\end{eqnarray}
where
\begin{eqnarray}\label{P_jm_alternative_form}
 P^m_j(x) \equiv \frac{(-1)^j}{2^jj!} \sqrt{p}^m(x) \smbr{\od{}{x}}^{j+m} p^j(x)
\end{eqnarray}
is an alternative expression of the associated Legendre function of $m$-th level besides that given in \eq{Pmj_1st_form}.

In \eq{Phi_jm_top_down_final} or \eq{Phijm_plus_alternative}, $\Phi_{jm}(x)$ is different from $P^m_j(x)$ by a normalization factor. It is noteworthy that, at each $m$-th level of intertwining action, if, instead of defining the raising recurrence relation as \eq{Phi_jm_raising_ladder} for  a unity-normed $\Phi_{jm}(x)$, we simply define the raising recurrence relation as $\Phi_{jm}(x) \equiv \mfk{L}^\dagger_m \Phi_{jm-1}(x)$, then the resultant $\Phi_{jm}(x)$ coincides with the usual associated Legendre function  $P^m_j(x)$.

We have built the associated Legendre \eq{eigen_equ_normalized_ascending_generic} of generic association level, with the associated Legendre operator coming out as the product of a Hermitian conjugate pair, i.e., $L_m={\mfk{L}_{m+1}}{\mfk{L}^\dagger_{m+1}}$. However, this factorization is still asymmetric. One may wonder what the explicit (standard) supersymmetric factorization of $L_m$ is. With in mind the lumped forms of ${\mfk{L}_{m+1}}$ given in \eq{mfkL_mplus1} and that of ${\mfk{L}^\dagger_{m+1}}$ in \eq{mfkLd_mplus1}, applying ${\cal S}_x$ to $L_m={\mfk{L}_{m+1}}{\mfk{L}^\dagger_{m+1}}$, we obtain the explicit supersymmetric factorization
\begin{eqnarray}\label{tldLm_symmetric_factorization}
 \tilde{L}_m = \tilde{\mfk{L}}_{m+1} \tilde{\mfk{L}^\dagger}_{m+1},
\end{eqnarray}
where
\begin{eqnarray}
 \tilde{L}_m\equiv {\cal S}_x\smbr{L_m},
\end{eqnarray}
\begin{eqnarray}\label{tldmfkLmplus1}
\begin{aligned}
 \tilde{\mfk{L}}_{m+1}&\equiv {\cal S}_x\smbr{\mfk{L}_{m+1}}={\cal S}^{-2\smbr{m+1}}_x\smbr{\tilde{\mfk{L}}_0},\\
                      &={\cal S}^{-\smbr{2m+1}}_x\smbr{\sqrt{p}\od{}{x}},\\
                      &=\sqrt{p}\od{}{x}+W^a_{m+1}(x),
 \end{aligned}
\end{eqnarray}
\begin{eqnarray}\label{tldmfkLdaggermplus1}
\begin{aligned}
 \tilde{\mfk{L}^\dagger}_{m+1}&\equiv {\cal S}_x\smbr{\mfk{L}^\dagger_{m+1}}={\cal S}^{2\smbr{m+1}}_x\smbr{\tilde{\mfk{L}}^\dagger_0},\\
                      &={\cal S}^{2m+1}_x\smbr{-\sqrt{p}\od{}{x}},\\
                      &=-\sqrt{p}\od{}{x}+W^a_{m+1}(x),
 \end{aligned}
\end{eqnarray}
and
\begin{eqnarray}
 W^a_m(x)\equiv (1-2m) W^a_0(x)=\smbr{m-\frac{1}{2}}\sqrt{p}^{(1)}
\end{eqnarray}
is the superpotential of the $m$-th association level.

Here we see that the same supersymmetrization transformation ${\cal S}_x$ that supersymmtrizes the asymmetric factorization $L_0=\mfk{L}^\dagger_0 \mfk{L}_0= \mfk{L}_1\mfk{L}^\dagger_1$ also supersymetrizes the asymmetric factorization $L_m={\mfk{L}_{m+1}}{\mfk{L}^\dagger_{m+1}}$. This is mainly because $L_m$ is equal to $L_0$ plus an ordinary function of $x$.

The standard supersymmetrically factorized Hamiltonian corresponding to the symmetric factorization \eqnm{tldLm_symmetric_factorization} is easily obtained by the momentum operator map \eqnm{momentum_operator_map2}  and going to $\theta$-coordinate, which will be addressed in Subsubsection \ref{direct_factorization_AL_equ_theta_coordinate}. Another form of standard supersymmetrically factorized Hamiltonian corresponding to the associated Legendre differential operator will be obtained in $y$ coordinate demonstrated in Subsubsection \ref{associated_Legendre_polynomials_y_coord}.

Applying ${\cal S}_x$ to the pattern invariance condition \eqnm{shape_inviance_condition_ascending_generic}, we obtain its incarnation in terms of  $\tilde{\mfk{L}}_m$ and $\tilde{\mfk{L}^\dagger}_m$ as
\begin{eqnarray}\label{shape_inviance_condition_ascending_standard}
 \tilde{\mfk{L}}^\dagger_n \tilde{\mfk{L}}_n-\tilde{\mfk{L}}_{n+1}\tilde{\mfk{L}}^\dagger_{n+1}= \delta_n =2n.
\end{eqnarray}
That is, the supersymmetrization transformation ${\cal S}_x$ establishes the 1-to-1 correspondence between the pattern invariance condition expressed in terms of the non-standard ladder operators and the standard ones. It is in this sense that we says that such transformation preserves the pattern invariance condition.

Now we summarize the SUSYQM algorithm built in this subsection. Once the initial \emph{defining} factorization $L_0\equiv L=\mathfrak{L}_1\mathfrak{L}^\dagger_1$,  which is an implicit supersymmetric factorization, is fixed,  the very existence of  associated Legendre equation of higher levels and its eigensolution are completely controlled by the shape invariance symmetry between any two neighboring levels, as this symmetry allows the supersymmetric factorization continue to proliferate to higher and higher association levels. In this sense, SUSY and shape invariance symmetry work together as the two dictating symmetry principles for the very existence of the associated Legendre functions. In the algorithm of proliferating all levels of associated Legendre equations and theire eigensolutions, the key operation flow consists of the repeated intertwining action and renormalization. The main steps of this inductive SUSYQM algorithm are summarized as follows. 1) prepare the prerenormalized eigenequation by acting the previous (last) level renormalized eigenequation with the raising operator for that level; 2) introduce the raising operator, raising recursion relation and eigenfunction for the present level; 3) prepare the shape invariance condition for renormalizing the eigenequation of present level; 4) renormalize the eigenequation of present level.

 Of course, with almost the same amount of efforts, one can implement the same SUSYQM algorithm for building  the associated Legendre equation and its eigensolution in the $\theta$-coordinate parametrization, starting with the initial factorized base Legendre equation \eq{Legendre_equ_theta_coord0}.

\subsection{Negative Association Levels}\label{Negative_Association_Levels}

In the previous subsection, we have just built a tower of associated Lgendre equations and their eigensolutions \emph{simultaneously} from the base Legendre equation in ascending order of the \emph{positve} association level $m$ ($m=1,\,2,\,\cdots,\,j$), by repeatedly perform the intertwining action and renormalization. Now we start with the same base Legendre equation and try to follow a similar inductive procedure to build a tower of associated Legendre equations of of \emph{negative} association levels and their eigensolutions simultaneously  but in descending (negatively ascending) order, assuming that the generic associated Legendre equation of negative association level is blind to us.

To this end, the initial factorization $L_0(x)\equiv L(x)=\mfk{L}^\dagger_0(x)\mfk{L}_0(x)$ in \eq{self_adjoint_form1} should be adopted, simply because the right-situated factor $\mfk{L}_0(x)$ serves as the lowering operator in our convention that \emph{lower} the associated level by one unit each time. The appropriateness of this adoption will be justified as we proceed. For convenience, we relabel eigenfunction $\Phi_j$ as $\Phi_{j,0}$ and eigenvalue $\lambda_j$ as $\lambda_{j,0}$, and the factorized version of base Legendre equation \eqnm{Legendre_equ_x_coord} reads
\begin{eqnarray}\label{L_L_eigen_equ1}
L_0\Phi_{j,0}\equiv \mathfrak{L}^\dagger_0 \mathfrak{L}_0 \Phi_{j,0}=\lambda_{j,0} \Phi_{j,0},
\end{eqnarray}
which will work as the startup eigenequation in its implicit supersymmetric factorization form for constructing the associated Legendre equations and their eigensolutions with negative association levels.

Upon intertwiningly acting \eq{L_L_eigen_equ1} with $\mfk{L}_0$, the right factor of the factorization $\mathfrak{L}^\dagger_0 \mathfrak{L}_0$ in \eq{L_L_eigen_equ1}, we obtain the superpartner eigenequation
\begin{eqnarray}\label{eigen_equ_1st_assoc_level_before_rnmlz}
   \mfk{L}_0  \mfk{L}^\dagger_0\Phi_{j,-1} &= \lambda_{j,0} \Phi_{j,-1},
\end{eqnarray}
where
\begin{eqnarray}\label{Phi_jminus1_preparing}
   \Phi_{j,-1} \propto  \mfk{L}_0  \Phi_{j,0},
\end{eqnarray}
Soon we will see that \eq{eigen_equ_1st_assoc_level_before_rnmlz} works as the breakthrough eigenequation for marching towards the rest hierarchy of eigen-equations with higher negative association levels and their eigensolutions.

Note that $\mathfrak{L}^\dagger_0$ and $\mathfrak{L}_0$ are Hermitian conjugate to each other. \eq{L_L_eigen_equ1} then implies that $\mathfrak{L}_0\Phi_{j,0}$ has the norm $\lambda^{1/2}_{j,0}$. This allows us to promote the proportionality relation \eqnm{Phi_jminus1_preparing} to the equation
\begin{eqnarray}\label{Phi_jminus1_equ}
   \Phi_{j,-1} =  \frac{\mfk{L}_0}{\lambda^{1/2}_{j0}}  \Phi_{j,0} = \frac{\sqrt{p}}{{\lambda^{1/2}_{j,0}}}\Phi^{(1)}_{j,0},
\end{eqnarray}
in which $\Phi_{j,-1}$ has a unity norm. Formally, this shows that  $\mathfrak{L}_0$ works as a lowering operator that lowers the association level by one unit.

To proceed, we need the following pattern invariance condition
\begin{eqnarray}\label{SI_minus1_level}
 \mathfrak{L}_0 \mathfrak{L}^\dagger_0  -  \mathfrak{L}^\dagger_{-1} \mathfrak{L}_{-1} = \delta_1
\end{eqnarray}
for a certain constant $\delta_1$ to be determined and the Hermitian conjugate pair of first-order differential operators $\mfk{L}^\dagger_{-1}$ and  $\mfk{L}_{-1}$ to be designed, in order to renormalize \eq{eigen_equ_1st_assoc_level_before_rnmlz} into the form
\begin{eqnarray}\label{eigen_equ_1st_assoc_level_after_rnmlz}
L_{-1}\Phi_{j,-1}&\equiv \mfk{L}^\dagger_{-1} \mfk{L}_{-1}  \Phi_{j,-1} = \lambda_{j,-1} \Phi_{j,-1},
\end{eqnarray}
where
\begin{eqnarray}
 \lambda_{j,-1} &\equiv \lambda_{j,0} -\delta_1.
\end{eqnarray}
\eq{eigen_equ_1st_assoc_level_after_rnmlz} exhibits the same pattern as the startup eigenequation \eqnm{L_L_eigen_equ1}, in particular, the daggered operator is situated on the left of the un-daggered one. Indeed, formally the former is  obtained from the latter simply by the subscript substitution $0\rightarrow-1$.

One can make a clever guess on how $\mfk{L}_{-1}$ and $\mfk{L}^\dagger_{-1}$ look like. Recall that in Eqs. \eqnm{eigen_equ_1st_assoc_level_before_rnmlz} and \eqnm{Phi_jminus1_equ}, we have claimed that $\mfk{L}_0$ works as a lowering operator. One may wonder what would happen if acting \eq{Phi_jminus1_equ} with $\mfk{L}_0\equiv \sqrt{p} d/dx$ one more time. It is straightforward to verify that
\begin{eqnarray}\label{for_def_L_plus_1}
\begin{aligned}
 \mfk{L}_0 \Phi_{j,-1} = \sqrt{p}^{(1)} \Phi_{j,-1} + \frac{\sqrt{p}^2}{\lambda^{1/2}_{j0}} \Phi^{(2)}_{j,0}.
\end{aligned}
\end{eqnarray}
where \eq{Phi_jminus1_equ} has been made us of. It is instructive to merge the two terms involving $\Phi_{j,-1}$ in the above equation and define $\mfk{L}_{-1}$ and its Hermitian conjugate $\mfk{L}^\dagger_{-1}$ as
\begin{eqnarray}\label{mfkL_minus1}
 \mfk{L}_{-1} &\equiv \mfk{L}_0-\sqrt{p}^{(1)},\quad \mfk{L}^\dagger_{-1} &\equiv \mfk{L}^\dagger_0-\sqrt{p}^{(1)},
\end{eqnarray}
so that \eq{for_def_L_plus_1} is rewritten as
\begin{eqnarray}\label{mfkL_minus1Phijminus1}
  \mfk{L}_{-1} \Phi_{j,-1} = \frac{\sqrt{p}^2}{\lambda^{1/2}_{j,0}}\Phi^{(2)}_{j,0}.
\end{eqnarray}
Here $\mfk{L}_{-1}$ is obtained by shifting $\mfk{L}_0$ \emph{downwards} by one unit of $\sqrt{p}^{(1)}$, which justifies why the subscript $-1$ has been assigned.

Combining the definition of $\mfk{L}_0$ in \eq{mfkL_0} and that of $\mfk{L}^\dagger_0$ in \eq{mfkLd_0}, we have $\mfk{L}^\dagger_0=-\mfk{L}_0 -\sqrt{p}^{(1)}$ and
\begin{eqnarray}\label{product_mfkL0_mfkLd0}
 \mathfrak{L}_0 \mathfrak{L}^\dagger_0=-\mfk{L}^2_0-\mfk{L}_0\sqrt{p}^{(1)}.
\end{eqnarray}
On the other hand, the product of $\mfk{L}^\dagger_{-1}$  and $\mfk{L}_{-1}$ defined in \eq{mfkL_minus1} is
\begin{eqnarray}\label{product_mfkLdm1_mfkLm1}
 \mathfrak{L}^\dagger_{-1} \mathfrak{L}_{-1}=-\mfk{L}^2_0-\mfk{L}_0\sqrt{p}^{(1)}+2\left[\mfk{L}_0,\,\sqrt{p}^{(1)}\right]+2\smbr{\sqrt{p}^{(1)}}^2,\nonumber \\
\end{eqnarray}
where $[\,,\,]$ is the usual commutator. Subtracting \eq{product_mfkLdm1_mfkLm1} from \eq{product_mfkL0_mfkLd0}, we indeed reach the pattern invariance condition \eqnm{SI_minus1_level} with the constant $\delta_1$ given by
\begin{eqnarray}
 \delta_1\equiv -2\left[\mfk{L}_0,\,\sqrt{p}^{(1)}\right]-2\smbr{\sqrt{p}^{(1)}}^2=-p^{(2)}=2.
\end{eqnarray}

The symmetry relations ${\mfk L}_0=-{\mfk L}^\dagger_1$ and ${\mfk L}^\dagger_0=-{\mfk L}_1$ (see \eq{mfkL_1_def}) together give us $\mathfrak{L}_0 \mathfrak{L}^\dagger_0={\mfk L}^\dagger_1{\mfk L}_1$. On the other hand, the definitions \eqnm{mfkL_minus1} for ${\mfk L}_{-1}$ and ${\mfk L}^\dagger_{-1}$  and the definitions \eqnm{mfkL2} for ${\mfk L}_2$ and ${\mfk L}^\dagger_2$ together yield symmetry relations of the same pattern
\begin{eqnarray}\label{symmetry_minus_plus_1st_level}
 {\mfk L}_{-1}=-{\mfk L}^\dagger_2,\quad {\mfk L}^\dagger_{-1}=-{\mfk L}_2,
\end{eqnarray}
and hence ${\mfk L}^\dagger_{-1}{\mfk L}_{-1}={\mfk L}_2{\mfk L}^\dagger_2$.
Therefore,
\begin{eqnarray}
 \mathfrak{L}_0 \mathfrak{L}^\dagger_0  -  \mathfrak{L}^\dagger_{-1} \mathfrak{L}_{-1}
 &=&\mathfrak{L}^\dagger_1 \mathfrak{L}_1  -  \mathfrak{L}_2 \mathfrak{L}^\dagger_2 =\delta_1.
\end{eqnarray}
This explains why we use the same notation $\delta_1$ in the two pattern invariance conditions \eqnm{SI_plus1_level} and \eqnm{SI_minus1_level}.

\eq{eigen_equ_1st_assoc_level_after_rnmlz} works as the renewed step-stone for building the eigenequation of the next association level $m=-2$. Intertwiningly acting this equation with $\mfk{L}_{-1}$, its superpartner eigenequation is obtained as
\begin{eqnarray}\label{eigen_equ_2nd_assoc_level_before_rnmlz}
  \mfk{L}_{-1} \mfk{L}^\dagger_{-1}  \Phi_{j,-2}=\lambda_{j,-1} \Phi_{j,-2},
\end{eqnarray}
where
\begin{eqnarray}\label{Phi_jminus2_preparing}
\Phi_{j,-2}\propto \mfk{L}_{-1} \Phi_{j,-1}.
\end{eqnarray}

\eq{eigen_equ_1st_assoc_level_after_rnmlz} implies that $\mfk{L}_{-1}  \Phi_{j,-1}$ has the norm $\lambda^{1/2}_{j,-1}$, in view that $\Phi_{j,-1}$ has a unity norm. This allows us to promote the proportionality relation \eqnm{Phi_jminus2_preparing} to the equation
\begin{eqnarray}\label{Phi_l_minus2_tentative_def}
 \Phi_{j,-2} \equiv \frac{\mfk{L}_{-1}}{\lambda^{1/2}_{j,-1}} \Phi_{j,-1} = \frac{\sqrt{p}^2}{\lambda^{1/2}_{j,-1}\lambda^{1/2}_{j,0}}\Phi^{(2)}_{j,0},
\end{eqnarray}
where \eq{mfkL_minus1Phijminus1} has been made use of to reach the second equality.
This lowering recurrence relation between $\Phi_{j,-2}$ and $\Phi_{j,-1}$ has the same pattern as the lowering relation \eqnm{Phi_jminus1_equ} between $\Phi_{j,-1}$ and $\Phi_{j,0}$.

To proceed, we need to appropriately devise the operator $\mfk{L}_{-2}$ and its Hermitian conjugate $\mfk{L}^\dagger_{-2}$.  These two operators are required to be subject to the pattern invariance condition
\begin{eqnarray}\label{SI_minus2_level}
 \mathfrak{L}_{-1} \mathfrak{L}^\dagger_{-1}  -  \mathfrak{L}^\dagger_{-2} \mathfrak{L}_{-2} =\delta_2
\end{eqnarray}
for some constant $\delta_2$ to be determined, which assumes the same pattern as the condition \eqnm{SI_minus1_level}. If this condition does hold, \eq{eigen_equ_2nd_assoc_level_before_rnmlz} is then renormalized into
\begin{eqnarray}\label{eigen_equ_2nd_assoc_level_after_rnmlz}
L_{-2} \Phi_{j,-2} \equiv \mfk{L}^\dagger_{-2} \mfk{L}_{-2}  \Phi_{j,-2} =\lambda_{j,-2} \Phi_{j,-2},
\end{eqnarray}
where
\begin{eqnarray}
  \lambda_{j,-2} \equiv \lambda_{j,-1} -\delta_2 =\lambda_{j,0} -\smbr{\delta_1 + \delta_2}.
\end{eqnarray}

Similar to what has been done around \eq{for_def_L_plus_1} for introducing $\mfk{L}^\dagger_{-1}$, now we act \eq{Phi_l_minus2_tentative_def} one more time with $\mfk{L}_{-1}$ given in \eq{mfkL_minus1}, in order to introduce $\mfk{L}_{-2}$. It is easy to check that
\begin{eqnarray}\label{mfkL_minus2_introduction}
 \mfk{L}_{-1} \Phi_{j,-2} = \sqrt{p}^{(1)} \Phi_{j,-2} +  \frac{\sqrt{p}^3}{\lambda^{1/2}_{j,-1}\lambda^{1/2}_{j,0}} \Phi^{(3)}_{j,0},
\end{eqnarray}
where \eq{Phi_l_minus2_tentative_def} has been made use of.
Upon  introducing $\mfk{L}_{-2}$ and its Hermitian conjugate $\mfk{L}^\dagger_{-2}$  as
\begin{eqnarray}\label{mkfL_minus2}
\begin{aligned}
 \mfk{L}_{-2} &\equiv \mfk{L}_{-1} - \sqrt{p}^{(1)}=\mfk{L}_0 - 2 \sqrt{p}^{(1)},\\
 \mfk{L}^\dagger_{-2} &\equiv \mfk{L}^\dagger_{-1} - \sqrt{p}^{(1)}=\mfk{L}^\dagger_0 - 2 \sqrt{p}^{(1)},
\end{aligned}
\end{eqnarray}
and merging the two terms involving $\Phi_{j,-2}$ into one, \eq{mfkL_minus2_introduction} is reorganized as
\begin{eqnarray}\label{Phil_minus_3}
 \mfk{L}_{-2} \Phi_{j,-2}= \frac{\sqrt{p}^3}{{\lambda^{1/2}_{j,-1}\lambda^{1/2}_{j,0}}} \Phi^{(3)}_{j,0}.
\end{eqnarray}

Making use of \eq{mfkL_minus1}, the product of $\mfk{L}_{-1}$ and $\mfk{L}^\dagger_{-1}$ is obtained as
\begin{eqnarray}\label{product_mfkLm1_mfkLdm1}
 \mathfrak{L}_{-1} \mathfrak{L}^\dagger_{-1} =-\mfk{L}^2_0-\mfk{L}_0\sqrt{p}^{(1)}-\left[\mfk{L}_0,\,\sqrt{p}^{(1)}\right]+2\smbr{\sqrt{p}^{(1)}}^2.\nonumber \\
\end{eqnarray}
On the other hand, making use of \eq{mkfL_minus2}, the product of $\mfk{L}^\dagger_{-2}$ and $\mfk{L}_{-2}$ is obtained as
\begin{eqnarray}\label{product_mfkLdm2_mfkLm2}
 \mathfrak{L}^\dagger_{-2} \mathfrak{L}_{-2} =-\mfk{L}^2_0-\mfk{L}_0\sqrt{p}^{(1)}+3\left[\mfk{L}_0,\,\sqrt{p}^{(1)}\right]+6\smbr{\sqrt{p}^{(1)}}^2.\nonumber \\
\end{eqnarray}
Subtracting \eq{product_mfkLdm2_mfkLm2} from \eq{product_mfkLm1_mfkLdm1}, we do reach the pattern invariance condition \eqnm{SI_minus2_level}, with the constant $\delta_2$ defined as
\begin{eqnarray}
 \delta_2 \equiv -4\left[\mfk{L}_0,\,\sqrt{p}^{(1)}\right]-4\smbr{\sqrt{p}^{(1)}}^2=-2 p^{(2)}=4.
\end{eqnarray}

Note that the two pattern invariance conditions \eqnm{SI_plus1_level} and \eqnm{SI_minus2_level} involve the \emph{same} $\delta_2$. This is not an accident. On the one hand, the relations in \eq{symmetry_minus_plus_1st_level} give $\mathfrak{L}_{-1} \mathfrak{L}^\dagger_{-1}=\mathfrak{L}^\dagger_2 \mathfrak{L}_2$. On the other hand,  the definitions of $\mfk{L}_{-2}$ and $\mfk{L}^\dagger_{-2}$ in \eq{mkfL_minus2} and those of ${\mfk L}_3$ and ${\mfk L}^\dagger_3$ in \eq{mfkL3} together give us the symmetry relations
\begin{eqnarray}\label{symmetry_mfkL_minus23}
 \mfk{L}_{-2}=-\mfk{L}^\dagger_3,\quad \mfk{L}^\dagger_{-2}=-\mfk{L}_3.
\end{eqnarray}
As a consequence, $\mathfrak{L}^\dagger_{-2} \mathfrak{L}_{-2}= \mathfrak{L}_3 \mathfrak{L}^\dagger_3$. Therefore, we have
\begin{eqnarray}
 \mathfrak{L}_{-1} \mathfrak{L}^\dagger_{-1}  -  \mathfrak{L}^\dagger_{-2} \mathfrak{L}_{-2}
 =\mathfrak{L}^\dagger_2 \mathfrak{L}_2-\mathfrak{L}_3 \mathfrak{L}^\dagger_3=\delta_2.
\end{eqnarray}

Up to now, we have successfully verified that $\Phi_{j, -2}$ defined in \eq{Phi_l_minus2_tentative_def} indeed satisfies the eigenequation given \eqnm{eigen_equ_2nd_assoc_level_after_rnmlz}, that has the \emph{same pattern} as the eigenequation \eqnm{eigen_equ_1st_assoc_level_after_rnmlz} satisfied by $\Phi_{j,-1}$.
This also justifies the appropriateness of the tentative definition of $\Phi_{j,-2}$ in \eq{Phi_l_minus2_tentative_def}.

Intertwiningly acting \eq{eigen_equ_2nd_assoc_level_after_rnmlz} with $\mfk{L}_{-2}$, we obtain its superpartner eigenequation
\begin{eqnarray}\label{L_2_superpartner_equ}
 \mfk{L}_{-2} \mfk{L}^\dagger_{-2}   \Phi_{j,-3} =\lambda_{j,-2} \Phi_{j,-3},
\end{eqnarray}
where
\begin{eqnarray}\label{Phi_jminus3_propto}
  \Phi_{j,-3} \propto \mfk{L}_{-2}\Phi_{j,-2}.
\end{eqnarray}

Because $\mfk{L}^\dagger_{-2}$ and $\mfk{L}_{-2}$ are Hermitian conjugate to each other, \eq{eigen_equ_2nd_assoc_level_after_rnmlz} implies that $\mfk{L}_{-2}\Phi_{j,-2}$ has the norm $\lambda^{1/2}_{j,-2}$, given that $\Phi_{j,-2}$ has been defined to have a unity norm. This fact allows us to lift the proportionality relation \eqnm{Phi_jminus3_propto} to the exact lowering relation
\begin{eqnarray}\label{Phi_jminus3_tentative_def}
  \Phi_{j,-3} \equiv \frac{\mfk{L}_{-2}}{\lambda^{1/2}_{j,-2}}\Phi_{j,-2} = \frac{\sqrt{p}^3}{\lambda^{1/2}_{j,-2}\lambda^{1/2}_{j,-1}\lambda^{1/2}_{j,0}} \Phi^{(3)}_{j,0},
\end{eqnarray}
where $\mfk{L}_{-2}\Phi_{j,-2}$ has been substituted for via \eq{Phil_minus_3} to reach the second equality. Such $\Phi_{j,-3}$ has a unity norm.

In order to renormalize \eq{L_2_superpartner_equ} into the eigenequation that assumes the same pattern as Eqs. \eqnm{eigen_equ_2nd_assoc_level_after_rnmlz} and \eqnm{eigen_equ_1st_assoc_level_after_rnmlz}, i.e.,
\begin{eqnarray}\label{eigen_equ_3nd_assoc_level_after_rnmlz}
L_{-3} \Phi_{j,-3} \equiv \mfk{L}^\dagger_{-3} \mfk{L}_{-3}  \Phi_{j,-3} =\lambda_{j,-3} \Phi_{j,-3}
\end{eqnarray}
for
\begin{eqnarray}
  \lambda_{j,-3} \equiv \lambda_{j,-2} -\delta_3 =\lambda_{j,0} -\smbr{\delta_1 + \delta_2+ \delta_3},
\end{eqnarray}
$\mfk{L}_{-3}$ and its Hermitian conjugate $\mfk{L}^\dagger_{-3}$ have to be appropriately devised in such a fashion that the following shape invariance condition
\begin{eqnarray}\label{SI_minus3_level}
 \mathfrak{L}_{-2} \mathfrak{L}^\dagger_{-2}  -  \mathfrak{L}^\dagger_{-3} \mathfrak{L}_{-3} =\delta_3
\end{eqnarray}
holds, where $\delta_3$ is a certain constant  as yet to be determined.

Acting $\Phi_{j,-3}$ in the form on the right hand side of the second equal sign in \eq{Phi_jminus3_tentative_def} with $\mfk{L}_{-2}$, we obtain
\begin{eqnarray}\label{mfkL_minus_Phi_jminus3}
 \mfk{L}_{-2} \Phi_{j,-3} =\sqrt{p}^{(1)}\Phi_{j,-3} + \frac{\sqrt{p}^4}{\lambda^{1/2}_{j,-2}\lambda^{1/2}_{j,-1}\lambda^{1/2}_{j,0}} \Phi^{(4)}_{j,0}.
\end{eqnarray}
Merging the two terms involving $\Phi_{j,-3}$ and introducing $\mfk{L}_{-3}$ and its Hermitian conjugate $\mfk{L}^\dagger_{-3}$ as
\begin{eqnarray}\label{mfkL_minus3}
\begin{aligned}
 \mfk{L}_{-3}&\equiv\mfk{L}_{-2}-\sqrt{p}^{(1)}=\mfk{L}_0 - 3 \sqrt{p}^{(1)},\\
 \mfk{L}^\dagger_{-3}&\equiv\mfk{L}^\dagger_{-2}-\sqrt{p}^{(1)}=\mfk{L}^\dagger_0 - 3 \sqrt{p}^{(1)},
 \end{aligned}
\end{eqnarray}
\eq{mfkL_minus_Phi_jminus3} is then reorganized as
\begin{eqnarray}
 \mfk{L}_{-3} \Phi_{j,-3} = \frac{\sqrt{p}^4}{\lambda^{1/2}_{j,-2}\lambda^{1/2}_{j,-1}\lambda^{1/2}_{j,0}} \Phi^{(4)}_{j,0}.
\end{eqnarray}

On the one hand, the product of $\mfk{L}_{-2}$ and $\mfk{L}^\dagger_{-2}$, which are defined in \eq{mkfL_minus2}, is computed as
\begin{eqnarray}\label{product_mfkLm2_mfkLdm2}
 \mathfrak{L}_{-2}\mathfrak{L}^\dagger_{-2} =-\mfk{L}^2_0-\mfk{L}_0\sqrt{p}^{(1)} -2\left[\mfk{L}_0,\,\sqrt{p}^{(1)}\right]+6\smbr{\sqrt{p}^{(1)}}^2.\nonumber \\
\end{eqnarray}
On the other hand, the product of $\mfk{L}^\dagger_{-3}$ and $\mfk{L}_{-3}$, which are defined  in \eq{mfkL_minus3}, is computed as
\begin{eqnarray}\label{product_mfkLdm3_mfkLm3}
 \mathfrak{L}^\dagger_{-3}\mathfrak{L}_{-3} = -\mfk{L}^2_0-\mfk{L}_0\sqrt{p}^{(1)} + 4\left[\mfk{L}_0,\,\sqrt{p}^{(1)}\right]+12\smbr{\sqrt{p}^{(1)}}^2.\nonumber \\
\end{eqnarray}
Subtracting \eq{product_mfkLdm3_mfkLm3} from \eq{product_mfkLm2_mfkLdm2}, we indeed reach the pattern invariance condition \eqnm{SI_minus3_level}, with constant $\delta_3$ given by
\begin{eqnarray}
 \delta_3 \equiv -6\left[\mfk{L}_0,\,\sqrt{p}^{(1)}\right]-6\smbr{\sqrt{p}^{(1)}}^2=-3 p^{(2)}=6.
\end{eqnarray}

In view of the relation ${\mfk L}^\dagger_0=-{\mfk L}_0-\sqrt{p}^{(1)}$, the definitions of $\mfk{L}_{-3}$ and $\mfk{L}^\dagger_{-3}$ in \eq{mfkL_minus3} and those of ${\mfk L}_4$ and ${\mfk L}^\dagger_4$ in \eq{mfkL4} together then give us the symmetry relations
\begin{eqnarray}
 \mfk{L}_{-3}=-\mfk{L}^\dagger_4,\quad \mfk{L}^\dagger_{-3}=-\mfk{L}_4.
\end{eqnarray}
As a consequence, $\mathfrak{L}^\dagger_{-3} \mathfrak{L}_{-3}=\mfk{L}_4\mfk{L}^\dagger_4$. This together with $\mathfrak{L}_{-2} \mathfrak{L}^\dagger_{-2}=\mathfrak{L}^\dagger_3 \mathfrak{L}_3$, which follows form the symmetry relations in \eq{symmetry_mfkL_minus23}, give us
\begin{eqnarray}
 \mathfrak{L}_{-2} \mathfrak{L}^\dagger_{-2}  -  \mathfrak{L}^\dagger_{-3} \mathfrak{L}_{-3}
 =\mathfrak{L}^\dagger_3 \mathfrak{L}_3-\mathfrak{L}_4 \mathfrak{L}^\dagger_4=\delta_3.
\end{eqnarray}
That is, $\delta_3$ in \eq{SI_minus3_level} is the same as that in \eq{delta_3}.

Now the generic pattern that evidently emerges from the first three turns of iteration can be induced as the following, when keeping the iteration going to $m$-th negative association level ($m>3$).

The associated Legendre equation of $-m$-th association level in its form of implicit supersymmetric factorization reads
\begin{eqnarray}\label{eigen_equ_mth_assoc_level_after_rnmlz}
 L_{-m}\Phi_{j,-m} \equiv \mfk{L}^\dagger_{-m} \mfk{L}_{-m}  \Phi_{j,-m} = \lambda_{j,-m} \Phi_{j,-m},
\end{eqnarray}
where the ladder operators are defined as
\begin{eqnarray}\label{mfkL_minusm}
\begin{aligned}
  \mfk{L}_{-m} &\equiv \mfk{L}_0 -m \sqrt{p}^{(1)},\\
               &=\sqrt{p}^m\smbr{\sqrt{p}\od{}{x}}\sqrt{p}^{-m}, \\
               &={\cal S}^{2m}_x\smbr{\mfk{L}_0},
\end{aligned}
\end{eqnarray}
\begin{eqnarray}\label{mfkLd_minusm}
 \begin{aligned}
  \mfk{L}^\dagger_{-m} &\equiv \mfk{L}^\dagger_0 -m \sqrt{p}^{(1)}, \\
                       &= \sqrt{p}^{-m-1}\smbr{-\sqrt{p}\od{}{x}}\sqrt{p}^{m+1},\\
                       &={\cal S}^{-2m}_x\smbr{\mfk{L}^\dagger_0},
 \end{aligned}
\end{eqnarray}
with ${\cal S}^{\pm 2m}_x$ being the $2m$-th power of ${\cal S}^{\pm1}_x$,
and the eigenvalue is
\begin{eqnarray}\label{lambda_jminusm}
  \lambda_{j,-m} &\equiv& \lambda_{j,0} -\sum^{m}_{n=1}\delta_n =\lambda_{j,0}-m(m+1).
\end{eqnarray}

The pattern invariance condition reads
\begin{eqnarray}\label{SI_nth_level}
\mfk{L}_{-m+1} \mfk{L}^\dagger_{-m+1} -\mfk{L}^\dagger_{-m}\mfk{L}_{-m}=-m p^{(2)} = 2m \equiv \delta_m.
\end{eqnarray}

The lowering recurrence relation is
\begin{eqnarray}\label{lower_ladder_relation_minus_branch}
\Phi_{j,-m} &\equiv& \frac{\mfk{L}_{-m+1}}{\lambda^{1/2}_{j,-m+1}} \Phi_{j,-m+1}.
\end{eqnarray}

Consecutively accumulating \eq{lower_ladder_relation_minus_branch} with $m\rightarrow m,\,m-1,\,\cdots,\,1\,\text{and}\,0$, we obtain the generic eigenfunction as
\begin{eqnarray}
\Phi_{j,-m} &=& \frac{\mfk{L}_{-m+1}}{\lambda^{1/2}_{j,-m+1}} \frac{\mfk{L}_{-m+2}}{\lambda^{1/2}_{j,-m+2}} \cdots \frac{\mfk{L}_{-1}}{\lambda^{1/2}_{j,-1}} \frac{\mfk{L}_0}{\lambda^{1/2}_{j,0}} \Phi_{j,0},\nonumber \\
            &=& \frac{\sqrt{p}^{m} }{\lambda^{1/2}_{j,-m+1}\lambda^{1/2}_{j,-m+2}\cdots \lambda^{1/2}_{j,-1}\lambda^{1/2}_{j,0}}\Phi^{(m)}_{j,0},\label{top_down_approach_Phjminusm}
\end{eqnarray}
where to shorten the operator product the lumped form of $\mfk{L}_{-m}$ in \eq{mfkL_minusm} for $m\rightarrow m-1,\,m-2,\,\cdots,\,\text{and}\, 1$ has been made use of.

In the following, we will derive more compact forms of $\Phi_{j,-m}$ without resort to the concrete form of $\Phi_{j,0}$, such as that has been obtained in Section \ref{SUSYQM_base_Legendre_operator}.

\eq{lambda_jminusm} allows us to determine maximum negative association level that is labeled as $m_{max}=j$ such that $\lambda_{j,-j}=0$. This gives us $\lambda_{j,0}=j(j+1)$. For this association level, \eq{eigen_equ_mth_assoc_level_after_rnmlz} implies that
\begin{eqnarray}\label{1st_order_equ_top_minus_association_level}
\mfk{L}_{-j} \Phi_{j,-j}=\sqbr{\sqrt{p}\od{}{x}-j\sqrt{p}^{(1)}} \Phi_{j,-j}=0,
\end{eqnarray}
which means that the lowering procedure according to \eq{lower_ladder_relation_minus_branch} terminates at $m=j$ and all $\Phi_{j,-j-n}$ vanish for integers $n\ge1$.

\eq{1st_order_equ_top_minus_association_level} is readily integrated as $\Phi_{j,-j}\propto \sqrt{p}^j$. Because $\mfk{L}_{-j}=\mfk{L}^\dagger_{j+1}$, \eq{1st_order_equ_top_minus_association_level} is essentially the same as \eq{annihilation_condition_top_m}. Thus, we can take $\Phi_{j,-j}$ as
\begin{eqnarray}\label{Phi_jminusj}
 \Phi_{j,-j}=(-1)^{j}\Phi_{jj}=\frac{\sqrt{(2j+1)!!}}{\sqrt{2^{j+1}j!} } \sqrt{p}^j,
\end{eqnarray}
The relative phase between $\Phi_{j,-j}$ and $\Phi_{jj}$ is chosen in such a way that it follows the generic phase relation \eqnm{phase_between_plus_and_minus}.

Substituting for ${\mfk L}_{-m}\Phi_{j,-m}$ in \eq{eigen_equ_mth_assoc_level_after_rnmlz} via \eq{lower_ladder_relation_minus_branch} with $m\rightarrow m+1$, we obtain the raising recurrence relation
\begin{eqnarray}\label{negatively_raising_relation}
 \Phi_{j,-m} = \frac{{\mfk L}^\dagger_{-m}}{\lambda^{1/2}_{j,-m}}\Phi_{j,-(m+1)}.
\end{eqnarray}

By consecutively applying relation \eqnm{negatively_raising_relation} with $m\rightarrow m,\, m+1,\cdots,\,j-2\; \text{and}\;j-1$, $\Phi_{j,-m}$ is related to $\Phi_{j,-j}$ in the form
\begin{eqnarray}\label{Phi_jminusm}
 \Phi_{j,-m} = \frac{{\mfk L}^\dagger_{-m}}{\lambda^{1/2}_{j,-m}} \frac{{\mfk L}^\dagger_{-m-1}}{\lambda^{1/2}_{j,-m-1}} \cdots \frac{{\mfk L}^\dagger_{-j+2}}{\lambda^{1/2}_{j,-j+2}} \frac{{\mfk L}^\dagger_{-j+1}}{\lambda^{1/2}_{j,-j+1}} \Phi_{j,-j}.
\end{eqnarray}

Making use of the lumped form of ${\mfk L}^\dagger_{-m}$ in the second line of \eq{mfkLd_minusm} with $m\rightarrow m,\, m+1,\cdots,\,j-2\, \text{and}\,j-1$, the operator product in \eq{Phi_jminusm} is compressed as
\begin{eqnarray}\label{numerator_of_Phi_jminusm}
 &&{\mfk L}^\dagger_{-m}{\mfk L}^\dagger_{-m-1}\cdots{\mfk L}^\dagger_{-j+2}{\mfk L}^\dagger_{-j+1}\nonumber \\
 &&=(-1)^{j-m}\sqrt{p}^{-m}\smbr{\od{}{x}}^{j-m}\sqrt{p}^j.
\end{eqnarray}

Making us of the fact that $\lambda_{j,-m}=\smbr{j-m}\smbr{j+m+1}$, which follows from the combination of \eq{lambda_jminusm} and $\lambda_{j,0}=j\smbr{j+1}$, the eigenvalue product in \eq{Phi_jminusm} can be shortened as
\begin{eqnarray}\label{denumerator_of_Phi_jminusm}
 &&\lambda_{j,-m}\lambda_{j,-m-1}\cdots \lambda_{j,-j+2} \lambda_{j,-j+1} \nonumber \\
 &&=\frac{(j-m)!}{\smbr{j+m}!}(2j-1)!!\, 2^j j!.
\end{eqnarray}

Making substitutions via Eqs. \eqnm{Phi_jminusj}, \eqnm{numerator_of_Phi_jminusm} and \eqnm{denumerator_of_Phi_jminusm}, $\Phi_{j,-m}$ in \eq{Phi_jminusm} is then simplified as
\begin{eqnarray}\label{Phi_jminusm_shortened}
\begin{aligned}
 \Phi_{j,-m}(x)&=\frac{(-1)^{j-m}}{2^jj!}\sqrt{\frac{2j+1}{2}}\sqrt{\frac{(j+m)!}{(j-m)!}}\sqrt{p}^{-m}\smbr{\od{}{x}}^{j-m} p^j. \\
 &=\sqrt{\frac{2j+1}{2}}\sqrt{\frac{(j-m)!}{(j+m)!}} P^m_j(x),
\end{aligned}
\end{eqnarray}
where
\begin{eqnarray}
 P^m_j(x)\equiv \frac{(-1)^{j-m}}{2^jj!}\sqrt{p}^{-m}\smbr{\od{}{x}}^{j-m} p^j,
\end{eqnarray}
which is the same as the usual associated Legendre function $P^m_j(x)$ of $m$-th level defined in \eq{Pmj_1st_form}.

It follows from \eq{Phi_jminusm_shortened} that
\begin{eqnarray}\label{Phi_jzero}
 \Phi_{j,0}(x)=\frac{(-1)^j}{2^jj!}\sqrt{\frac{2j+1}{2}}\frac{d^j p^j}{d x^j} =\sqrt{\frac{2j+1}{2}} P_j(x),
\end{eqnarray}
which is the same as $\Phi_{j0}$ given in \eq{Phi_j0_top_down}. Here $P_j(x)\equiv P^0_j(x)$ is the base Legendre polynomial of $j$-th rank.

Making us of the fact that $\lambda_{j,-m}=\smbr{j-m}\smbr{j+m+1}$ with $m\rightarrow m-1,\, m-2,\cdots,\,1\; \text{and}\;0$, we obtain
\begin{eqnarray}\label{lambda_product2}
\hspace{-12pt} \lambda_{j,-m+1}\lambda_{j,-m+2}\cdots \lambda_{j,-1}\lambda_{j,0}=\frac{(j+ m)!}{(j-m)!}.
\end{eqnarray}

Substituting for $\Phi_{j,0}$ via \eq{Phi_jzero} and for $\lambda_{j,-m+1}\lambda_{j,-m+2}\cdots \lambda_{j,-1}\lambda_{j,0}$ via \eq{lambda_product2}, $\Phi_{j,-m}$ in  \eq{top_down_approach_Phjminusm} is simplified as
\begin{eqnarray}\label{Phi_jminusm_alternative}
\begin{aligned}
\Phi_{j,-m}(x)&=\frac{(-1)^j}{2^jj!}\sqrt{\frac{2j+1}{2}}\sqrt{\frac{(j-m)!}{(j+m)!}}\sqrt{p}^m \smbr{\od{}{x}}^{j+m} p^j,\\
           &=\sqrt{\frac{2j+1}{2}}\sqrt{\frac{(j-m)!}{(j+m)!}} P^m_j(x),
\end{aligned}
\end{eqnarray}
where $P^m_j(x)$ is the alternative form of the associated Legendre function of $m$-th association level given in \eq{P_jm_alternative_form}.

\subsection{Symmetries between Positive and Negative Association Levels}

Comparing \eq{mfkL_minusm} with \eq{mfkLd_mplus1} and \eq{mfkLd_minusm} with \eq{mfkL_mplus1} and keeping in mind that $\mfk{L}^\dagger_0=-\mfk{L}_0-\sqrt{p}^{(1)}$, we have the symmetry relations between the ladder operators of $-m$-th level and those of $(m+1)$-th level in the following forms
\begin{eqnarray}\label{ladder_operator_plus_minus_relation}
\begin{aligned}
\mfk{L}_{-m}&=-\mfk{L}^\dagger_{m+1} =\sqrt{p}\od{}{x}- m \sqrt{p}^{(1)}={\cal S}^{2m}_x\smbr{\mfk{L}_0},\\
 \mfk{L}^\dagger_{-m}&=-\mfk{L}_{m+1} =-\sqrt{p}\od{}{x}-\smbr{m+1}\sqrt{p}^{(1)}={\cal S}^{-2m}_x\smbr{\mfk{L}^\dagger_0}.
 \end{aligned}
\end{eqnarray}
Consequently, the symmetry relation between the associated Legendre operators $L_{-m}\equiv \mfk{L}^\dagger_{-m} \mfk{L}_{-m}$ of $-m$-th and $L_m\equiv \mfk{L}_{m+1} \mfk{L}^\dagger_{m+1}$ of $m$-th association levels comes out to be
\begin{eqnarray}
 L_{-m} =L_m ={\cal S}^{-2m}_x\smbr{\mfk{L}^\dagger_0}{\cal S}^{2m}_x\smbr{\mfk{L}_0}.
\end{eqnarray}

The pattern invariance conditions of the $m$-th positive and negative association level have the symmetry relation
\begin{eqnarray}\label{pattern_invariance_symmetry_two_branches}
 &&{\mfk L}_{-m+1}{\mfk L}^\dagger_{-m+1}-{\mfk L}^\dagger_{-m} {\mfk L}_{-m}\\
 &&={\mfk L}^\dagger_m {\mfk L}_m-{\mfk L}_{m+1} {\mfk L}^\dagger_{m+1}=2 m.\nonumber
\end{eqnarray}
Here we just juxtaposed together Eqs. \eqnm{shape_inviance_condition_ascending_generic} and \eqnm{SI_nth_level}, but this relation can be checked directly by making use of the two equations in \eqnm{ladder_operator_plus_minus_relation} and  \eq{delta1}.

The symmetry relations in \eq{ladder_operator_plus_minus_relation} and \eq{pattern_invariance_symmetry_two_branches} together eventually lead to the degeneracy of the eigenvalues $\lambda_{j,-m}$ and $\lambda_{j,m}$ in that
\begin{eqnarray}\label{eigen_value_degeneracy}
 \lambda_{j,-m}=\lambda_{jm}=j\smbr{j+1}-m\smbr{m+1}.
\end{eqnarray}
In fact, this equality is obvious from the combination of \eq{lambda_jminusm} and \eq{lambda_jm}.

Comparing $\Phi_{j,-m}(x)$ given by \eq{Phi_jminusm_shortened} or \eq{Phi_jminusm_alternative} with $\Phi_{j,m}(x)$ by \eq{Phi_jm_top_down_final} or \eq{Phijm_plus_alternative}, we have the relation
\begin{eqnarray}\label{phase_between_plus_and_minus}
 \Phi_{j,-m}(x)=(-1)^m \Phi_{j,m}(x).
\end{eqnarray}
This phase relation originates as the accumulated $m$ copies of the sign difference appearing in the symmetry relation $\mfk{L}_{-n+1}=-\mfk{L}^+_n$, by comparing how $\Phi_{j,m}(x)$ is consecutively raised from $\Phi_{j0}(x)$, shown in \eq{Phi_jm_bottom_up}, with how $\Phi_{j,-m}(x)$ is consecutively lowered from $\Phi_{j0}(x)$, shown in \eq{top_down_approach_Phjminusm}, and noticing the eigenvalue degeneracy relation \eqnm{eigen_value_degeneracy}.

All the above listed symmetry relations can be traced back to the fundamental symmetry relations
\begin{eqnarray}
\mfk{L}_1=-\mfk{L}^\dagger_0,\quad  \mfk{L}^\dagger_1=-\mfk{L}_0.
\end{eqnarray}
Though these are introduced at first as the definitions of $\mfk{L}_1$ and $\mfk{L}^\dagger_1$ given in \eq{mfkL_1_def}, they allow us to factorize the base Legendre equation in two slightly different forms (\eq{Legendre_equ_right_dagger} and \eq{L_L_eigen_equ1}), one proliferates into the associated Legendre equation of all positive higher levels and the other into that of all negative higher levels.

From \eq{ladder_operator_plus_minus_relation}, it is obvious that all the operators $\mfk{L}^\dagger_{-m}$, $\mfk{L}_{-m}$, $\mfk{L}_{m+1}$ and $\mfk{L}^\dagger_{m+1}$ are odd in $x$ under spatial reflection. Thus, the products $L_{-m}\equiv \mfk{L}^\dagger_{-m} \mfk{L}_{-m}$ and $L_m\equiv \mfk{L}_{m+1} \mfk{L}^\dagger_{m+1}$ are even, then the eigenfunctions of $L_{\pm m}$ have definite parities. In fact, \eq{Phi_jm_top_down_final}/\eq{Phijm_plus_alternative} and \eq{Phi_jminusm_shortened}/\eq{Phi_jminusm_alternative} show respectively that
\begin{eqnarray}\label{Phijm_parity_properties}
\begin{aligned}
 \Phi_{jm}(-x)&=(-1)^{j\pm m}\Phi_{jm}(x),\\
 \Phi_{j,-m}(-x)&=(-1)^{j\pm m}\Phi_{j,-m}(x).
 \end{aligned}
\end{eqnarray}

Moreover, it follows also from \eq{ladder_operator_plus_minus_relation} that
\begin{eqnarray}
\begin{aligned}
 \mfk{L}_{-m}(-x) &=\mfk{L}^\dagger_{m+1}(x),\\
 \mfk{L}^\dagger_{-m}(-x) &= \mfk{L}_{m+1}(x).
\end{aligned}
\end{eqnarray}
So $L_{-m}(-x)=L_{-m}(x)=L_m(x)=L_m(-x)$. This tells us that the combination of spatial reflection $x\rightarrow -x$, which is associated with $\theta \rightarrow \pi -\theta$ ($\cos\theta\rightarrow -\cos\theta$),  with the sign flip operation $m\rightarrow -m$, which is associated with $\phi\rightarrow \pi +\phi$ and $e^{i m\phi}\rightarrow - e^{i m\phi}$, amounts to going from the minus branch to the plus branch.

With $\mfk{L}^\dagger_0=- \mfk{L}_0 -\sqrt{p}^{(1)}$, $\mfk{L}^\dagger_{-(m+1)} = \mfk{L}^\dagger_0+(m+1) \sqrt{p}^{(1)}$ and $\mfk{L}^\dagger_m = \mfk{L}^\dagger_0-m \sqrt{p}^{(1)}$, it is easy to check that $\mfk{L}_{-(m+1)}^\dagger  \mfk{L}_{-(m+1)}=\mfk{L}_m  \mfk{L}^\dagger_m$. Thus, $\Phi_{j,-m}(x)$ and $\Phi_{j,m}(x)$ are the two degenerate eigen-states of $\mfk{L}_m  \mfk{L}^\dagger_m$ with common eigenvalue $\lambda_{jm}=j(j+1)-m(m+1)$.

Note that $J_z=-i \hbar \partial/\partial \varphi$ and $J_z\ket{jm}=m\hbar \ket{jm}$. Taking the complex conjugate, we have  $J_z\ket{jm}^*=-m\hbar \ket{jm}^*$. Comparing it with $J_z\ket{jm}=m\hbar \ket{jm}$, we see that $\ket{jm}^*\propto \ket{j,-m}$. This is consistent with the fact that $\braket{\theta,\varphi|jm}\propto \Theta_{jm}(\theta) e^{i m\varphi}$, where $\Theta_{jm}(\theta)$ is a real function of $\theta$. That is, we can pass from the eigenfunction of the positive association level to that of the equally negative one through taking the complex conjugate. To be consistent with the relation \eqnm{phase_between_plus_and_minus}or \eqnm{Theta_relation_plus_and_minus}, by choosing a particular phase factor, this proportionality is refined to the equality
\begin{eqnarray}\label{complex_conjugate_jm}
 \ket{j,-m}=(-1)^m\ket{j,m}^*.
\end{eqnarray}

Conversely, with the relation \eqnm{complex_conjugate_jm} in mind, the parity property of $\ket{jm}\propto \Theta_{jm}(\theta) e^{i m\varphi}$ then implies the relation \eqnm{phase_between_plus_and_minus}or \eqnm{Theta_relation_plus_and_minus}.

Now some comments are in order. The inductive bottom-up SUSYQM algorithm of building the associated Legendre equation as well as its eigensolution simultaneously is just the specialization of the same algorithm for the general hypergeometric-like differential operators, where the latter algorithm is detailed in \cite{Zhou_SUSY_hypergeometric_operator_2024}. The main reason we  include this algorithm for Legendre differential operator here is for the self-containment. The other reason for working out details of this algorithm here lies in that, the definitions of the positive and negative branches are switched, comparing to those in \cite{Zhou_SUSY_hypergeometric_operator_2024}, in order to be consistent with the usual definitions of the positive and negative branches for associated Legendre functions.

The inductive algorithms demonstrated in this section is different from those, e.g., in \cite{jafarizadeh_SUSY_and_diff_equs_1997, Lorente_2001, Cotfas_SI_raising_lowering_hypergeometric_equations_2002}, mainly in that, there the associated hypergeometric-type differential equation of $m$-th association level is derived first by taking $m$-fold derivatives of the base hypergeometric-type differential equation, as traditionally does, the supersymmetric factorization of this derived equation is pursued next, and the associated hypergeometric-type eigenfunction comes thirdly. In contrast, in our algorithm, staring with the only input \text{--} the base Legendre equation, the associated Legendre equation and the associated Legendre function are then built \emph{simultaneously} level by level through an iterative SUSYQM algorithm, and at each level, the associated Legendre equation comes out \emph{automatically in its (implicit) supersymmetrically factorized form}.

\section{Constructing the SUSYQM Algorithms by Directly Factorizing the Associated Legendre Equation}\label{direct_factorization_of_associated_Legendre}

In Section \ref{associated_SUSY_x_coordinate}, we actually have already constructed the SUSYQM algorithm of the associated Legendre equation in $x$ coordinate in an inductive approach. In particular, Eqs. \eqnm{eigen_equ_normalized_ascending_generic} and \eqnm{eigen_equ_mth_assoc_level_after_rnmlz} can be viewed as the results of directly supersymmetrically factorizing the associated Legendre equation in $x$ coordinate in its ``ill-organized'' hypergeometric-like form that is not shown.

In many contexts of classical physics, the associated Legendre functions come as the polar eigenfunctions of angular Laplacian, and in quantum mechanics they are the polar eigenfunctions of angular momentum operator for those central potentials for which the angular momenta are conserved.

In spherical coordinates $(r,\theta,\phi)$, as usual, the square of the quantum angular momentum operator is defined as $J^2=-\vartriangle_\Omega$ in the unit $\hbar=1$, where $\vartriangle_\Omega$ is the angular part of the Laplacian,
\begin{eqnarray}\label{angular_Laplacian}
\vartriangle_\Omega=\frac{1}{\sin\theta}\pd{}{\theta} \sin\theta \pd{}{\theta} + \frac{1}{\sin^2\theta} \frac{\partial^2}{\partial \phi ^2},
\end{eqnarray}
and $J^2$ has the spherical harmonics $Y_{jm}\smbr{\theta,\phi}$ as its eigen-functions with eigenvalues $\lambda_j$, i.e.,
\begin{eqnarray}\label{J_square_eigenequation}
 J^2 Y_{jm}\smbr{\theta,\phi}=\lambda_jY_{jm}\smbr{\theta,\phi}.
\end{eqnarray}

$Y_{jm}\smbr{\theta,\phi}$ actually takes the separated form $Y_{jm}\smbr{\theta,\phi}=\Theta_{jm}\smbr{\theta}\Phi_m\smbr{\phi}$, and it is easy to check that ${\Phi}_m (\phi)$ has the azimuthal harmonic solution, i.e., ${\Phi}_m (\phi)\propto e^{i m\phi}$ for any integer $m$. Then it follows from Eq.(\ref{J_square_eigenequation}) that the eigenequation for $\Theta_{jm}\smbr{\theta}$ -- the associated Legendre equation -- takes the form
\begin{eqnarray}\label{associated_Legendre_equ_theta_coord}
\sqbr{J_0\smbr{\theta} + \frac{m^2}{\sin^2\theta} } \Theta_{jm}\smbr{\theta}=\lambda_j \Theta_{jm}\smbr{\theta},
\end{eqnarray}
where $J_0\smbr{\theta}$ is given in \eq{J0_theta}.

In this section, we will construct the SUSYQM algorithms for exhausting the eigenspectra of the associated Legendre operator appearing in the square brackets of \eq{associated_Legendre_equ_theta_coord}, by directly supersymmetrically factorizing the associated Legendre differential equation represented in $\theta$ and $y$ coordinates, and along with this procedure we will see two more examples on \tqt{\emph{supersymmetrization and momentum operator map at work}}.

\subsection{The Iterative SUSYQM Algorithm of Associated Legendre Operator Represented in $y$ Coordinate}\label{ALE_y_coord}

\subsubsection{Directly Factorizing the Associated Legendre Equation in $y$-coordinate}\label{associated_Legendre_polynomials_y_coord}

Note that $J_0(\theta)$ already comes with the natural factorization $J_0(\theta)=-[\sin^{-1}\theta d/d\theta]\cdot[ \sin\theta d/d\theta]$, though \emph{asymmetric}. In order to obtain a momentum/kinetic energy operator map directly from such factorization and quickly pass to a certain standard Schr\"odinger equation directly from  \eq{associated_Legendre_equ_theta_coord}, similar to what have been done at the beginning of Subsections \ref{Hermitian_factorization} and \ref{second_momentum_map} and exactly the same as what has been done in Appendix \ref{theta_coord_J0}, we {\it symmetrize} the two factorial layers of $J_0(\theta)$ by multiplying both sides of \eq{associated_Legendre_equ_theta_coord} with $\sin^2\theta$ and turn $J_0(\theta)$  into $-\smbr{\sin\theta d/d\theta}^2$.  \eq{associated_Legendre_equ_theta_coord} is then rewritten as
\begin{eqnarray}\label{symmetrized_AL_equ_theta_coord}
&&\sqbr{-\smbr{\sin\theta\od{}{\theta} }^2 - \lambda_j \sin^2\theta +E_j}\Theta_{jm}\smbr{\theta},\nonumber \\
&&=\smbr{E_j-m^2} \Theta_{jm}\smbr{\theta},
\end{eqnarray}
where, as usual, a constant $E_j$ has been added for completing the further supersymmetric factorization. We see that this symmetrization simply switches the positions of the  “potential" term and the eigenvalue term in \eq{associated_Legendre_equ_theta_coord}.

It is easy to check that, the operator inside the square brackets in \eq{symmetrized_AL_equ_theta_coord} admits the following supersymmetric factorization
\begin{eqnarray}\label{AL_operator_SUSY_factorization_theta}
 &&-\smbr{\sin\theta\od{}{\theta} }^2 - \lambda_j \sin^2\theta + E_j \nonumber \\
 &=&\sqbr{-\sin\theta \od{}{\theta} - e_j \cos\theta}\sqbr{\sin\theta \od{}{\theta} - e_j \cos\theta}
\end{eqnarray}
for
\begin{eqnarray}
 \lambda_j=e_j\smbr{e_j+1},\quad E_j=e^2_j,
\end{eqnarray}
and the common function $-e_j\cos\theta$ being the superpotential.

\eq{symmetrized_AL_equ_theta_coord} can also be viewed as the result of firstly writing \eq{associated_Legendre_equ_theta_coord} in the form
\begin{eqnarray}\label{associated_Legendre_equ_theta_coord1}
 \widehat{J}_j(\theta)\Theta_{jm}\equiv \sqbr{J_0(\theta)+ \frac{E_j}{\sin^2\theta}-\lambda_{jm}}\Theta_{jm}=\frac{E_j-m^2}{\sin^2\theta}\Theta_{jm} \nonumber \\
\end{eqnarray}
and then multiplying it with $\sin^2\theta$. So \eq{AL_operator_SUSY_factorization_theta} is the supersymmetric factorization of the operator $\sin^2\theta \widehat{J}_j(\theta)$.

Now the common first-order differential operator $\sin\theta d/d\theta$ appearing in the two factors in \eq{AL_operator_SUSY_factorization_theta} is identified as the  standard momentum operator $\hat{p}_y \equiv -i d/dy$ in $y$ coordinate multiplied by $i$, i.e., we have the momentum operator map
\begin{eqnarray}\label{momentum_operator_map_1A}
 \sin\theta \od{}{\theta} = \od{}{y},
\end{eqnarray}
where $y$ is a new coordinate. This momentum operator map is equivalent to the differential equation $dy=  d\theta/\sin\theta$, which has the solution
\begin{eqnarray}\label{theta_y_coord_transformation}
 y(\theta)&= \ln \tan \frac{\theta}{2}\equiv - \arctanh(\cos\theta),\quad \cos\theta =-\tanh y. \nonumber \\
\end{eqnarray}
where $y\in(-\infty,\infty)$. We see that this type of coordinate transformations map the circular trigonometric algebra/calculus into the hyperbolic trigonometric algebra/calculus.

Meanwhile, the common function $- e_j\cos\theta$ appearing in the two factors in \eq{AL_operator_SUSY_factorization_theta} is identified as the superpotential ${\cal W}_j(y)$ in $y$ coordinate, i.e.,
\begin{eqnarray}
 {\cal W}_j(y)\equiv -e_j \cos\theta = e_j \tanh y.
\end{eqnarray}

The supersymmetric factorization \eqnm{AL_operator_SUSY_factorization_theta} in $\theta$ coordinate is then mapped into the supersymmetric factorization
of the Hamiltonian ${\widehat H}_j(y)\equiv \sin^2\theta \widehat{J}_j$ in $y$ coordinate as
\begin{eqnarray}\label{Hamiltonian_Hly}
{\widehat H}_j(y) &\equiv \widehat{A}^\dagger_j(y)\widehat{A}_j(y),
\end{eqnarray}
where the two factorial operators are defined as
\begin{eqnarray}
\begin{aligned}
 \widehat{A}^\dagger_j(y)&\equiv -\od{}{y} + {\cal W}_j(y)=-\sin\theta \od{}{\theta} - e_j \cos\theta, \\
 \quad \widehat{A}_j(y)&\equiv \od{}{y} + {\cal W}_j(y)=\sin\theta \od{}{\theta} - e_j \cos\theta.
\end{aligned}
\end{eqnarray}

Obviously, $\widehat{A}^\dagger_j(y)$ and $\widehat{A}_j(y)$ are adjoint to each other with respect to the inner product that has the integral measure $dy$, and the above supersmmetric factorization of ${\widehat H}_j(y)$ is a self-adjoint factorization with respect to this inner product. Then the eigenvalue of ${\widehat H}_j(y)$ must be non-negative, that is, $e^2_j\ge m^2$.

\eq{symmetrized_AL_equ_theta_coord} is then mapped into the supersymmetric factorization of a standard Schr\"odinger equation
\begin{eqnarray}\label{H_associated_eigen_equ_etal_state}
 {\widehat H}_j(y) \widehat{\Phi}_{jm}\smbr{y} = \smbr{e^2_j-m^2} \widehat{\Phi}_{jm}\smbr{y},
\end{eqnarray}
where $\widehat{\Phi}_{jm}\smbr{y}\equiv \Theta_{jm}(\theta(y))$ and  $\theta(y)=2\arctan\smbr{e^y}$.

We note that the other momentum operator map $-\sin\theta\, d/d\theta =d/dy$ also works for this purpose.

In the above, the supersymmetric factorization \eqnm{AL_operator_SUSY_factorization_theta} was achieved first within $\theta$ coordinate and the passage to $y$ coordinate was conducted later by introducing the momentum operator map \eqnm{momentum_operator_map_1A}. Of course, one can go to the $y$ coordinate first by identifying the square form in \eq{symmetrized_AL_equ_theta_coord} as a kinetic energy operator, i.e.,
\begin{eqnarray}\label{kinetic_operator_map}
-\smbr{\sin\theta \od{}{\theta}}^2 = - \frac{d^2}{d y^2},
\end{eqnarray}
which is consistent with the momentum operator map \eqnm{momentum_operator_map_1A}, and perform the supersymmetric factorization afterwards in $y$ coordinate.

It follows from the second equation in \eq{theta_y_coord_transformation} that $\sin^2\theta=1-\tanh^2y$. Therefore, \eq{symmetrized_AL_equ_theta_coord} is mapped into a Schr\"odinger equation in $y$ coordinate as
\begin{eqnarray}\label{symmetrized_AL_equ_theta_coord_y_coord}
 -\frac{d^2 \widehat{\Phi}_{jm}\smbr{y}}{d y^2} +\lambda_j  \smbr{\tanh^2 y-1}\widehat{\Phi}_{jm}\smbr{y}=-m^2 \widehat{\Phi}_{jm}\smbr{y}.
\end{eqnarray}

 In order to  achieve the supersymmetric factorization of \eq{symmetrized_AL_equ_theta_coord_y_coord} in the form
\begin{eqnarray}\label{supersymmetrized_AL_equ_theta_coord_y_coord}
 \sqbr{-\od{}{y}+{\cal W}_j(y)}\sqbr{\od{}{y}+{\cal W}_j(y)}\widehat{\Phi}_{jm}=\smbr{E_j-m^2}\widehat{\Phi}_{jm},
\end{eqnarray}
where the term $E_j \widehat{\Phi}_{jm}(y)$ has been added to both sides of \eq{symmetrized_AL_equ_theta_coord_y_coord} for completing this factorization, with $E_j$ some factorization constant, we have to pin down the superpotential ${\cal W}_j(y)$ associated with the potential
 \begin{eqnarray}\label{VJy_original}
  {\cal V}_j(y)\equiv  \lambda_j  \smbr{\tanh^2 y-1} + E_j
 \end{eqnarray}
via the Ricatti equation
 \begin{eqnarray}\label{VJy}
  {\cal V}_j(y) &\equiv -\od{{\cal W}_j(y)}{y} + {\cal W}^2_j(y).
 \end{eqnarray}

Furthermore, the identity
 \begin{eqnarray}\label{sin_tanh}
 \od{\tanh (y)}{y}=1-\tanh^2 (y)
 \end{eqnarray}
 reminds us that \eq{VJy} can hold in such a way that the three terms ${\cal V}_j(y)$, $d{\cal W}_j(y)/dy$ and ${\cal W}^2_j(y)$ are all binomials of $\tanh y$ without terms linear in $\tanh y$. Thus, ${\cal W}_j(y)$ can be chosen as
 \begin{eqnarray}\label{Wjy}
 {\cal W}_j(y)=e_j \tanh y
 \end{eqnarray}
 for some constant coefficient $e_j$.

 Substituting for ${\cal W}_j(y)$ in \eq{VJy} and equating the two forms of ${\cal V}_j(y)$ in Eqs. \eqnm{VJy_original} and \eqnm{VJy} then lead to
 \begin{eqnarray}\label{Vjy}
   e_j(e_j+1)\tanh^2y-e_j= \lambda_j \tanh^2y-\lambda_j+E_j.
 \end{eqnarray}
Thus, we obtain
\begin{eqnarray}\label{lambda_j_e_j_relation}
 \lambda_j =e_j\smbr{e_j +1},\quad E_j =e^2_j,
\end{eqnarray}
and \eq{supersymmetrized_AL_equ_theta_coord_y_coord} recovers \eq{H_associated_eigen_equ_etal_state}.

 The coordinate transformation \eqnm{theta_y_coord_transformation} from $\theta$ to $y$ is consistent with that from $x$ to $y$  given in \eq{x_to_y_trans1} once we take $x=-\cos\theta$. Note that this is not a coincidence. In fact, the operator inside the square brackets in \eq{symmetrized_AL_equ_theta_coord} is just the operator $p(x)L(x)-\lambda_j p(x)$ in \eq{Legendre_equ_x_coord1}, and the kinetic energy operator $-(p d/dx)^2$ in  $p(x)L(x)$ is the same as the kinetic energy operator $-\smbr{\sin\theta d/d\theta}^2$ in \eq{symmetrized_AL_equ_theta_coord}, hence the underlying two momentum operator maps in Eqs. \eqnm{momentum_operator_map_1A} and \eqnm{momentum_operator_map1} from $\theta$ and $x$ coordinates respectively to the \emph{same} $y$ coordinate, which we juxtapose here,
\begin{eqnarray}\label{three_types_momentum_operator_maps}
 \sin\theta \od{}{\theta} = \od{}{y}=p(x) \od{}{x}.
\end{eqnarray}

Since the momentum operator map \ref{momentum_operator_map_1A} for supersymmetrically factorizing the associated Legendre operator $J_0(\theta)+m^2/\sin^2\theta$ in \eq{associated_Legendre_equ_theta_coord} and that for supersymmetrically factorizing the base Legendre operator $L(x)$ are essentially the same, the SUSYQM algorithms for these two operators, in particular the designs of ladder operators, are envisaged to be the same or similar, too. However, the two momentum operator maps in \eq{three_types_momentum_operator_maps} are drastically different from (though consistent with) the momentum operator map $\sqrt{p}d/dx=\pm d/d\theta$ (which is \eq{momentum_operator_map2}),  the supersymmetric factorization of $J_0(\theta)+m^2/\sin^2\theta$ in $y$ coordinate is then expected to be sharply different from $L_m \equiv {\mfk L}_{m+1}{\mfk L^\dagger}_{m+1}$ (see \eq{eigen_equ_normalized_ascending_generic}).

 Here we see another example of quickly turning an eigen-differential equation of second order into a standard Schr\"odinger equation by the method of preparing {\it momentum operator map} \eqnm{momentum_operator_map_1A}, which is just the $\theta$-coordinate incarnation of the $x$-coordinate momentum map \eqnm{momentum_operator_map1}. The caveat is, do not break $J_0(\theta)$ apart into the second-order and first-order differential pieces, instead, start with the lumped form of $J_0(\theta)$, which is an asymmetric product of two first-order differential operators, and then prepare the kinetic energy operator by symmetrizing the lumped $J_0(\theta)$ into a square form of a first-order differential operator through simply multiplying the factor $\sin^2\theta$.

 The traditional analytical method, which is shown in the second part of Appendix \ref{traditional_analytical_method}, targets at eliminating the explicit first-order $\theta$-derivative in \eq{associated_Legendre_equ_theta_coord}, by looking for the coordinate transformation $\theta=\theta(y)$ via solving another differential equation of second order for $\theta(y)$. In this method the broken form $J_0(\theta)=-d^2/d\theta^2-\cot\theta d/d\theta$ is adopted from very beginning so that the first-order $\theta$-derivative of $\Theta_{jm}(\theta)$ is singled out.

This traditional approach is not severely complex in this particular example, because the resultant second-order differential equation for the coordinate transformation function $\theta(y)$ turns out to be particularly easy to integrate. However, the most unfavorable side of this method lies in that the adopted expanded form obscures not only the obvious correspondence between $\pm i\sin\theta d/d\theta$ and the momentum operator $\hat{p}_y\equiv-i d/dy$, as manifested in \eq{kinetic_operator_map}, but also the sought-after implicit supersymmetric factorization that is only one-step away from the lumped form of $J_0(\theta)$, hence such analytical method can not guide us to quickly reach the expected supersymmetric factorization \eqnm{Hamiltonian_Hly}.

We think our algebraic approach of reaching the supersymmetric factorization \text{--} passing from the raw eigenequation to the corresponding standard Schr\"odinger equation in its supersymmetrically factorized form through a simple multiplier and the momentum operator map \text{--} is much more efficient and conceptually simpler than the traditional analytical approach, because our approach is full of the flavor of algebraic quantum mechanics, to which the SUSYQM belongs, hence it is especially useful for constructing the SUSYQM algorithms for resolving the spectra of a large class of linear differential operators such as the Legendre-type operators. So our approach is worthy of being advocated and amplified.

Acting \eq{H_associated_eigen_equ_etal_state} with ${\widehat A}_j(y)$, we obtain the superpartner eigenequation
\begin{eqnarray}\label{superpartner_H_associated_eigen_equ_etal_stat}
 {\widehat A}_j(y) \widehat{A}^\dagger_j (y) \widehat{\Phi}_{j-1,m}\smbr{y}=\smbr{e_j^2-m^2}\widehat{\Phi}_{j-1,m}\smbr{y},
\end{eqnarray}
where
\begin{eqnarray}\label{lowering_recurrence_relation_theta_coord}
 \widehat{\Phi}_{j-1,m}\smbr{y} = \frac{1}{\sqrt{e^2_j-m^2}} {\widehat A}_j(y)\widehat{\Phi}_{j,m}(y),
\end{eqnarray}
which shows that ${\widehat A}_j(y)$ works as a lowering operator that lowers the index $j$ by one unit while keeping $m$ fixed. Here $\widehat{\Phi}_{j-1,m}\smbr{y}$ has been normalized to unity with respect to the inner product with unity weight, in view that \eq{H_associated_eigen_equ_etal_state} implies that ${\widehat A}_j(y)\widehat{\Phi}_{j,m}$ has the norm $\smbr{e^2_j-m^2}^{1/2}$.

Substituting for ${\widehat A}_j(y)\widehat{\Phi}_{j,m}$ in \eq{H_associated_eigen_equ_etal_state} via \eq{lowering_recurrence_relation_theta_coord}, we obtain the raising recurrence relation
\begin{eqnarray}\label{raising_recurrence_relation_theta_coord}
 \widehat{\Phi}_{j,m}\smbr{y} = \frac{1}{\sqrt{e^2_j-m^2}} {\widehat A}^\dagger_j(y)\widehat{\Phi}_{j-1,m}(y),
\end{eqnarray}
which says that $\widehat{A}^\dagger_j (y)$ works as a raising operator that raises the index $j$ by on unit while keeping the index $m$ fixed.  Identifying Eqs. \eqnm{lowering_recurrence_relation_theta_coord} and \eqnm{raising_recurrence_relation_theta_coord} respectively as the lowering and raising recurrence relations will be justified soon if there exists appropriate pattern invariance condition and with it \eq{superpartner_H_associated_eigen_equ_etal_stat} can be renormalized to the $(j-1)$-th version of \eq{H_associated_eigen_equ_etal_state}.

The action of  $\widehat{A}^\dagger_j (y)$ or $\widehat{A}_j (y)$ on $\widehat{\Phi}_{jm}$ leaves the index $m$ intact since the potential ${\cal V}_j(y)$, or $\widehat{A}^\dagger_j (y)$/$\widehat{A}_j (y)$ is independent of $m$. This feature is the same as that for factorizing the operator $p(x)L(x)-\lambda_j p(x)+E_j$ in $y$ coordinate shown in Subsection \ref{Hermitian_factorization}.

There exists the \emph{downward-connected} pattern invariance condition
\begin{eqnarray}\label{SI_Al}
 {\widehat A}_j(y) \widehat{A}^\dagger_j (y)-\widehat{A}^\dagger_{j-1}(y)\widehat{A}_{j-1}(y)=\Delta_j
\end{eqnarray}
for a constant $\Delta_j$ that will be determined soon.

\eq{superpartner_H_associated_eigen_equ_etal_stat} is then renormalized into
\begin{eqnarray}
 \widehat{A}^\dagger_{j-1} (y) {\widehat A}_{j-1}(y) \widehat{\Phi}_{j-1,m}\smbr{y}=\smbr{e^2_{j-1}-m^2}\widehat{\Phi}_{j-1,m}\smbr{y},
\end{eqnarray}
where $e^2_{j-1}\equiv e^2_j -\Delta_j$.  This equation has the same pattern as \eq{H_associated_eigen_equ_etal_state}, indeed it is just the $e_{j-1}$ version of the latter.

Substituting for ${\widehat A}_j(y)$, ${\widehat A}_{j-1}(y)$, ${\widehat A}^\dagger_j(y)$ and ${\widehat A}^\dagger_{j-1}(y)$ via the two lower equations in \eq{Hamiltonian_Hly}, and for ${\cal W}_j(y)$ via \eq{Wjy}, the above condition becomes
\begin{eqnarray}
 \smbr{e_j+e_{j-1}}\sqbr{\od{\tanh(y)}{y}+\smbr{e_j-e_{j-1}}\tanh^2(y)}=\Delta_j.
\end{eqnarray}
By further substituting for $d\tanh(y)/dy$ via \eq{sin_tanh}, we arrive at
\begin{eqnarray}
 \smbr{e_j+e_{j-1}}\sqbr{1+ \smbr{e_j-e_{j-1}-1}\tanh^2(y)}=\Delta_j.
\end{eqnarray}
As $\Delta_j$ is a constant by design, the coefficient of $\tanh^2(y)$ must vanish. We then obtain
\begin{eqnarray}\label{e_j_and_Delta_j}
 \begin{aligned}
  e_j&=e_{j-1}+1,\\
  \Delta_j&= e_j+ e_{j-1}.
 \end{aligned}
\end{eqnarray}
On the other hand, the eigenfunction $\Phi_{00}(\theta)$ of \eq{associated_Legendre_equ_theta_coord}  for $j=0$ and $m=0$ can be a constant, which requires that $\lambda_0=0$. Consequently, \eq{lambda_j_e_j_relation} implies that $e_0=0$. Thus, with these initial data, it follows from \eq{e_j_and_Delta_j} and \eq{lambda_j_e_j_relation} that
\begin{eqnarray}
\begin{aligned}
 e_j&=j, \\
 \Delta_j&=2j-1,\\
 \lambda_j &=j\smbr{j+1},\\
 E_j&=j^2.
\end{aligned}
\end{eqnarray}

Up to this point, we may notice that the algebraic algorithms for pinning down the parameters of supersymmetric factorization, the pattern invariance condition and the ladder relations represented in $y$ coordinate for the \emph{associated} Legendre equation in this subsubsection are almost the same as those for the \emph{base} Legendre equation shown in Subsections \ref{Hermitian_factorization} and \ref{detailed_factorization_scheme_y_coordinate}. One can say that such high similarity is a manifest of the universality of the algorithms themselves. But what is more is that almost the details of these algorithms are the same, e.g., even the notations ${\widehat H}_j(y)$, $\widehat{A}_j(y)$, $\widehat{A}^\dagger_j(y)$, ${\cal V}_j$, ${\cal W}_j$, $E_j$, $e_j$ and $\Delta_j$ have been chosen delicately to be exactly the same as those in Subsection \ref{Hermitian_factorization} and particularly in Subsubsection \ref{detailed_factorization_scheme_y_coordinate}. This is not an accident. The basic reason for doing so lies in that the two Schr\"odinger equations \eqnm{H_associated_eigen_equ_etal_state} and \eqnm{type1_S_Equ_standard}  involve the \emph{same} Hamiltonian in $y$ coordinate, in view that the potential ${\cal V}_j(y)$ in \eq{Vjy} is the same as that in \eq{Vj_definition1}, though the eigenvalues are different. In fact, \eq{type1_S_Equ_standard} is the $m=0$ case of \eq{H_associated_eigen_equ_etal_state}, that is, by making the two replacements $\widehat{\Phi}_j\smbr{y}\rightarrow \widehat{\Phi}_{jm}\smbr{y}$ and $E_j=j^2\rightarrow e^2_j-m^2$ in \eq{type1_S_Equ_standard}, we obtain \eq{H_associated_eigen_equ_etal_state}. These are the advantages of parametrizing the base as well as the associated Legendre operators in the \emph{same} $y$-coordinate, i.e., the base as well as the associated Legendre equations are supersymmetrically factorized in the same fashion, the pattern invariance condition is the same as well, and we will see in the next subsubsection the iterative procedure of building the eigenfunctions are highly similar, too, as a consequence of the fact that the lowering/raising operators $\widehat{A}^\dagger_j(y)$ and $\widehat{A}_j(y)$ only change the base index $j$ and keep the association index $m$ intact, as shown in Eqs. \eqnm{lowering_recurrence_relation_theta_coord} and \eqnm{raising_recurrence_relation_theta_coord}.

Although the main part of the procedure of the supersymmetric factorization in this subsubsection is the same as in Subsection \ref{Hermitian_factorization}, in order to maintain the integrity of the whole algorithm in this subsubsection and keep the smoothness of the exposition of this algorithm  and to reduce cross-references to Subsection \ref{Hermitian_factorization} as less as possible, we have delicately repeated the procedure of supersymmetric factorization here.

\subsubsection{Iteratively Constructing the Eigenfunctions in $y$ Coordinate in a Bottom-up Approach}\label{direct_factorization_AL_equ_y_coord}
\eq{lowering_recurrence_relation_theta_coord} can be rearranged as
\begin{eqnarray}\label{lowering_recursion_associated_Legendre1}
 \widehat{A}_j (y) \widehat{\Phi}_{j,m} \smbr{y} =\sqrt{j^2-m^2} \widehat{\Phi}_{j-1,m}\smbr{y},
\end{eqnarray}
where $j\ge \abs{m}$.

Upon making use of the lumped form of $\widehat{A}_j (y)$ in \eq{ladder_operators_hat_Aj_and_hat_Aj} and setting $j=\abs{m}$, \eq{lowering_recursion_associated_Legendre1} reduces to the annihilation condition
\begin{eqnarray}\label{equ_lowest_l}
 \widehat{A}_{\abs{m}} (y) \widehat{\Phi}_{\abs{m},m} =  \cosh^{-{\abs{m}}} y \od{}{y} \cosh^{\abs{m}} y\, \widehat{\Phi}_{\abs{m},m}= 0.
\end{eqnarray}
This has the solution
\begin{eqnarray}\label{hat_Theta_bottom_level}
\widehat{\Phi}_{\abs{m},m}\smbr{y} = (-1)^{\abs{m}} \sqbr{\frac{(2\abs{m}+1)!!}{2^{\abs{m}+1} \abs{m}!}}^{1/2} \cosh^{-\abs{m}} y, \nonumber \\
\end{eqnarray}
where the normalization constant and phase factor have been chosen in such a way that this solution is just the $y$-coordinate incarnation of \eqnm{Theta_jj} after the substitutions $j\rightarrow \abs{m}$ and $\sin\theta \rightarrow \cosh^{-1} y$ have been made.

Along with the momentum operator map \eqnm{momentum_operator_map_1A}, there is the measure map
\begin{eqnarray}
 \sin\theta d\theta =\sech^2y dy.
\end{eqnarray}
Therefore, to be consistent with the normalization in $\theta$ coordinate that is weighted by $\sin\theta$, in $y$ coordinate the normalization should be performed with the weight $\sech^2 y$. In fact, it is easy to check that $\widehat{\Phi}_{\abs{m},m}\smbr{y}$ given in \eq{hat_Theta_bottom_level} is normalized to unity only with respect to the measure $\sech^2 dy$, instead of $dy$, by noticing that $\int^\infty_{-\infty} \cosh^{-2\abs{m}}y \sech^2y dy=\smbr{2^{\abs{m}+1} \abs{m}!}/\smbr{2\abs{m}+1}!!$.

The repeated application of \eq{raising_recurrence_relation_theta_coord} for $e_j=j \rightarrow j,\;j-1,\; \cdots,\abs{m}+2,\;\text{and}\; \abs{m}+1$ accumulates to
\begin{eqnarray}\label{Theta_m+nn}
 \widehat{\Phi}_{j,m}\smbr{y} &=&\frac{1}{\sqrt{j^2-m^2} }\frac{1}{\sqrt{\smbr{j-1}^2-m^2} }\cdots \frac{1}{\sqrt{\smbr{\abs{m}+1}^2-m^2} }  \nonumber \\ 
 &&\quad \cdot \widehat{A}^\dagger_j (y) \widehat{A}^\dagger_{j-1} (y)\cdots \widehat{A}^\dagger_{\abs{m}+1} (y) \widehat{\Phi}_{\abs{m},m}\smbr{y}.
\end{eqnarray}

Making use of the lumped form of $\widehat{A}^\dagger_j (y)$ given in \eq{ladder_operators_hat_Aj_and_hat_Aj} for $j \rightarrow j,\;j-1,\; \cdots,\; \abs{m}+2,\;\text{and}\; \abs{m}+1$, the operator product above is compressed as
\begin{eqnarray}\label{raising_operators_product}
 &&\widehat{A}^\dagger_j (y) \widehat{A}^\dagger_{j-1} (y)\cdots \widehat{A}^\dagger_{\abs{m}+1} (y) \nonumber  \\
 &=&\smbr{-1}^{j-\abs{m}} \cosh^j y \smbr{\od{}{y} \cosh^{-1}y}^{j-\abs{m}} \cosh^{-\abs{m}} y.\nonumber  \\
\end{eqnarray}

Inserting $\widehat{\Phi}_{\abs{m},m}$ via \eq{hat_Theta_bottom_level} and the operator product via \eq{raising_operators_product}, \eq{Theta_m+nn} becomes
\begin{eqnarray}\label{Theta_m+nn_2}
 \widehat{\Phi}_{j,m}\smbr{y}&=&(-1)^{j} \frac{\sqrt{(2\abs{m}+1)!!}}{\sqrt{ 2 \smbr{j+\abs{m}}!(j-\abs{m})!}}  \\
 &&\; \cdot \cosh^{j}y \smbr{\od{}{y}\cosh^{-1}y}^{j-\abs{m}} \sech^{2\abs{m}} y. \nonumber
\end{eqnarray}

Setting $m=0$ in \eq{Theta_m+nn_2}, we obtain $\widehat{\Phi}_{j,0}$ \text{--} the base Legendre polynomial of $j$-th level \text{--} in its compact form
\begin{eqnarray}
 \widehat{\Phi}_{j,0}\smbr{y}&=&\frac{(-1)^{j}}{\sqrt{2} j!} \cosh^{j}y \smbr{\od{}{y}\cosh^{-1}y}^j \cdot 1,
\end{eqnarray}
which is just $\widehat{\Phi}_j(y)$ given in \eq{L_polynomial_lth_y_coord}.

In this subsection, the SUSYQM algorithm for resolving the eigen spectrum of the associated Legendre operator has been successfully built in the standard $y$ coordinate. We see that, in this parameterization the actions of the ladder operators $\widehat{A}^\dagger_j(y)$ and $\widehat{A}_j(y)$ on the eigenfunctions $\widehat{\Phi}(y)$ change the principal index $j$ while do not change the association index $m$. One may wonder if there are other coordinate parametrizations in which the corresponding SUSYQM algorithms can also be built and the ladder operators in those algorithms change the index $m$ while keeping the index $j$ intact. In fact, we have met one of such algorithms for building simultaneously the associated Legendre equation as well as its eigensolution in $x$ coordinate, in which the associated Legendre operator appears as a hypergeometric-like operator, as shown in Subsections \eqnm{Positive_Association_Levels} and \eqnm{Negative_Association_Levels}. Running that algorithm conversely, we obtain a SUSYQM algorithm of factorizing the associated Legendre equation and resolving its eigenspectrum in $x$ coordinate. Similar SUSYQM algorithm can also be constructed in $\theta$ coordinate, which we are coming to in the next subsection.

\subsection{The Iterative SUSYQM Algorithm Represented Directly in $\theta$ Coordinate}\label{direct_SUSYQM_AL_theta_coord}

Besides going to $y$ coordinate for directly factorizing the associated Legendre differential equation into its explicit supersymmetric form and then constructing the SUSYQM algorithm, as has been done in Subsection \ref{ALE_y_coord}, we can also stay within $\theta$ representation of this equation and directly perform its supersymmetric factorization and then complete the SUSYQM algorithm for resolving its eigen spectrum.

\subsubsection{Directly Factorizing the Associated Legendre Equation in $\theta$ Coordinate}\label{direct_factorization_AL_equ_theta_coordinate}

In view of the structure of the associated Legendre differential operator inside \eq{associated_Legendre_equ_theta_coord}, we should look for the supersymmetric factorization of an operator $J_m(\theta)$ that takes the form
\begin{eqnarray}\label{Jm_unfactorized_form}
 J_m(\theta)=J_0(\theta)+\frac{m^2}{\sin^2\theta}+ C_m=J_0(\theta)+m^2\cot^2\theta+\widetilde{C}_m,\nonumber \\
\end{eqnarray}
where $\widetilde{C}_m=C_m+m^2$,  both $C_m$ and $\widetilde{C}_m$ are constants added for completing the supersymmetric factorization of $J_m(\theta)$, which are $m$-dependent in such a way that $C_0=\widetilde{C}_0=0$ so that this $J_m(\theta)$ goes back to $J_0(\theta)$ in the case $m=0$.

We can find out the supersymmetric factorization of $J_m(\theta)$ \emph{within $\theta$ coordinate} (without going into $y$ coordinate) in two steps. Firstly, as shown in \ref{Legendre_operator_theta_coordinate}, we start with the lumped form of $J_0(\theta)=-\sin^{-1}\theta(d/d\theta)\sin\theta (d/d\theta)$ and make use of the active supersymmetrization transformation ${\cal S}_\theta(\text{--})\equiv \sin^{1/2}\theta(\text{--})\sin^{-1/2}\theta$ to supersymmetrize $J_0(\theta)$ into $\tilde{J}_0(\theta)$ given in \eq{J0_theta_SUSY_factorization}. Through such \emph{partial} supersymmetrization of $J_m(\theta)$,  the first-order $\theta$  differential operator in $J_m(\theta)$ is automatically gotten rid of. At the same time, the two non-differential terms inside $J_m(\theta)$ are kept intact and they will works to shift the superpotential ${\cal W}_0(\theta)$ for $\tilde{J}_0(\theta)$ of the $0$th association level into the overall superpotential ${\cal W}_m(\theta)$ of the generic association level $m$. The operator $-d^2/d\theta^2$ then works as the kinetic energy operator of the resultant Hamiltonian. Thus, applying ${\cal S}_\theta$ to \eq{Jm_unfactorized_form}, we obtain the \emph{partial} supersymmetric factorization of $J_m(\theta)$ in the form
\begin{eqnarray}
 {\cal S}_\theta\smbr{J_m(\theta)}&=&\smbr{\od{}{\theta}+\frac{1}{2}\cot\theta}\smbr{-\od{}{\theta}+\frac{1}{2}\cot\theta}\nonumber \\
 &&+\smbr{-m\cot\theta}^2+\widetilde{C}_m.
\end{eqnarray}
Secondly, by choosing $\widetilde{C}_m=m$ and absorbing one copy of $-m\cot\theta$ in $\smbr{-m\cot\theta}^2$ into each factor of the product, we then reach the \emph{overall} supersymmetric factorization of $J_m(\theta)$ in the form
\begin{eqnarray}\label{Jm_theta_supersymmetric_factorization}
 {\tilde J}_m(\theta) \equiv {\cal S}_\theta\smbr{J_m(\theta)}=\tilde{\mfk{J}}^+_m(\theta)\tilde{\mfk{J}}^-_m(\theta),
\end{eqnarray}
where
\begin{eqnarray}\label{tildeJ_mpm_theta}
\begin{aligned}
 \tilde{\mfk{J}}^+_m(\theta)&\equiv \od{}{\theta}+\smbr{\frac{1}{2}-m}\cot\theta,\\
 \tilde{\mfk{J}}^-_m(\theta)&\equiv -\od{}{\theta}+\smbr{\frac{1}{2}-m}\cot\theta.
 \end{aligned}
\end{eqnarray}
That is, in $\theta$ coordinate ${\tilde J}_m(\theta)$ is a standard Hamiltonian with
the momentum operator $\hat{p}_\theta = -i d/d\theta$ and the kinetic energy operator $-d^2/d\theta^2$, and it has the supersymmetric factorization \eqnm{Jm_theta_supersymmetric_factorization} with the superpotential
\begin{eqnarray}\label{Wam_superpotential}
 {\cal W}^a_m(\theta)\equiv \smbr{\frac{1}{2}-m} \cot\theta.
\end{eqnarray}

Applying the inverse transformation ${\cal S}^{-1}_\theta(-)\equiv \sin^{-1/2}\theta(-)\sin^{1/2}$ to \eq{Jm_theta_supersymmetric_factorization}, we obtain the asymmetric factorization of $J_m(\theta)$ in the form
\begin{eqnarray}\label{Jm_asymmetric_factorization1}
J_m(\theta)=\mfk{J}^+_m(\theta) \mfk{J}^-_m(\theta)\equiv{\cal S}^{-1}_\theta\smbr{{\tilde J}_m(\theta)},
\end{eqnarray}
where
\begin{eqnarray}\label{mfkJpm_m}
\begin{aligned}
 \mfk{J}^+_m(\theta)&\equiv {\cal S}^{-1}_\theta\smbr{\tilde{\mfk{J}}^+_m(\theta)},\\
 \mfk{J}_m(\theta)&\equiv {\cal S}^{-1}_\theta\smbr{\tilde{\mfk{J}}_m(\theta)},
\end{aligned}
\end{eqnarray}
or, equivalently,
\begin{eqnarray}\label{Jpm_m_relations}
\begin{aligned}
 \tilde{\mfk{J}}^+_m(\theta)&={\cal S}_\theta(\mfk{J}^+_m(\theta)),\\
 \tilde{\mfk{J}}^-_m(\theta)&={\cal S}_\theta(\mfk{J}^-_m(\theta)).
 \end{aligned}
\end{eqnarray}
Of course, the factorization \eqnm{Jm_asymmetric_factorization1} of $J_m(\theta)$ can be reached directly from its broken form in \eq{Jm_unfactorized_form}.

It is easy to check directly that
\begin{eqnarray}\label{essential_roles_Spm1}
\begin{aligned}
 {\cal S}_\theta\smbr{\pm \od{}{\theta}} &=\pm \od{}{\theta} \mp \frac{1}{2}\cot\theta,\\
 {\cal S}^{-1}_\theta\smbr{\pm \od{}{\theta}} &= \pm \od{}{\theta} \pm \frac{1}{2}\cot\theta,
\end{aligned}
\end{eqnarray}
which also follow from \eq{essential_roles_Spm}.
We see that, the essential roles of ${\cal S}_\theta$ are to shift $d/d\theta$ downward by $\cot\theta/2$ and to shift $-d/d\theta$ upward by $\cot\theta/2$, and the essential roles of ${\cal S}^{-1}_\theta$ on $\pm d/d\theta$ are just the opposite.

With the shifting (asymmetrizing) roles of ${\cal S}^{-1}_\theta(-)$ in mind, \eq{mfkJpm_m} is rewritten as
\begin{eqnarray}\label{mfkJpm_def0}
 \begin{aligned}
  \mfk{J}^+_m(\theta) &\equiv \od{}{\theta}+\smbr{1-m}\cot\theta = \tilde{\mfk{J}}^+_m(\theta) + \frac{1}{2}\cot\theta,\\
  \mfk{J}^-_m(\theta) &\equiv -\od{}{\theta}-m\cot\theta = \tilde{\mfk{J}}^-_m(\theta)  - \frac{1}{2}\cot\theta.
 \end{aligned}
\end{eqnarray}

Another more efficient way of building $\tilde{J}_m(\theta)/J_m(\theta)$  is to construct the two factorial operators $\tilde{\mfk{J}}^\pm_m(\theta)/\mfk{J}^\pm_m(\theta)$, either directly from the factorial operators $\tilde{\mfk{J}}^\pm_0(\theta)$ / $\mfk{J}^\pm_0(\theta)$ \text{--} the $m=0$ case of $\tilde{\mfk{J}}^\pm_m(\theta)/\mfk{J}^\pm_m(\theta)$, or directly from the two most basic differential operators $\pm d/d\theta$, by applying certain appropriate \emph{bilateral transformations}. Soon we will see that such transformations can be easily pinned down as the powers of the more basic transformations ${\cal S}_\theta$ and ${\cal S}^{-1}_\theta$, by keeping in mind the essential \emph{shifting roles} of these transformations, as shown in \eq{essential_roles_Spm1}. The advantage of this approach lies in that the reached $\tilde{J}_m(\theta)/J_m(\theta)$  is automatically  in its explicit/implicit supersymmetrically factorized form.

\eq{essential_roles_Spm1} just define the \emph{one-fold} shifting roles of ${\cal S}_\theta$ and ${\cal S}^{-1}_\theta$ when they are applied to $\pm d/d\theta$ once. When these bilateral transforms are applied to $\pm d/d\theta$  \emph{multiple folds}, $\pm d/d\theta$ will be shifted multiple folds. Indeed, we can define $\tilde{\mfk{J}}^+_m(\theta)$ as the result of applying ${\cal S}_\theta$  to $d/d\theta$ \emph{$(2m-1)$ folds}, which produce $2m-1$ copies of $\cot \theta/2$ down-shift, i.e.,
\begin{eqnarray}\label{(2m-1)_fold_S_theta}
 \tilde{\mfk{J}}^+_m(\theta) \equiv {\cal S}^{2m-1}_\theta\smbr{\od{}{\theta}} 
                      =\od{}{\theta} - \smbr{2m-1}\frac{\cot\theta}{2}.
\end{eqnarray}
This can also be understood as the result of applying ${\cal S}^{-1}_\theta$ to $d/d\theta$ once  for producing up-shift $(1/2)\cot\theta$ first, followed by  applying ${\cal S}_\theta$ to $d/d\theta$ \emph{$2m$ folds} for producing down-shift $m\cot\theta$. One of the reasons for choosing the fold number $2m-1$ in defining $\tilde{\mfk{J}}^+_m(\theta)$ is that such $\tilde{\mfk{J}}^+_m(\theta)$ collapses correctly into $\tilde{\mfk{J}}^+_0(\theta)$ defined in \eq{tldmfkJpm0} for $m=0$.

Meanwhile, we can define $\tilde{\mfk{J}}^-_m(\theta)$ to be the result of applying ${\cal S}^{-1}_\theta$ to $-d/d\theta$  $(2m-1)$ folds, which produces $(2m-1)$ copies of $\cot\theta/2$ downward-shift, i.e.,
\begin{eqnarray}\label{(2m-1)_fold_inverse_S_theta}
 \tilde{\mfk{J}}^-_m(\theta) \equiv {\cal S}^{-(2m-1)}_\theta\smbr{-\od{}{\theta}}
                    =-\od{}{\theta}-\smbr{2m-1}\frac{\cot\theta}{2}.\nonumber \\
\end{eqnarray}
This can also be understood as the result of applying ${\cal S}_\theta$ to $-d/d\theta$ once  for producing the positive shift $(1/2)\cot\theta$ first, followed by  applying ${\cal S}^{-1}_\theta$ to $-d/d\theta$  \emph{$2m$-fold} for producing the negative shift $-m\cot\theta$. $\tilde{\mfk{J}}^-_m(\theta)$ defined this way collapses correctly into $\tilde{\mfk{J}}^-_0(\theta)$ defined in \eq{tldmfkJpm0} for $m=0$.

$\tilde{J}_m(\theta)$ is then defined as the product of $\tilde{\mfk{J}}^+_m(\theta)$ in \eq{(2m-1)_fold_S_theta} and $\tilde{\mfk{J}}^-_m(\theta)$ in \eq{(2m-1)_fold_inverse_S_theta},
\begin{eqnarray}\label{tJm_standard_SUSY_factorization}
 \tilde{J}_m(\theta)&\equiv&\tilde{\mfk{J}}^+_m(\theta) \tilde{\mfk{J}}^-_m(\theta),\\
                    &=&\smbr{\od{}{\theta}+{\cal W}^a_m(\theta)}\smbr{-\od{}{\theta}+{\cal W}^a_m(\theta)}. \nonumber
\end{eqnarray}
This supersymmetric factorization of the standard Hamiltonian $\tilde{J}_m(\theta)$ is exactly the same as that obtained in \eq{Jm_theta_supersymmetric_factorization}.

At this point, it is quite clear that why we have delicately chosen the fold number of ${\cal S}_\theta$ on $d/d\theta$ in \eq{(2m-1)_fold_S_theta} and that of ${\cal S}^{-1}_\theta$ on $-d/d\theta$ in \eq{(2m-1)_fold_inverse_S_theta} to be the \emph{same} integer $2m-1$. In so doing, the \emph{same} amount of total down-shift ${\cal W}^m_a(\theta)$ is then reached so that this \emph{common shift} works as the desired superpotential for the explicit supersymmetric factorization of the Hamiltonian $\tilde{J}_m(\theta)$.

Note that $\mfk{J}^+_m(\theta)$ defined in \eq{mfkJpm_def0} is the down-shift of $\mfk{J}^+_0(\theta)\equiv d/d\theta+ \cot\theta$ by $m\cot\theta$, which can be implemented by applying the transformation ${\cal S}_\theta$ to $\mfk{J}^+_0(\theta)$ $2m$ folds.  This amounts to the direct down-shift $(m-1)\cot\theta$ of $d/d\theta$, implemented by applying $2(m-1)$ folds ${\cal S}_\theta$ to $d/d\theta$. That is,
\begin{eqnarray}\label{mfkJpm_def}
 \mfk{J}^+_m(\theta)&\equiv&{\cal S}^{2m}_\theta\smbr{\mfk{J}^+_0(\theta)}={\cal S}^{2(m-1)}_\theta\smbr{\od{}{\theta}},
\end{eqnarray}
which correctly collapses into $\mfk{J}^+_0(\theta)={\cal S}^{-2}_\theta\smbr{d/d\theta}$ in the case $m=0$.

Similarly, $\mfk{J}^-_m(\theta)$ defined in \eq{mfkJpm_def0} is the result of applying the transformation ${\cal S}^{-1}_\theta$ on $\mfk{J}^-_0(\theta)=-d/d\theta$ by the \emph{same} $2m$ folds,
\begin{eqnarray}\label{mkfJmm_def}
 \mfk{J}^-_m(\theta)&\equiv& {\cal S}^{-2m}_\theta\smbr{\mfk{J}^-_0(\theta)}={\cal S}^{-2m}_\theta\smbr{-\od{}{\theta}},
\end{eqnarray}
which correctly collapses into $\mfk{J}^-_0(\theta)=-d/d\theta$ in the case $m=0$.

$J_m(\theta)$ is then identified with the product of $\mfk{J}^+_m(\theta)$ in \eq{mfkJpm_def} and $\mfk{J}^-_m(\theta)$ in \eq{mkfJmm_def},
\begin{eqnarray}\label{Jm_asymmetric_factorization}
 J_m(\theta)&\equiv&\mfk{J}^+_m(\theta)\mfk{J}^-_m(\theta),\nonumber\\
            &=&{\cal S}^{2(m-1)}_\theta\smbr{\od{}{\theta}}{\cal S}^{-2m}_\theta\smbr{-\od{}{\theta}},\nonumber \\
            &=&-\sin^{m-1}\theta \od{}{\theta}\sin^{-2m+1}\theta \od{}{\theta}\sin^m\theta,\nonumber\\
            &=&\sqbr{\od{}{\theta}+\smbr{1-m}\cot\theta}\cdot\sqbr{-\od{}{\theta}-m\cot\theta},\nonumber \\
            &=&J_0\smbr{\theta}+\frac{m^2}{\sin^2\theta}-m\smbr{m-1}.
\end{eqnarray}
The $J_m(\theta)$ constructed this way is the same as what has been defined in \eq{Jm_unfactorized_form}. Here we see that, the primary purpose of delicately choosing the \emph{same} fold number $2m$ for applying the two basic transformations ${\cal S}^{\pm1}_\theta$ to $\mfk{J}^\pm_0(\theta)$ respectively is to achieve the \emph{same} amount of shift $-m \cot\theta$, so that the  operator $J_0(\theta)+m^2/\sin^2\theta$ (the first two terms in the last line of \eq{Jm_asymmetric_factorization}) coming with the associated Legendre equation \eqnm{associated_Legendre_equ_theta_coord} is recovered in the product $\mfk{J}^+_m(\theta)\mfk{J}^-_m(\theta)$, so that this product can be defined as the associated Legendre differential operator $J_m(\theta)$.

We note that the same supersymmetrization transformation ${\cal S}_\theta$ that supersymetrizes the asymmetric factorization $J_0(\theta)=\mfk{J}^+_0(\theta)\mfk{J}^-_0(\theta)$ also supersymetrizes the asymmetric factorization $J_m(\theta)=\mfk{J}^+_m(\theta)\mfk{J}^-_m(\theta)$ in \eq{Jm_asymmetric_factorization}. This is because $J_0(\theta)$ is the $m=0$ case of $J_m(\theta)$, which implies that the superpotentials ${\cal W}^a_0(\theta)$ and ${\cal W}^a_m(\theta)$ are of the same type of functions (indeed, ${\cal W}^a_0(\theta)$ is the $m=0$ case of ${\cal W}^a_m(\theta)$), and ${\cal S}_\theta$ is independent of $m$.

In terms of the factorization of $J_m(\theta)$ in \eq{Jm_asymmetric_factorization1}, \eq{associated_Legendre_equ_theta_coord} is factorized as
\begin{eqnarray}\label{associated_Legendre_asymmetric_factorization}
J_m\smbr{\theta}\Theta_{jm}\smbr{\theta}=\mfk{J}^+_m \mfk{J}^-_m\Theta_{jm}\smbr{\theta}=\smbr{c^-_{jm}}^2 \Theta_{jm}\smbr{\theta},
\end{eqnarray}
where
\begin{eqnarray}\label{cmjm_def}
 c^-_{jm}\equiv \sqbr{\lambda_j-m\smbr{m-1}}^{1/2}.
\end{eqnarray}

Multiplying $\sin^{1/2}\theta$ for preparing ${\cal S}_\theta$ and making use of \eq{Jpm_m_relations}, the explicit supersymmetric factorized version of \eq{associated_Legendre_asymmetric_factorization} is
\begin{eqnarray}\label{associated_Legendre_supersymmetric_factorization}
 \widetilde{\mfk{J}}^+_m \widetilde{\mfk{J}}^-_m \widetilde{\Theta}_{jm}(\theta) = \smbr{c^-_{jm}}^2 \widetilde{\Theta}_{jm}(\theta),
\end{eqnarray}
where $\widetilde{\Theta}_{jm}(\theta)\equiv \sin^{1/2}\theta \Theta_{jm}(\theta)$.

The rest iterative SUSYQM algorithm for resolving the associated Legendre function $\Theta_{jm}\smbr{\theta}$ and the eigenvalue $\lambda_j$ can be based either on \eq{associated_Legendre_asymmetric_factorization} (that comes with the implicit supersymmetric factorization) or on \eq{associated_Legendre_supersymmetric_factorization} (that comes with the explicit supersymmetric factorization), as these two equations are equivalent and are related to each other simply by the unitary transformation ${\cal S}_\theta$ and its inverse. In the next subsubsection, \eq{associated_Legendre_asymmetric_factorization} is chosen to start with.

It is worthwhile of pointing out that the operators $J_m(\theta)$, $\tilde{J}_m(\theta)$ and their supersymmetric factorial operators $\mfk{J}^\pm_m(\theta)$ and $\tilde{\mfk{J}}^\pm_m(\theta)$  introduced in this subsubsection are not restricted to non-negative integer $m$. In fact, $m$ can also be a negative integer. This point is particularly clear for $\mfk{J}^\pm_m(\theta)$ introduced by multiple-fold shifts of $\pm d/d\theta$. Flipping the sign of $m$ amounts to the change of the directions of such shifts.

\subsubsection{Iteratively Constructing the Eigensolutions of the Associated Legendre Oprator in $\theta$ Coordinate}\label{eigen_solutions_associated_equ_theta_coord}

Intertwiningly acting \eq{associated_Legendre_asymmetric_factorization} with $\mfk{J}^-_m$, we obtain the superpartner eigenequation of that equation as
\begin{eqnarray}\label{SP_associated_Legendre_equ1}
\mfk{J}^-_m\mfk{J}^+_m \Theta_{jm-1}\smbr{\theta}=\smbr{c^-_{jm}}^2 \Theta_{jm-1}(\theta),
\end{eqnarray}
where we have assumed that
\begin{eqnarray}\label{Thetajmminus1}
\Theta_{jm-1}\smbr{\theta} \propto \mfk{J}^-_m \Theta_{jm}\smbr{\theta},
\end{eqnarray}
which will be justified soon.

With the forms of $\mfk{J}^\pm_m$ in \eq{mfkJpm_def0}, we obtain
\begin{eqnarray}\label{J_minus_times_J_plus}
 \mfk{J}^-_m \mfk{J}^+_m=J_0\smbr{\theta} + \frac{(m-1)^2}{\sin^2\theta}-m\smbr{m-1}.
\end{eqnarray}
Subtracting the $(m-1)$-version of \eq{Jm_asymmetric_factorization} from \eq{J_minus_times_J_plus} to cancel the two common terms $J_0(\theta)$ and $(m-1)^2/\sin^2\theta$, we obtain the pattern invariance symmetry relation
\begin{eqnarray}\label{shape_invariance_condition_theta_coord_mfkJ}
 \mfk{J}^-_m \mfk{J}^+_m-\mfk{J}^+_{m-1} \mfk{J}^-_{m-1} = -2(m-1).
\end{eqnarray}
This relation is then made use of to renormalize \eq{SP_associated_Legendre_equ1} into the following normally ordered one,
\begin{eqnarray}\label{SP_associated_Legendre_equ_normal_order}
\mfk{J}^+_{m-1}\mfk{J}^-_{m-1} \Theta_{jm-1}\smbr{\theta}=\smbr{c^-_{jm-1}}^2 \Theta_{jm-1}\smbr{\theta},
\end{eqnarray}
where $\smbr{c^-_{jm-1}}^2\equiv \smbr{c^-_{jm}}^2+2(m-1)=\lambda_j-\smbr{m-1}\smbr{m-2}$.

\eq{SP_associated_Legendre_equ_normal_order} says that $\mfk{J}^-_m \Theta_{jm}\smbr{\theta}$ is the eigenfunction of $J_{m-1}=\mfk{J}^+_{m-1}\mfk{J}^-_{m-1}$, and $\mfk{J}^-_m$ works as the lowering operator that lowers the index $m$ by one unit. Thus, the relation \eqnm{Thetajmminus1} is justified.

\eq{associated_Legendre_asymmetric_factorization} says that $\mfk{J}^-_m \Theta_{jm}\smbr{\theta}$ has the norm $c^-_{jm}$. Thus,  the proportionality relation  \eqnm{Thetajmminus1} can be refined to the lowering recurrence relation
\begin{eqnarray}\label{lowering_recurrence_prop_fixed}
 \mfk{J}^-_m \Theta_{jm}\smbr{\theta} =C^-_{jm} \Theta_{jm-1}\smbr{\theta}.
\end{eqnarray}

Substituting for $\mfk{J}^-_m \Theta_{jm}\smbr{\theta}$ in \eq{associated_Legendre_asymmetric_factorization} via \eq{lowering_recurrence_prop_fixed}, we then obtain the raising recurrence relation
\begin{eqnarray}\label{raising_recurrence_relation}
 \mfk{J}^+_m \Theta_{jm-1}\smbr{\theta} =C^-_{jm} \Theta_{jm}\smbr{\theta},
\end{eqnarray}
or
\begin{eqnarray}\label{raising_recurrence_relation1}
 \mfk{J}^+_{m+1} \Theta_{jm}\smbr{\theta} =C^-_{jm+1} \Theta_{jm+1}\smbr{\theta}.
\end{eqnarray}

In \eq{raising_recurrence_relation1}, for a given $j$, there exists a maximum of $m$, denoted by $m_{max}$, such that $c^-_{j,\,m_{max}+1}=0$ and this raising procedure terminates in the sense that $\mfk{J}^+_{m_{max}+1} \Theta_{jm_{max}}\smbr{\theta}=0$. According to the definition \eqnm{cmjm_def} for $c^-_{j,\,m+1}$, the condition $c^-_{j,\,m_{max}+1}=0$  implies that $\lambda_j=m_{max}(m_{max}+1)$. Since up to now, the quantum number $j$ is free to be specified, it does not bother us to set $j=m_{max}$, we then have
\begin{eqnarray}\label{lambbaj}
\lambda_j=j(j+1).
\end{eqnarray}

In a bottom-up approach, $\Theta_{jm}(\theta)$ can be built from
\begin{eqnarray}\label{Thetaj0}
\Theta_{j0}(\theta)=\sqrt{\frac{2j+1}{2}}P_j(\cos\theta),
\end{eqnarray}
which is the normalized Legendre polynomial of $j$-th order, by repeatedly applying the raising recursion relation \eqnm{raising_recurrence_relation} for  $m\rightarrow m,\; m-1,\;\cdots,\;2 ,\; 1$, i.e.,
\begin{eqnarray}\label{Theta_tower_raising_0}
 \Theta_{jm}(\theta) &=&\frac{\mfk{J}^+_m}{c^-_{jm}}\frac{\mfk{J}^+_{m-1}}{c^-_{jm-1}}\cdots \frac{\mfk{J}^+_1}{c^-_{j1}} \Theta_{j0}(\theta).
\end{eqnarray}

In order to obtain $\Theta_{jm}$ in a more compact form, we utilize the lumped form $\mfk{J}^+_m=-\sin^m\theta(d/d\cos\theta)\sin^{-(m-1)}\theta$ (see \eq{mfkJpm_def} for this) to shorten the operator product in \eq{Theta_tower_raising_0} in the form
\begin{eqnarray}\label{mfkJp_m_product}
\mfk{J}^+_m\mfk{J}^+_{m-1}\cdots \mfk{J}^+_2 \mfk{J}^+_1 &=& (-1)^m \sin^m\theta \frac{d^m}{d(\cos\theta)^m}.
\end{eqnarray}

The combination of  \eq{cmjm_def} and \eq{lambbaj} yields
\begin{eqnarray}\label{cmjm_def1}
 c^-_{jm}=\smbr{j-m +1}^{1/2}\smbr{j+m}^{1/2}.
\end{eqnarray}

Making use of \eq{cmjm_def1} for $c^-_{jm}$ with $m \rightarrow m,\,m-1,\, \cdots,\, 2,\,1$, we obtain
\begin{eqnarray}\label{c_plus_product}
 &&c^-_{jm}c^-_{jm-1}\cdots c^-_{j2}c^-_{j1}=\sqbr{\frac{(j+m)!}{(j-m)!}}^{1/2}.
\end{eqnarray}
Thus, by substitution via Eqs. \eqnm{mfkJp_m_product} and \eqnm{c_plus_product} and making use of \eq{Thetaj0}, \eq{Theta_tower_raising_0} reduces to
\begin{eqnarray}\label{Theta_tower_raising_1}
 \Theta_{jm}(\theta) &=& (-1)^m \sqrt{\frac{2j+1}{2}}\sqrt{\frac{(j-m)!}{(j+m)!}} P^m_j(\cos\theta),
\end{eqnarray}
where
\begin{eqnarray}\label{Pjm_1st_def}
 P^m_j(\cos\theta)\equiv \sin^m\theta \frac{d^m  P_j(\cos\theta)}{d(\cos\theta)^m}
\end{eqnarray}
is the usual associated Legendre function of $m$-th level.

Substituting for $\Theta_{jm}$ and $\Theta_{jm\pm1}$ via \eq{Theta_tower_raising_1} in the ladder relations \eqnm{J_tilde_raising} and \eqnm{J_tilde_lowering}, we obtain the ladder relations for $P^m_j(\cos\theta)$,
\begin{eqnarray}
 \mfk{J}^+_{m+1}P^m_j(\cos\theta) &=& - P^{m+1}_j(\cos\theta), \label{ladder_raising_Pjm}\\
  \mfk{J}^-_m P^m_j(\cos\theta) &=& -\smbr{c^-_{jm}}^2 P^{m-1}_j(\cos\theta). \label{ladder_lowering_Pjm}
\end{eqnarray}
In fact, the raising relation \eqnm{ladder_raising_Pjm} can be directly proved by making use of the lumped form of $\mfk{J}^+_{m+1}$ in \eq{raising_operator} and the definition \eqnm{Pjm_1st_def} for $P^m_j$, and the lowering relation \eqnm{ladder_lowering_Pjm} can be directly proved by making use of the lumped form of $\mfk{J}^-_m$ in \eq{lowering_operator} and the definition \eqnm{Pjm_2nd_def} for $P^m_j$

Adding Eqs. \eqnm{ladder_raising_Pjm} and \eqnm{ladder_lowering_Pjm} up leads to the three-term algebraic recurrence relation
\begin{eqnarray}
  2 m \cot\theta P^m_j(\cos\theta) &=& P^{m+1}_j(\cos\theta) \\
                                    && + \smbr{j+m}\smbr{j-m+1}P^{m-1}_j(\cos\theta).\nonumber
  \end{eqnarray}

Subtracting \eq{ladder_lowering_Pjm} from \eq{ladder_raising_Pjm} leads to the three-term differential recurrence relation
\begin{eqnarray}
  2 \od{P^m_j(\cos\theta)}{\theta}  &=& -P^{m+1}_j(\cos\theta) \\
                                    && + \smbr{j+m}\smbr{j-m+1}P^{m-1}_j(\cos\theta).\nonumber
\end{eqnarray}

To get a more compact form of $\Theta_{jm}(\theta)$, the above bottom-up approach of building $\Theta_{jm}$ relies on the input of a concrete form of $\Theta_{j0}(\theta)$ such as that given in \eq{Thetaj0}, or the concrete form of $P_j(\cos\theta)$ from other approaches, e.g., such as those demonstrated in Subsection \ref{SUSYQM_base_Legendre_operator} and Appendix \ref{theta_coord_J0}. In the following, we will build $\Theta_{jm}$ in a top-down approach. In such approach we will see that even $\Theta_{j0}(\theta)$ also comes out naturally within this approach in a certain compact form, instead of resort to some other algorithms that are outside of this approach.

For the top association level $m_{max}=j$, as a result of $C^-_{jj+1}=0$, \eq{associated_Legendre_asymmetric_factorization} or \eq{raising_recurrence_relation1} reduces to
 \begin{eqnarray}
  \mfk{J}_{j+1}\Theta_{jj}=\smbr{\od{}{\theta} - j \cot \theta} \Theta_{jj}=0,
 \end{eqnarray}
which has the solution $\Theta_{jj} \propto\sin^j\theta$. Making use of the fact that $\int^\pi_0 d\theta \sin^{2j+1}\theta = 2^{j+1} j!/(2j+1)!!$, the normalized $\Theta_{jj}$ is then obtained as
\begin{eqnarray}\label{Theta_jj}
\Theta_{jj}(\theta) = (-1)^j \sqbr{\frac{(2j+1)!!}{2^{j+1} j!}}^{1/2}\sin^j\theta,
\end{eqnarray}
where the normalization has been taken with respect to the inner product $(g_1(\theta),g_2(\theta))\equiv\int^\pi_0d\theta \sin\theta g_1(\theta)g_2(\theta)$, and the extra phase factor $(-1)^j$ is chosen for this $\Theta_{jj}$ to coincide with the $m=j$ case of $\Theta_{jm}(\theta)$ given \eq{Theta_jm_top_down}.

Starting with the top level solution \eqnm{Theta_jj}, one can use the lower recurrence relation \eqnm{lowering_recurrence_prop_fixed} consecutively to construct all other $\Theta_{jm}(\theta)$ for $0\le m<j$ in the form
\begin{eqnarray}\label{Theta_tower_lowering}
 \Theta_{jm}(\theta)=\frac{\mfk{J}^-_{m+1}}{c^-_{jm+1}}\frac{\mfk{J}^-_{m+2}}{c^-_{jm+2}} \cdots \frac{\mfk{J}^-_{j-1}}{c^-_{jj-1}} \frac{\mfk{J}^-_j}{c^-_{jj}} \Theta_{jj}(\theta).
\end{eqnarray}

Making use of the lumped form $\mfk{J}^-_m(\theta)=\sin^{-(m-1)}\theta (d/d\cos\theta)\sin^m \theta$ that follows from \eq{mkfJmm_def}, the operator product in \eq{Theta_tower_lowering} is compressed as
\begin{eqnarray}
\mfk{J}^-_{m+1} \mfk{J}^-_{m+2}\cdots \mfk{J}^-_{j-1} \mfk{J}^-_j = \sin^{-m}\theta \frac{d^{j-m}}{d (\cos\theta)^{j-m}}\sin^j\theta.
\end{eqnarray}
On the other hand, with $c^-_{jm}$ in the form given in \eq{cmjm_def1} for $m\rightarrow m+1,\, \cdots,\, j-1$ and $j$, we have
\begin{eqnarray}\label{c_minus_product}
 && c^-_{jm+1}c^-_{jm+2}\cdots c^-_{jj-1}c^-_{jj}=\sqrt{\frac{(j-m)!}{(j+m)!}}\sqrt{(2j)!}.
\end{eqnarray}

The combination of Eqs. \eqnm{Theta_jj} through \eq{c_minus_product} then yields the expression for $ \Theta_{jm}\smbr{\theta}$ as
\begin{eqnarray}
 \Theta_{jm}\smbr{\theta}&=&(-)^j \sqrt{\frac{2j+1}{2}} \frac{1}{2^j j!}\sqrt{\frac{(j+m)!}{(j-m)!}}  \label{Theta_jm_top_down}\\
 &&\quad \cdot \sin^{-m}\theta \frac{d^{j-m} \sin^{2j}\theta}{d (\cos\theta)^{j-m}}, \nonumber\\
 &=&(-1)^m \sqrt{\frac{2j+1}{2}}\sqrt{\frac{(j-m)!}{(j+m)!}} P^m_j(\cos\theta), \label{Theta_jm_final}
\end{eqnarray}
that is an alternative to the form in \eq{Theta_tower_raising_1}, where
\begin{eqnarray}\label{Pjm_2nd_def}
 P^m_j(\cos\theta) \equiv \frac{(-1)^{j+m}}{2^j j!}\frac{(j+m)!}{(j-m)!} \sin^{-m}\theta \frac{d^{j-m} \sin^{2j}\theta}{d (\cos\theta)^{j-m}}
\end{eqnarray}
is the usual alternative definition of associated Legendre function of $m$-th level.

Since $\Theta_{jm}(\theta)$ comes with a unity norm by design, we actually have obtained the normalization constant of $P^m_j(\cos\theta)$, which is just the overall coefficient in front of $P^m_j(\cos\theta)$ in \eq{Theta_jm_final}, in an approach completely different from the traditional one.
In particular, for $m=0$, the associated Legendre function \eqnm{Theta_jm_top_down}  reduces to the base Legendre polynomial with unity norm
\begin{eqnarray}\label{Thetaj01}
 \Theta_{j0}\smbr{\theta}&=&\sqrt{\frac{2j+1}{2}}P_j(\cos\theta),
\end{eqnarray}
where
\begin{eqnarray}\label{jth_Legendre_polynomial}
 P_j(\cos\theta)&\equiv& (-1)^j \frac{1}{2^j j!} \frac{d^j \sin^{2j}\theta}{d (\cos\theta)^j}
\end{eqnarray}
is the base Legendre function expressed in its Rodriguez' formula.

\eq{Thetaj0} tells us that $P_j(\cos\theta)$ is normalized to unity by the coefficient $\sqrt{(2j+1)/2}$. Traditionally, this normalization constant is obtained by perform multiple folds of integration by parts in the integral $\int^\pi_0 d\theta \sin\theta P^2_j(\cos\theta)$.

With $P_j(\cos\theta)$ given in \eq{jth_Legendre_polynomial}, the definition \eqnm{Pjm_1st_def} of $P^m_j(\cos\theta)$ becomes
\begin{eqnarray}\label{Pjm_1st_def1}
 P^m_j(\cos\theta)\equiv (-1)^j \frac{1}{2^j j!} \sin^m\theta \frac{d^{j+m}  \sin^{2j}\theta}{d(\cos\theta)^{j+m}}.
\end{eqnarray}

The consistency between the two definitions \eqnm{Pjm_2nd_def} and \eqnm{Pjm_1st_def1} for $P^m_j(\cos\theta)$ can be checked as follows. We note that the definition \eqnm{Pjm_2nd_def} essentially follows from the top-down construction \eqnm{Theta_tower_lowering} for $\Theta_{jm}(\theta)$, and that the definition \eqnm{Pjm_1st_def1} essentially follows from the bottom-up construction \eqnm{Theta_tower_raising_0} for $\Theta_{jm}(\theta)$. By design, these two constructions for the same $\Theta_{jm}(\theta)$ is consistent. In other words, one way to check the consistency of the two definitions \eqnm{Pjm_2nd_def} and \eqnm{Pjm_1st_def1} of $P^m_j(\cos\theta)$ is to turn the definition \eqnm{Pjm_2nd_def} into \eq{Theta_jm_top_down} for $\Theta_{jm}(\theta)$ and turn the definition \eqnm{Pjm_1st_def1} into \eq{Theta_tower_raising_1} for the same $\Theta_{jm}(\theta)$. Another simpler way around is to show that the consistency between these two definitions of $P^m_j(\cos\theta)$ directly follows from the property
\begin{eqnarray}\label{Theta_relation_plus_and_minus}
 \Theta_{j,-m}(\theta)=(-1)^m \Theta_{j,m}(\theta),
\end{eqnarray}
whose $x$-coordinate representation is \eq{phase_between_plus_and_minus}, which has been verified already.

Most of the statements made in this subsubsection are made for non-negative $m$. There exists a highly similar algorithm for building $\Theta_{jm}(\theta)$ for $m<0$.

\section{The SUSYQM of the Associated Legendre Operator Inherited from  the Angular Momentum Algebra}\label{angular_momentum_algebra}

As pointed out at the very beginning of Section \ref{direct_factorization_of_associated_Legendre}, the associated Legendre equation can result from the eigenequation of the squared angular momentum operator by separating the azimuthal harmonics. In Subsections \ref{direct_factorization_AL_equ_theta_coordinate} and \ref{eigen_solutions_associated_equ_theta_coord}, we started with the `\emph{ill-organized}' form  of such resultant associated Legendre operator and then did a little hard work to pursue its supersymmetric factorization a `\emph{well-organized}' form. In doing so, we completely ignored the fact that the spectrum of the associated Legendre operator is closely related to the representation theory of another elegant  algebra \text{--} the $so(3)$ Lie algebra of angular momentum, whose algebraic relations are far more than \emph{well-organized}. In fact, we  have only invoked the ill-organized form of the polar angle part of the Casimir operator of this Lie algebra \text{--} the square of the angular momentum operator \text{--} in Subsections \ref{direct_factorization_AL_equ_theta_coordinate} and \ref{eigen_solutions_associated_equ_theta_coord}, but do not take advantage of the full algebra. A natural question arising is, why not try to derive the well-organized supersymmetric structure from the pre-existing elegant $so(3)$ algebraic relations.  In this section, we will see  that, this approach is not only possible but also mathematically more appealing. The basic reason for the success of this approach lies in the fact that, there is a 2-parameter ${\cal N}=2$ $su(1,1)$ algebraic structure intrinsic to the $so(3)$ algebra, and the former is easily reduced into a 1-parameter ${\cal N}=2$ $su(1,1)$ algebra in systematic ways. This 1-variable $su(1,1)$ algebraic structure is exactly the SUSQM structure carried by the associated Legendre operator. Therefore, following this procedure of algebra reduction to construct the SUSYQM of the associated Legendre operator, separated extra attempts made for finding out the supersymmetric factorization of the ``ill-organized'' associated Legendre operator in Susubsection \ref{eigen_solutions_associated_equ_theta_coord} become unnecessary. In fact, corresponding to every relation in the angular momentum algebra and its representation, there is a reduced counterpart parametrized by $\theta$ coordinate. In this sense, the resultant SUSY is \emph{inherited} from or the residual of the original SUSY owned by the ($so(3)$) angular momentum algebra.

Of course, the reduction procedure still involves peeling off the azimuthal harmonics. We will see that this can be done systematically in elegant ways. The key point is that the SUSY algebraic structures manifestly remain against such peeling operation if one starts with the $so(3)$ algebra or its equivalent ${\cal N}=2$ $su(1,1)$ algebra.

\subsection{A Brief Review of the Representation of Angular Momentum Algebra and Its Representation}

Appearing as the Lie algebra of the symmetry group $SO(3)$ or $SU(2)$, the angular momentum algebra $so(3)$ and its irreducible representation can be found in many texts of quantum mechanics or of Lie algebras, see, e.g., \cite{Merzbacher_QM_1998, Georgi_Lie_algebras_1999}. Here we give a brief review on these in order to introduce the basic relations and notations that will be used in later subsections.

The three Cartesian components $(J_1, J_2, J_3)$ of the angular momentum operator ${\boldsymbol J}$  satisfy the algebraic relations
\begin{eqnarray}\label{basic_algebra}
 \sqbr{J_i, J_j}=i \varepsilon_{ijk}J_k,\quad i,\;j,\;k \in\cubr{1,\; 2, \;3},
\end{eqnarray}
where $\varepsilon_{ijk}$ is the Levi-Civita complete antisymmetric tensor, and, for simplicity, we adopt the units in which $\hbar =1$.

It follows from \eq{basic_algebra} that the Casimir operator $J^2=J^2_1+J^2_2+J^2_3$ commutes with $J_3$. Therefore,  $J^2$ and $J_3$ have the simultaneous eigenstates $\ket{j m}$,
\begin{eqnarray}\label{Jz_eigenequation}
  J^2\ket{j m}=\lambda_j\ket{j m},\quad J_3|j m>=m \ket{j m}.
\end{eqnarray}
The basis $\{\ket{jm}\}$ span the representation space of such Lie algebra. The first equation above is just \eq{J_square_eigenequation}.

In order to construct the ladder relations inside the representation of $so(3)$, it is more instructive to translate the algebraic relations in \eq{basic_algebra} into those expressed in terms of the ladder operators
\begin{eqnarray}\label{J_plus_minus}
 J^\pm \equiv J_1 \pm iJ_2
\end{eqnarray}
and $J_3$.

It follows from \eqnm{basic_algebra} that
\begin{eqnarray}
 J_3 J^\pm &=& J^\pm\smbr{J_3\pm1}, \label{J3Jpm}\\
J^\pm J^\mp &=& J^2-J^2_3 \pm J_3. \label{Casimir_operators}
\end{eqnarray}

From \eq{Casimir_operators}, by subtraction, we obtain
\begin{eqnarray}\label{shape_invariance_operator_relation}
 J^-J^+ - J^+J^-=-2\, J_3.
\end{eqnarray}

The relations \eqnm{J3Jpm} and \eqnm{shape_invariance_operator_relation} can be respectively rewritten into the following commutator relations,
\begin{eqnarray}\label{alternative_so3_algebra}
 [J_3,\; J^\pm]=\pm J^\pm,\quad [J^-,\;J^+]=-2\, J_3.
\end{eqnarray}
These work as an alternative defining algebraic relations for the $so(3)$ algebra that are equivalent to those in \eq{basic_algebra}. These algebraic relations are \emph{almost} the same as those for $su(1,\,1)$ Lie algebra expressed in terms of its two ladder operators and a $J_3$-like operator, except for the \emph{minus} sign on the right hand side of the second commutator above, which distinguishes this algebra of $\{J^\pm,\,J_3\}$ from the usual $su(1,\,1)$ Lie algebra.

Acting on $\ket{j m}$ with \eq{J3Jpm} and making use of the second equation in \eqnm{Jz_eigenequation}, we obtain
\begin{eqnarray}
 J_3 \smbr{J^\pm \ket{j m}}=\smbr{m\pm1} \smbr{J^\pm \ket{j m}}.
\end{eqnarray}
This shows that $J^\pm \ket{j m}$ are also the eigenstates of $J_3$ yet with eigenvalues $m\pm1$, so they can be labeled by these eigenvalues as
\begin{eqnarray}\label{raising_lowering_relations}
 J^\pm \ket{j m}= c^\pm_{jm} \ket{j m\pm1},
\end{eqnarray}
where $c^\pm_{jm}$ are two constants as yet to be determined, and following the convention in quantum mechanics community, we choose $c^\mp_{jm}$ to be real to avoid extra phase factors. Because of such raising and lowering roles on the eigenvalue $m$, $J^\pm$ are notoriously called raising and lowering operators.

The relations \eqnm{Casimir_operators} show that $J^\pm J^\mp$ commute with both $J^2$ and $J_3$. So $J^\pm J^\mp$ share the eigen-states $\ket{j m}$ with $J^2$ and $J_3$. In fact, acting on $\ket{j m}$ with \eq{Casimir_operators}, we obtain
\begin{eqnarray}\label{Casimir_eigen_equ_m}
J^\pm J^\mp \ket{j m}=\sqbr{\lambda_j-m\smbr{m\mp1}} \ket{j m}.
\end{eqnarray}

Projecting the two equations in \eq{Casimir_eigen_equ_m} onto $\bra{jm}$, we obtain the norms of $J^\mp\ket{j m}$ as
\begin{eqnarray}\label{norms}
\abs{J^\mp\ket{j m}}=\sqbr{\lambda_j-m\smbr{m\mp1}}^{1/2}.
\end{eqnarray}
Thus, $c^\mp_{jm}$ in \eq{raising_lowering_relations} can be chosen as
\begin{eqnarray}\label{coefficients}
 c^\mp_{jm} = \sqbr{\lambda_j-m\smbr{m\mp 1}}^{\frac{1}{2}}.
\end{eqnarray}
Obviously, there exists the symmetry relation
\begin{eqnarray}\label{c_plus_minus_relation}
 c^+_{jm-1}=c^-_{jm}.
\end{eqnarray}

\eq{Casimir_eigen_equ_m} is then rewritten as
\begin{eqnarray}\label{Casimir_eigen_equ_m1}
J^\pm J^\mp \ket{j m}=\smbr{c^\mp_{jm}}^2 \ket{j m}.
\end{eqnarray}

\subsection{The ${\cal N}=2$ $su(1,\,1)$ Superalgebra behind the $so(3)$ Algebra}\label{2D_supersymmetry}

\subsubsection{The Supercharge Formulation of the $su(1,\,1)$ Superalgebra Associated with the Super-Hamiltonian $H^{ss}_J$}\label{su(11)_intuitive}

The $(m-1)$-th version of the lower equation in \eq{Casimir_eigen_equ_m1} reads
\begin{eqnarray}\label{Casimir_eigen_equ_m_minus_1}
J^- J^+ \ket{jm-1}=\smbr{c^-_{jm}}^2 \ket{jm-1}.
\end{eqnarray}
where the symmetry relation \eqnm{c_plus_minus_relation}  has been made use of for substituting for $c^+_{jm-1}$.  Since the upper equation in \eqnm{Casimir_eigen_equ_m1} and \eq{Casimir_eigen_equ_m_minus_1} have the common eigenvalue $\smbr{c^-_{jm}}^2$, these two eigenequations can be packed into the super-eigenequation
\begin{eqnarray}\label{super_eigen_equ_spherical_harmonics}
H^{ss}_J \left(\begin{array}{c}
                                                               \ket{jm} \\
                                                               \ket{jm-1}
                                                              \end{array}\right)=\smbr{c^-_{jm}}^2\left(\begin{array}{c}
                                                                                                                            \ket{jm} \\
                                                                                                                            \ket{jm-1}                                                                                                                    \end{array}\right),
\end{eqnarray}
where the super-Hamiltonian $H^{ss}_J$ is a $2\times$ matrix in which the two component Hamiltonians $H_\pm\equiv J^\pm J^\mp$ are diagonally arranged, i.e.,
\begin{eqnarray}\label{super_H_J}
H^{ss}_J\equiv \left(\begin{array}{cc}
                     J^+ J^-  & 0 \\
                      0 & J^- J^+
                   \end{array}\right)=\left(\begin{array}{cc}
                     H^+  & 0 \\
                      0 & H^-
                   \end{array}\right).
\end{eqnarray}

$H^{ss}_J$ has the factorization
\begin{eqnarray}\label{super_H_J1}
 H^{ss}_J = Q^2,
\end{eqnarray}
 where $Q$ assumes the \emph{anti-diagonal} form
\begin{eqnarray}\label{antidiagonal_matrix}
 Q\equiv \left(\begin{array}{cc}
                      0  & J^+ \\
                      J^- & 0
                   \end{array}\right).
\end{eqnarray}
This $Q$ has the natural splitting
\begin{eqnarray}\label{Q_natural_splitting}
 Q&= Q^+ + Q^-,
\end{eqnarray}
with
\begin{eqnarray}\label{supercharge_Q_pm}
 Q^+ \equiv \left(\begin{array}{cc}
                           0 & J^+\\
                           0  & 0
                        \end{array}\right) = \Sigma^+\cdot J^+,\quad
Q^-\equiv \left(\begin{array}{cc}
                           0 & 0 \\
                           J^- & 0
                        \end{array}\right) = \Sigma^-\cdot J^-, \nonumber \\
\end{eqnarray}
and
\begin{eqnarray}\label{Sigma_pm}
 \Sigma^+ \equiv\left(\begin{array}{cc}
                            0 & 1 \\
                            0 & 0 \\
                      \end{array}\right),\quad
\Sigma^- \equiv \left(\begin{array}{cc}
                           0 & 0 \\
                           1 & 0 \\
                      \end{array}\right).
\end{eqnarray}

It is easy to check that  $\Sigma^\pm$ own the nil-potent (fermionic) property $\smbr{\Sigma^\pm}^2=0$, and  $Q^\pm$ inherit these properties, i.e., $\smbr{Q^\pm}^2=0$. The latter can be rewritten as
\begin{eqnarray}\label{Qpm_nilpotent}
 \cubr{Q^+, Q^+}=\cubr{Q^-,\; Q^-}=0.
\end{eqnarray}

With the properties \eqnm{Qpm_nilpotent}, only the cross terms survive in the expanded form of $Q^2$. Therefore, \eq{super_H_J1} is then rewritten as the anticommutator
\begin{eqnarray}\label{Hs_J_anticommutator}
        H^{ss}_J            &=&\cubr{Q^+,\; Q^-}.
\end{eqnarray}

It follows from the properties \eqnm{Qpm_nilpotent} and \eq{Hs_J_anticommutator} that $Q^\pm$ are conserved quantities in the sense that
\begin{eqnarray}\label{Qpm_conservation}
 \sqbr{Q^\pm,\; H^{ss}_J}=0.
\end{eqnarray}

Eqs. \eqnm{Hs_J_anticommutator}, \eqnm{Qpm_nilpotent} and \eqnm{Qpm_conservation} together define the standard ${\cal N}=2$ $\emph{su}$ (1,1) supersymmetry algebra \cite{Witten_SUSY_breaking_1981, Witten_1982} satisfied by the operator set $\{H^{ss}_J,\,Q^\pm\}$, and the two generators $Q^\pm$ are called surpercharges of this algebra. These two supercharges can be read off directly from  $H^{ss}_J$, once the latter is expressed as the square of the anti-diagonal matrix $Q$. We see that it is the specific diagonal structure of the form of $H^{ss}_J$ that leads to the standard ${\cal N}=2$ $su(1,\,1)$ supersymmetry owned by the operator set $\{H^{ss}_J,\,Q^\pm\}$, without any detailed algebraic properties of $J^\pm$ invoked. However, the super eigenequation \eqnm{super_eigen_equ_spherical_harmonics} does come out of the representation of the $so(3)$ algebra satisfied by the operator set $\{J^\pm,\,J_3\}$.

The supercharge conservation in \eq{Qpm_conservation} implies the eigenvalue degeneracy between the superstate $(\ket{jm},\;\ket{jm-1})^T$ and the superstates $Q^\pm (\ket{jm},\;\ket{jm-1})^T$, which is a reflection of this supersymmetry. \eq{Qpm_conservation} also implies the superstate transformations
\begin{eqnarray}\label{eigenstate_transmutation}
 \left(\begin{array}{c}
   \ket{jm} \\
      0
     \end{array}\right) \quad  \xrightleftharpoons[Q^+/c^-_{jm}]{Q^-/c^-_{jm}}  \left(\begin{array}{c}
     0 \\
    \ket{j m-1}
     \end{array}\right).
\end{eqnarray}
The latter are the incarnations of the raising and lowering recursion relations \eqnm{raising_lowering_relations}. These transformations tell us that the supercharges $Q^\pm$ switch the bosonic and fermionic states (if one is viewed as bosonic state and then the other as the fermionic one), while the energy remains degenerate.

\subsubsection{Constructing the $2$-parameter $su(1,1)$ Superalgebra Directly from the $so(3)$ Algebra through Spin Matrix Maps}\label{superalgebra_spin_matrix_maps}

In Subsubsection \ref{su(11)_intuitive}, the supercharge formulation of the ${\cal N}=2$ supersymmetry algebra $su(1,1)$ behind the \emph{so}(3) algebra has been obtained by intuitively splitting the $2\times2$ matrix $Q$ or the super-Hamiltonian $H^{ss}_J$, on the basis of the structure of $H^{ss}_J$ appearing in the super eigenequation \eqnm{super_eigen_equ_spherical_harmonics}. This leaves us an impression that such supercharge formulation of the superalgebra $su(1,1)$ is based on the spectral representation of the ${so}(3)$ Lie algebra, because \eq{super_eigen_equ_spherical_harmonics} is reached at almost the final stage of searching for this spectral representation, although the supercharges $Q^\pm$ are directly derived from the very structure of the super-Hamiltonian $H^{ss}_J$ itself, without resort to any particular algebraic property of $J^\pm$. One may suspect that the supercharges $Q^\pm$ and the super-Hamiltonian $H^{ss}_J$ come at first place and even the whole super algebra $su(1,1)$ be constructed more straightforwardly from the ${so}(3)$ algebra, \emph{without} falling into any detail of the spectral representation of the \emph{so}(3) algebra, in particular, without reference to the super eigenequation \eqnm{super_eigen_equ_spherical_harmonics}.

$Q^\pm $ defined in \eq{supercharge_Q_pm} can be viewed as the (first-order) \emph{perpendicular spin matrix maps/representations} of $J^\pm$ in the sense that
\begin{eqnarray}\label{Sigmapm_maps}
 \Sigma^\pm:\,J^\pm \longmapsto \Sigma^\pm \cdot J^\pm =\frac{1}{2} \sigma^\pm \cdot J^\pm\equiv Q^\pm,
\end{eqnarray}
where the usual raising and lowering Pauli matrices $\sigma^\pm$ are defined as
\begin{eqnarray}\label{sigma_pm}
\begin{aligned}
 \sigma^+  \equiv \sigma_1 + i \sigma_2=\left(\begin{array}{cc}
                                               0&2\\
                                               0&0
                                              \end{array}
                                              \right),\quad \sigma^- \equiv \sigma_1 - i \sigma_2=\left(\begin{array}{cc}
                                               0&0\\
                                               2&0
                                              \end{array}
                                              \right)
\end{aligned}\nonumber \\
\end{eqnarray}
for the Pauli matrices
\begin{eqnarray}
 \sigma_1\equiv \left(\begin{array}{cc}
                           0 & 1 \\
                           1 & 0
                        \end{array}\right),\quad \sigma_2\equiv \left(\begin{array}{cc}
                           0 & -i \\
                           i & 0
                        \end{array}\right).
\end{eqnarray}

On the other hand, \eq{antidiagonal_matrix} shows that $Q$ is a linear combination of the two components $J_1$ and $J_2$ of the angular momentum operator $\boldsymbol{J}$, with the combination coefficients involving Pauli matrices. In other words, $Q$ can be introduced as the following \emph{perpendicular spin matrix map} of $\boldsymbol{J}_\perp$,
\begin{eqnarray}\label{perpendicular_spin_matrix_map1}
                     \boldsymbol{\sigma}^*_\perp: \boldsymbol{J}_\perp \longmapsto  Q \equiv \boldsymbol{\sigma}^*_\perp \cdot \boldsymbol{J}_\perp=\sigma^*_1J_1+\sigma^*_2J_2,
\end{eqnarray}
where  $\boldsymbol{\sigma}_\perp\equiv \smbr{\sigma_1,\;\sigma_2,\; 0}$ and $\boldsymbol{J}_\perp\equiv \smbr{J_1,\;J_2,\;0}$ and the star denotes complex conjugate. Alternatively, in terms of the \emph{full spin matrix map/representation} of $\boldsymbol{J}$,
\begin{eqnarray}
 \boldsymbol{\sigma}^*:\,{\bf J}\longmapsto \boldsymbol{\sigma}^*\cdot \boldsymbol{J},
\end{eqnarray}
$Q$ can be also introduced as
\begin{eqnarray}\label{Q_spin_matrix_form}
 Q\equiv\boldsymbol{\sigma}^*\cdot \boldsymbol{J}- J_3 \sigma^*_3,
\end{eqnarray}
where $\boldsymbol{\sigma}\equiv\smbr{\sigma_1,\;\sigma_2,\;\sigma_3}$, and
\begin{eqnarray}
\sigma_3 =\begin{pmatrix}
 1&0\\
 0&-1
\end{pmatrix}
\end{eqnarray}
is the third Pauli matrix.

Upon inserting $\sigma_1=\smbr{\sigma^+ +\sigma^-}/2$ and $\sigma_2=\smbr{\sigma^+ -\sigma^-}/2i$ and making use of the definitions \eqnm{J_plus_minus} for $J^\pm$, $Q$ defined in \eq{perpendicular_spin_matrix_map1}
is naturally decomposed as
\begin{eqnarray}\label{Q_alternative_form}
 Q=Q^-+Q^+=\left(\begin{array}{cc}
                           0 & J^+ \\
                           J^- & 0
                        \end{array}\right),
\end{eqnarray}
where
\begin{eqnarray}\label{Q_pm}
 Q^\pm \equiv\frac{1}{2}\sigma^\pm \cdot J^\pm
\end{eqnarray}
are the two sought-after supercharges, which are the same as those defined directly through the perpendicular spin matrix maps \eqnm{Sigmapm_maps}.

The super-Hamiltonian $ H^{ss}_J$ associated with $Q$ is defined as $H^{ss}_J\equiv Q^2$ (the same as that in \eq{super_H_J1}). It can be expanded as
\begin{eqnarray}\label{superHamiltonian_HsJ}
 H^{ss}_J=Q^+Q^- + Q^-Q^+=\left(\begin{array}{cc}
                      J^+J^-  & 0 \\
                      0 & J^-J^+
                      \end{array}\right),
\end{eqnarray}
where the crossing terms in $Q^2$ vanish because of the nil-potent properties $\smbr{Q^\pm}^2=0$, which follow from the nil-potent properties $\smbr{\sigma^\pm}^2=0$.

The three defining equations \eqnm{Qpm_nilpotent}, \eqnm{Hs_J_anticommutator} and \eqnm{Qpm_conservation} of the ${\cal N}=2$ superalgebra $su(1,1)$ satisfied by the operator set $\left\{H^{ss}_J,\;Q^\pm \right\}$ can be derived exactly the same way as in Subsubsection \ref{su(11)_intuitive}.

So far in this subsubsection, no algebraic property of $J^\pm$ or that of the $so(3)$ algebra has been invoked. This means that the construct of such ${\cal N}=2$ $su(1,1)$ superalgebra applies to many other Lie algebras, too, with $J^\pm$ being a pair of Hermite conjugate generators.

Next, to search for the representation of such superalgebra,  the contact between $H^{ss}_J$ and the operator set $\{J^2, J_3\}$ should be established, in view that the latter carries the irreducible representation of $so(3)$ algebra, and the defining algebraic relations of  $so(3)$ in \eq{basic_algebra} or \eq{alternative_so3_algebra}  is then invoked. In fact, we can prove that
\begin{eqnarray}\label{HsJ_J_relation}
H^{ss}_J = J^2-J^2_3 + J_3 \sigma_3,
\end{eqnarray}
 by directly computing $Q^2$ with $Q$ in the form \eqnm{perpendicular_spin_matrix_map1}, with the aids of the relations \eqnm{basic_algebra} and $\sigma_i \sigma_j =\delta_{ij}I+i\epsilon_{ijk} \sigma_k$, where the second set of relations results from the combination of the Lie algebra as well as the Clifford algebra satisfied by the three Pauli matrices. On the other hand, \eq{HsJ_J_relation} is nothing but the matrix operator equation obtained by diagonally packing the two relations in \eq{Casimir_operators} for $J^+J^-$ and $J^- J^+$ , with $J^+J^-$ in the upper-left corner and $J^- J^+$ in the lower-right one, recalling that these two relations are equivalent to the defining algebraic relations of $so(3)$ in \eq{basic_algebra}.

In the above we have demonstrated that the $so(3)$ algebra always has an ${\cal N}=2$ $su(1,1)$ superalgebra structure, with the supercharges given by the spin matrix maps in \eq{Sigmapm_maps} or in \eq{Q_spin_matrix_form} and the super-Hamiltonian in  \eq{HsJ_J_relation}. Consequently, the representation of such $su(1,1)$ superalgebra is equivalent to the representation of the original $so(3)$ algebra.

It looks a bit `odd' that in Eqs. \eqnm{perpendicular_spin_matrix_map1} and \eqnm{Q_spin_matrix_form}  $Q$ is defined as the full spin matrix map of $\boldsymbol{J}$ in terms of the complex conjugate of spin matrix $\boldsymbol{\sigma}$ map, instead of $\boldsymbol{\sigma}$ itself. One wonders if a supercharge  can be defined in terms of $\boldsymbol{\sigma}$ itself. We will show this is completely OK.

We define the overall supercharge $\bar{Q}$ as the perpendicular spin matrix map
\begin{eqnarray}\label{perpendicular_spin_matrix_map2}
                     \boldsymbol{\sigma}_\perp: \boldsymbol{J}_\perp \longmapsto  \bar{Q} \equiv \boldsymbol{\sigma}_\perp \cdot \boldsymbol{J}_\perp=\sigma_1J_1+\sigma_2J_2,
\end{eqnarray}
or
\begin{eqnarray}\label{Qbar_spin_matrix_form}
 \bar{Q}\equiv\boldsymbol{\sigma}\cdot \boldsymbol{J}- J_3 \sigma_3.
\end{eqnarray}
Here the full spin matrix map $\boldsymbol{\sigma}\cdot \boldsymbol{J}$ maps the vector operator $\boldsymbol{J}$ into a rank-2 spin tensor operator.

The idea of spin matrix representation of a vector has been used in other contexts. A famous example is the spin map that maps a 3-D position vector $\boldsymbol{x}$ into the spin matrix $X\equiv \boldsymbol{\sigma} \cdot \boldsymbol{x}$, a rank-2 spin tensor. One of the advantages of such spin matrix representation of a vector $\boldsymbol{x}$ lies in that the Euclidean rotation of $\boldsymbol{x}$ into $\boldsymbol{x}'$ by the angle $\theta$ around the axis $\boldsymbol{n}$ is easily realized by the conjugate transformation of the spin matrix  $X$, according to ${\boldsymbol{\sigma}\cdot\boldsymbol{x}'}\equiv X'=R(\theta) X R^*(\theta)$, with the spin matrix representation of the rotation matrices $R(\theta)\equiv e^{-i\boldsymbol{\sigma}\cdot \boldsymbol{n}/2}$ and $R^*(\theta)\equiv e^{i\boldsymbol{\sigma}\cdot \boldsymbol{n}/2}$. See the detail of this example on pages 1140-1142 in \cite{MTW_2017}.

Upon changing the basis $\left\{\sigma_1,\;\sigma_2\right\}$ to the basis $\left\{\sigma^+,\;\sigma^-\right\}$ through $\sigma_1=\smbr{\sigma^+ +\sigma^-}/2$ and $\sigma_2=\smbr{\sigma^+ -\sigma^-}/2i$ and making use of the definitions \eqnm{J_plus_minus} for $J^\pm$, $Q$ in \eq{perpendicular_spin_matrix_map2}
is decomposed as
\begin{eqnarray}\label{Qbar_alternative_form}
 \bar{Q}=\left(\begin{array}{cc}
                           0 & J^- \\
                           J^+ & 0
                        \end{array}\right)=\bar{Q}^-+\bar{Q}^+,
\end{eqnarray}
where
\begin{eqnarray}\label{Qbar_pm}
 \bar{Q}^+ \equiv\left(\begin{array}{cc}
                  0&0\\
                  J^+&0
                 \end{array}\right)
                 =\frac{1}{2}\sigma^- \cdot J^+,\quad
 \bar{Q}^- \equiv\left(\begin{array}{cc}
                  0&J^-\\
                  0&0
                 \end{array}\right)
                 =\frac{1}{2}\sigma^+ \cdot J^-.\nonumber \\
 \end{eqnarray}

Here we see that the two supercharges $\bar{Q}^\pm$ defined in \eqnm{Qbar_pm} are just the transposes of $Q^\pm$ defined in \eq{supercharge_Q_pm}. Such transposing operation can be implemented by the bilateral transformation $\sigma_1\smbr{\cdots}\sigma_1$, in view that the left multiplication with $\sigma_1$ amounts to switching the two rows of a $2\times2$ matrix and the right multiplication with $\sigma_1$ amounts to switching the two columns of a $2\times2$ matrix. Therefore, we have the relations
\begin{eqnarray}
 {\bar Q}^\pm \equiv \sigma_1 Q^\pm\sigma_1,\quad {\bar Q} \equiv \sigma_1 Q\sigma_1.
\end{eqnarray}

The super-Hamiltonian $\bar{H}^{ss}_J$ is defined as

\begin{eqnarray}\label{barHs_J_J}
 \bar{H}^{ss}_J &\equiv&{\bar{Q}}^2  =\left(\begin{array}{cc}
                      J^-J^+ & 0 \\
                      0 &  J^+J^-
                      \end{array}\right), \nonumber\\
                      &=& J^2-J^2_3 - J_3 \sigma_3,
\end{eqnarray}
where the third equality has been reached by diagonally packing the two operator equations in \eq{Casimir_operators}, with $J^- J^+$ in the upper-left corner and $J^+J^-$ in the lower-right corner.
This $\bar{H}^{ss}_J$ is related to $H^{ss}_J$ through
\begin{eqnarray}
 \bar{H}^{ss}_J=\sigma_1 H^{ss}_J \sigma_1.
\end{eqnarray}

In fact, resort to the representation of $so(3)$ algebra, the matrix form of thesuper Hamiltonian $\bar{H}^{ss}_J$ in \eq{barHs_J_J} is easily read off from the super eigenequation
\begin{eqnarray}\label{super_eigen_equ_barH_J}
\begin{aligned}
{\bar H}^{ss}_J \left(\begin{array}{c}
                                                               \ket{jm-1} \\
                                                               \ket{jm}
                                                              \end{array}\right)&=\left(\begin{array}{cc}
                      J^-J^+ & 0 \\
                      0 & J^+J^-
                   \end{array}\right) \left(\begin{array}{c}
                                                               \ket{jm-1} \\
                                                               \ket{jm}
                                                              \end{array}\right)\\
                                                              &=\smbr{c^-_{jm}}^2\left(\begin{array}{c}
                                                                                                                            \ket{jm-1}\\
                                                                                                                            \ket{jm}
                                                                                                                    \end{array}\right),
\end{aligned}
\end{eqnarray}
which is just the upside-down package of the two eigen-equations in the super eigenequation \eqnm{super_eigen_equ_spherical_harmonics}.

It is easy to check that the operator set $\{{\bar Q}^\pm,\; \bar{H}^{ss}_J\}$ satisfy the same ${\cal N}=2$ $su(1,1)$ superalgebra as the operator set $\{Q^\pm,\;H^{ss}_J\}$.

To my best knowledge, the elucidation of the explicit $su(1,1)$ SUSY intrinsic to the $so(3)$ Lie algebra  at such formal level, based on the idea of spin matrix map, has not been reported before in literature.

Although the $so(3)$ algebraic relations in \eq{alternative_so3_algebra} indicates that the $so(3)$ Lie algebra fails to be the  $su(1,\,1)$ \emph{plain} Lie algebra, it is quite surprising that there always exists an ${\cal N}=2$ $su(1,\,1)$ \emph{super} Lie algebra associated with this $so(3)$ Lie algebra. Of course, a subsequent key issue is the spectral representation of such ${\cal N}=2$ $su(1,1)$ superalgebra.

 We have seen that the intimate connection of the abstract ${\cal N}=2$  $su(1,1)$ superalgebra with the $so(3)$ algebra can be established trough the spin matrix maps, and  neither does such connection necessarily rely on the defining algebraic relations of $so(3)$ in \eq{basic_algebra} or \eq{alternative_so3_algebra}, nor on the representation of $so(3)$. However, just as the relevance of the representation of the $so(3)$ algebra itself, a \emph{ concrete representation} of this ${\cal N}=2$  $su(1,1)$ algebra \emph{does} involve the properties of $J^\pm$, such as the commutator relations between $J^\pm$ and  $J^2$ or $J_3$. In particular, we will see that the expression \eqnm{HsJ_J_relation} for the super-Hamiltonian $H^{ss}_J$, or the expression \eqnm{barHs_J_J} for ${\bar H}^{ss}_J$, plays the key role in the representation of such superalgebra. We also need to impose the super-eigenequation \eqnm{super_eigen_equ_spherical_harmonics} for the super-Hamiltonian $H^{ss}_J$, or the super-eigenequation \eqnm{super_eigen_equ_barH_J} for the super-Hamiltonian ${\bar H}^{ss}_J$. On imposing these super eigenequations, one may wonder why the particular combination $\smbr{\ket{jm},\ket{jm-1}}^T$ for  $H^{ss}_J$ or $\smbr{\ket{jm-1},\ket{jm}}^T$ for  ${\bar H}^{ss}_J$ is chosen as the super eigenstate. One of the reasons for choosing this type of combination, e.g., for $H^{ss}_J$, lies in that the two component states should be subject to the raising/lowering recurrence relations \eqnm{eigenstate_transmutation}, in view of the specific algebraic properties \eqnm{Hs_m_anticommutator}, \eqnm{anti_commutators_Qm} and \eqnm{Qpmm_conservation} satisfied by $Q^\pm$ and $H^{ss}_J$.

As far as constructing a \emph{consecutive} SUSYQM series for exhausting the eigenspectra of $J^2$ and $J_3$, a sort of finding out the consecutive representation of the ${\cal N}=2$ super algebra $su(1,1)$, is concerned, the so-called pattern invariance condition for passing from the $(m-1)$-th level of SUSYQM to that of the $m$-th level is needed. In the following, we will show that the operator relation \eqnm{shape_invariance_operator_relation} works as such required pattern invariance condition.

As customarily done in SUSYQM, intertwiningly acting \eq{Casimir_eigen_equ_m_minus_1} with the right-situated operator $J^+$ and assuming that $J^+\ket{jm-1}\propto \ket{jm}$, we obtain the superpartner eigenequation of \eq{Casimir_eigen_equ_m_minus_1} as
\begin{eqnarray}\label{upper_eigen_equ}
 J^+J^- \ket{jm} = \smbr{c^-_{jm}}^2\ket{j m},
\end{eqnarray}
which is just the upper eigenequation for $J^+J^-$ in \eq{super_eigen_equ_barH_J}.

It follows from Eqs. \eqnm{coefficients} and \eqnm{c_plus_minus_relation}, or from the expectation value of \eq{shape_invariance_operator_relation} with respect the state $\ket{jm}$, that
\begin{eqnarray}\label{c_coeffecient_relation}
\smbr{c^-_{jm}}^2-2m=\smbr{c^-_{jm+1}}^2.
\end{eqnarray}

Substituting for $J^+J^-$ via \eq{shape_invariance_operator_relation} and making use of \eq{c_coeffecient_relation} and $J_z\ket{jm}=m\ket{jm}$, \eq{upper_eigen_equ} is renormalized into
\begin{eqnarray}\label{lower_eigen_equ_m}
 J^- J^+\ket{jm} = \smbr{c^-_{jm+1}}^2\ket{jm},
\end{eqnarray}
which is just the $m$-th version of \eq{Casimir_eigen_equ_m_minus_1}. It is in this sense that the operator relation \eqnm{shape_invariance_operator_relation} works as an operator form of pattern invariance condition, though the right hand side of \eq{shape_invariance_operator_relation} is not a constant, unlike in a usual pattern invariance condition.

\subsection{The Reduced 1-parameter ${\cal N}=2$ $su(1,1)$ Superalgebra and the Corresponding SUSYQM Algorithm for Constructing the Associated Legendre Functions}\label{1-variable_inherited_susyQM}

The ${\cal N}=2$ $su(1,1)$ superalgebra behind the $so(3)$ Lie algebra addressed in Subsection \ref{2D_supersymmetry}  can be parametrized concretely by the two spherical coordinates ($\theta$, $\phi$), hence it is termed as 2-parameter ${\cal N}=2$ $su(1,1)$ superalgebra from now on. We will first show that some intimately related $1$-parameter ${\cal N}=2$ $su(1,1)$ superalgebra satisfied by a set of operators parameterized by the single coordinate $\theta$ can be reduced from such 2-parameter superalgebra, by entering into the concrete spherical coordinate representation of this $2$-parameter $su(1,1)$ superalgebra. This type of superalgebraic relations is dubbed \emph{SUSY inherence} in this paper. One obvious way of deriving such 1-parameter superalgebra is through projecting out (stripping off) the azimuthal harmonics and computing the non-vanishing matrix elements of the operators $J^\pm$ \text{--} the generators of the $so(3)$ algebra \text{--} between a certain pair of such harmonics. We will see that, along this line of derivation, a SUSYQM algorithm for constructing the associated Legendre functions will be developed from the resultant associated Legendre equation and its superpartner equation, both of which come out in their \emph{well-organized} forms \text{--} the expected supersymmetrically factorized forms.

The eigenstates $\ket{jm}$ of $J^2$ ($m\ge 0$) represented in the coordinates ($\theta$, $\phi$) are called spherical harmonics, which take the form
\begin{eqnarray}
\braket{\theta, \phi|jm}&=\Theta_{jm}(\theta) \braket{\phi|m},
\end{eqnarray}
where
\begin{eqnarray}\label{Theta_jm_definition}
\Theta_{jm}(\theta) \equiv \braket{m|jm}
\end{eqnarray}
is the normalized associated Legendre polynomial, and is identified with the function appearing in \eq{symmetrized_AL_equ_theta_coord}, and
\begin{eqnarray}
 \braket{\phi|m}&\equiv \frac{e^{i m\phi}}{\sqrt{2\pi}}
\end{eqnarray}
is the normalized azimuthal harmonic.

In spherical coordinates $(\theta, \phi)$, $J^\pm$ are represented as \cite{Merzbacher_QM_1998}
\begin{eqnarray}
 J^\pm=\pm e^{\pm i\phi}\smbr{\pd{}{\theta} \pm i \cot\theta \pd{}{\phi}}
\end{eqnarray}
in the units in which $\hbar =1$. Thus, the matrix entries of $J^\pm$ between two azimuthal harmonics $\ket{m'}$ and $\ket{m}$ are
\begin{eqnarray}\label{J_plus_minus_matrices}
 \bra{m'}J^\pm\ket{m}=\pm \delta_{m',\,m\pm1} \smbr{\od{}{\theta} \mp m \cot\theta }.
\end{eqnarray}
Explicitly the non-vanishing matrix entries of $J^\pm$, that involve $\theta$ variable only, are denoted as
\begin{eqnarray}
\mfk{J}^+_m (\theta) &\equiv& \bra{m} J^+\ket{m-1} = \od{}{\theta} - (m-1) \cot\theta, \label{raising_operator} \\
\mfk{J}^-_m (\theta) &\equiv& \bra{m-1} J^-\ket{m} = -\od{}{\theta} - m \cot\theta. \label{lowering_operator}
\end{eqnarray}
Notice that the two operators $\mfk{J}^\pm_m (\theta)$ introduced this way are the same as those introduced in \eq{mfkJpm_def}. It is easy to check that $\mfk{J}^-_m(\theta)$ and $\mfk{J}^+_m(\theta)$ are Hermitian conjugate to each other with respect to the inner product $(g_1, g_2)_{\sin\theta}\equiv\int^\pi_0 d\theta \sin\theta g_1(\theta)g_2(\theta)$. $\mfk{J}^\pm_m (\theta)$ come with the symmetry relations
\begin{eqnarray}
\mfk{J}^+_m (\theta)=- \mfk{J}^-_{1-m}(\theta),\quad \mfk{J}^-_m (\theta)=- \mfk{J}^+_{1-m}(\theta).
\end{eqnarray}

The eigenequation satisfied by $\Theta_{jm}(\theta)$ can be obtained by {\it peeling off the azimuthal harmonics $\ket{m}$} in the upper equation in \eqnm{Casimir_eigen_equ_m}. Concretely, by projecting that equation onto $\bra{m}$, and inserting the completeness relation $\sum_{k\in \mathbb{Z}}\ket{k}\bra{k}=1$ into each gap among $J^+$, $J^-$ and $\ket{jm}$ on the left-hand side of that equation, we obtain
\begin{eqnarray}\label{Hermitian_factorization_theta_coord}
 \mfk{J}^+_m \mfk{J}^-_m \Theta_{jm}(\theta)=\smbr{c^-_{jm}}^2 \Theta_{jm}(\theta).
\end{eqnarray}
This is nothing but the supersymmetrically factorized associated Legendre equation \eq{associated_Legendre_asymmetric_factorization}, that has been reached through directly factorizing the ``ill-organized'' raw associated Legendre equation.

Similarly, upon projecting the lower equation in \eqnm{Casimir_eigen_equ_m} onto $\bra{m-1}$ and inserting the completeness relation $\sum_{k\in \mathbb{Z}}\ket{k}\bra{k}=1$ into each gap among $J^+$, $J^-$ and $\ket{jm}$ on the left-hand side of that equation, we obtain the superpartner eigenequation of \eq{Hermitian_factorization_theta_coord} as
\begin{eqnarray}\label{superpartner_eigenequ_theta_coord}
 \mfk{J}^-_m \mfk{J}^+_m  \Theta_{jm-1}(\theta)=\smbr{c^-_{jm}}^2 \Theta_{jm-1}(\theta).
\end{eqnarray}

Here we have seen that advantage of starting with the `elegantly-organized' Lie algebraic relations of $so(3)$ and its representation \eqnm{Casimir_eigen_equ_m}. That is, the resultant 1-parameter associated Legendre equation \eqnm{Hermitian_factorization_theta_coord} and its superparnter equation \eqnm{superpartner_eigenequ_theta_coord}, come out automatically in their supersymmetrically factorized forms, and extra effort of performing the supersymmetric factorization of the associated Legendre equation, that is needed for construct the SUSYQM algorithm of the associated Legendre operator if one starts with the `ill-organized' form \eqnm{associated_Legendre_equ_theta_coord} of the associated Legendre equation, is not needed any more. The deep reason for the existence of such economy lie in that: 1) the $so(3)$ Lie algebra intrinsically has a 2-parameter ${\cal N}=2$ $su(1,1)$ superalgebraic structure, as revealed in Subsection \ref{2D_supersymmetry}; 2) there exists the corresponding a 1-parameter ${\cal N}=2$ $su(1,1)$, that are already embodied by the pair of equations \eqnm{Hermitian_factorization_theta_coord} and \eqnm{superpartner_eigenequ_theta_coord} that are superpartners to each other, and will be written out explicitly immediately.

The two eigenequations \eqnm{Hermitian_factorization_theta_coord} and \eqnm{superpartner_eigenequ_theta_coord} can be packed into the following super-eigenequation
\begin{eqnarray}\label{matrix_eigen_equ_1D}
H^{ss}_m\left(\begin{array}{c}
                                                               \Theta_{jm} \\
                                                               \Theta_{jm-1}
                                                              \end{array}\right)=\smbr{c^-_{jm}}^2 \left(\begin{array}{c}
                                                                                                                            \Theta_{jm} \\
                                                                                                                            \Theta_{jm-1}
                                                                                                                    \end{array}\right)
\end{eqnarray}
with the diagonal super-Hamiltonian
\begin{eqnarray}
 H^{ss}_m \equiv \left(\begin{array}{cc}
                      \mfk{J}^+_m \mfk{J}^-_m & 0 \\
                      0 & \mfk{J}^-_m \mfk{J}^+_m
                   \end{array}\right).
\end{eqnarray}

$H^{ss}_m$ can be rewritten as
\begin{eqnarray}\label{Hs_m_anticommutator}
H^{ss}_m =Q^2_m=\left\{Q^+_m,\; Q^-_m\right\},
\end{eqnarray}
where
\begin{eqnarray}\label{Qbarm}
 Q_m\equiv \left(\begin{array}{cc}
                       0 & \mfk{J}^+_m \\
                      \mfk{J}^-_m & 0
                   \end{array}\right)
\end{eqnarray}
and the two supercharges $Q^\pm_m$ are
\begin{eqnarray}\label{Qpmm0}
Q^+_m\equiv \frac{1}{2}\sigma^+ \cdot \mfk{J}^+_m,\quad Q^-_m\equiv \frac{1}{2}\sigma^-\cdot \mfk{J}^-_m.
\end{eqnarray}

The nilpotent properties $\smbr{Q^\pm_m}^2=0$ can be rewritten as
\begin{eqnarray}\label{anti_commutators_Qm}
 \left\{Q^\pm_m,\;Q^\pm_m\right\}=0.
\end{eqnarray}
It then follows that
\begin{eqnarray}\label{Qpmm_conservation}
 \left[Q^\pm_m, \; H^{ss}_m\right]=0.
\end{eqnarray}

Eqs. \eqnm{Hs_m_anticommutator}, \eqnm{anti_commutators_Qm} and \eqnm{Qpmm_conservation}  have the same structures respectively as Eqs. \eqnm{Qpm_nilpotent}, \eqnm{Hs_J_anticommutator}  and \eqnm{Qpm_conservation}, the latter define the ${\cal N}=2$ $su(1,1)$ superalgebra satisfied by the operator set $\{H^{ss}_J,Q^\pm\}$. Thus, the operator set $\{H^{ss}_m,\; Q^\pm_m\}$ also satisfy the \emph{same} ${\cal N}=2$ $su(1,1)$ superalgebra, and the latter superalgebra is inherited from the former. In other words, when peeling off the azimuthal harmonics, the ${\cal N}=2$ $su(1,1)$ superalgebraic structure remains the same.  The only difference is that the former is parameterized by a single variable $\theta$, while the latter by two variables $\theta$ and $\phi$.

Upon making use of \eq{J_plus_minus_matrices}, the diagonal matrix element of the operator form of the pattern invariance condition \eqnm{shape_invariance_operator_relation} with respect to the state $\ket{m-1}$ becomes the following downward-connected pattern invariance relation
\begin{eqnarray}\label{shape_invariance_condition_theta_coord_mfkJ1}
 \mfk{J}^-_m \mfk{J}^+_m-\mfk{J}^+_{m-1} \mfk{J}^-_{m-1} = -2(m-1).
\end{eqnarray}
Of course, this condition can also be directly checked, by using the explicit forms of $\mfk{J}^\pm_m(\theta)$ given in \eqnm{raising_operator} and \eqnm{lowering_operator}, or by subtract the $(m-1)$ version of \eq{Hermitian_factorization_theta_coord} from \eq{superpartner_eigenequ_theta_coord}.

Intertwiningly acting \eq{Hermitian_factorization_theta_coord} with $\mfk{J}^-_m$, we obtain the superpartner eigenequation \eqnm{superpartner_eigenequ_theta_coord} if $\mfk{J}^-_m\Theta_{jm}(\theta)\propto \Theta_{jm-1}(\theta)$. Making substitution for $\mfk{J}^-_m\mfk{J}^+_m$ via the condition \eqnm{shape_invariance_condition_theta_coord_mfkJ1} and making use of the definition \eqnm{coefficients} for $c^-_{jm}$, we obtain
\begin{eqnarray}\label{renormalized_eigen_eq}
 \mfk{J}^+_{m-1}\mfk{J}^-_{m-1}\sqbr{\mfk{J}^-_m\Theta_{jm}(\theta)}=\smbr{c^-_{jm-1}}^2\sqbr{\mfk{J}^-_m\Theta_{jm}(\theta)}.
\end{eqnarray}
This says that $\mfk{J}^-_m\Theta_{jm}(\theta)$ is the eigenstate of $\mfk{J}^+_{m-1}\mfk{J}^-_{m-1}$ with eigenvalue $\smbr{c^-_{jm-1}}^2$. Thus, $\mfk{J}^-_m\Theta_{jm}(\theta)$ indeed corresponds to the eigenstate $\Theta_{jm-1}(\theta)$ and we can set it to be the lowering recurrence relation
\begin{eqnarray}\label{J_tilde_lowering}
\mfk{J}^-_m \Theta_{jm}(\theta) &=& c^-_{jm} \Theta_{jm-1}(\theta).
\end{eqnarray}
Consequently, \eq{renormalized_eigen_eq} is just the $(m-1)$-version of \eq{Hermitian_factorization_theta_coord}. In other words,
the condition \eqnm{shape_invariance_condition_theta_coord_mfkJ1} is the \emph{genuine} pattern invariance condition that works to renormalize the superpartner eigenequation \eqnm{superpartner_eigenequ_theta_coord} of the eigenequation \eqnm{Hermitian_factorization_theta_coord} into the {\it $(m-1)$-version} of the latter eigenequation.

The combination of Eqs. \eqnm{Hermitian_factorization_theta_coord} and \eqnm{J_tilde_lowering} yields the raising recurrence relation
\begin{eqnarray} \label{J_tilde_raising}
 \mfk{J}^+_m\Theta_{jm-1}(\theta) &=& c^-_{jm} \Theta_{jm}(\theta).
\end{eqnarray}

The raising and lowering recursion relations \eqnm{J_tilde_raising} and \eqnm{J_tilde_lowering} can also be obtained by appropriately peeling off the azimuthal harmonics $e^{i (m\pm 1)\phi}$ from the two recurrence relations in \eq{raising_lowering_relations}. Concretely,  upon inserting $\sum_{k\in \mathbb{Z}}\ket{k}\bra{k}=1$ between $J^\pm$ and $\ket{m\pm1}$, projecting the upper and the lower equations in \eqnm{raising_lowering_relations} to $\braket{m\pm1|\phi}\equiv e^{-i(m\pm1)\phi}/\sqrt{2\pi}$ respectively, and in view of the definitions \eqnm{raising_operator} and \eqnm{lowering_operator},
we obtain the one-parameter raising and lowering recurrence relations \eqnm{J_tilde_raising} and \eqnm{J_tilde_lowering}.

Note that, up to now, we have recovered all the necessary ingredients in Subsubsection \ref{eigen_solutions_associated_equ_theta_coord}, such as the ladder operators $\mfk{J}^\pm_m$, the raising/lowering relations, the two partner eigen-equations \eqnm{Hermitian_factorization_theta_coord} and \eqnm{superpartner_eigenequ_theta_coord}, the pattern invariance condition, etc. This means that the rest iterative SUSYQM algorithm of obtaining the eigensolution  $\Theta_{jm}(\theta)$ and $\lambda_j$ is completed by continuing the same procedure as that developed in Subsubsection \ref{eigen_solutions_associated_equ_theta_coord}, and it does not need to be duplicated here.

\subsection{Constructing the 1-parameter ${\cal N}=2$ $su(1,1)$ Superalgebra Directly through Higher-order Spin Matrix Maps}

In the middle of Subsection \ref{1-variable_inherited_susyQM}, the explicit ${\cal N}=2$ $su(1,1)$ superalgebra satisfied by the operator set $\left\{H^{ss}_m,\;Q^\pm_m \right\}$ has been derived with reference to the super eigenequation \eqnm{matrix_eigen_equ_1D}, which has been derived from the two eigenequations in \eq{Casimir_eigen_equ_m} by peeling off the azimuthal harmonics. Note that those two eigenequations in \eq{Casimir_eigen_equ_m} are the key representation equations for the $so(3)$ Lie algebra. So the derivation of the 1-parameter ${\cal N}=2$ $su(1,1)$ superalgebra is really through the representation of the $so(3)$ Lie algebra. How to derive such superalgebra without resort to the eigenequation \eqnm{matrix_eigen_equ_1D}, i.e., without falling into the representation of $so(3)$? Recall that, in Subsubsection \ref{superalgebra_spin_matrix_maps}, we have revealed the \emph{2-parameter} ${\cal N}=2$ $su(1,1)$ superalgebraic structures satisfied by the operator set $\left\{ H^{ss}_J,\; Q^\pm \right\}$ or $\left\{ {\bar H}^s_J,\; \bar{Q}^\pm \right\}$ through spin matrix maps, instead of resort to the representation of $so(3)$. In this subsection, we hope to start from there and derive the \emph{1-parameter} superalgebra satisfied by $\left\{H^{ss}_m,\;Q^\pm_m \right\}$  through spin matrix maps, too.

\eq{Qpmm0} shows that the two supercharges $Q^\pm_m$ can be obtained from $\mfk{J}^\pm_m$ defined in Eqs. \eqnm{raising_operator} and \eqnm{lowering_operator} through the \emph{plain} (first-order) perpendicular spin matrix maps
\begin{eqnarray}
 &&\frac{1}{2}\sigma^+: \mfk{J}^+_m \longmapsto Q^+_m\equiv\frac{1}{2}\sigma^+ \mfk{J}^+_m=\bra{m}Q^+\ket{m-1}, \label{spin_matrix_map_plus} \\
 &&\frac{1}{2}\sigma^-: \mfk{J}^-_m \longmapsto Q^-_m\equiv \frac{1}{2}\sigma^- \mfk{J}^-_m=\bra{m-1}Q^-\ket{m}.\label{spin_matrix_map_minus}
\end{eqnarray}
Based on these $Q^\pm_m$, one can introduce $Q_m\equiv Q^+_m+Q^-_m$ and $H^{ss}_m \equiv Q^2_m$ and then check that $H^{ss}_m$ and $Q^\pm_m$ satisfies the 1-parameter ${\cal N}=2$ $su(1,1)$ superalgebra defined by Eqs. \eqnm{anti_commutators_Qm}, \eqnm{Hs_m_anticommutator} and \eqnm{Qpmm_conservation}. However, such superalgebra cannot be neatly  reduced from the 2-parameter superalgebra by taking the average of the latter over the one-component azimuthal harmonic $\ket{m}$ or $\ket{m-1}$. Indeed, neither $Q^\pm_m$ nor $H^{ss}_m$ is such type of diagonal average (element).

Next, one may want to try the average of the 2-parameter superalgebra over the two-component superstate $\smbr{\ket{m},\;\ket{m-1}}^T$. However, a little thinking tells us that this would not work.  Because forming the matrix entry between any two such two-component spinors  reduces the dimension of a matrix by a half, if we directly take the average of a $2\times2$ matrix operator over such two-component spinor, the result is a $1\times1$ matrix operator. In order to result in the operator set $\left\{H^{ss}_m,\;Q^\pm_m \right\}$ that are  $2\times$ matrix operators, we have to firstly design a set of operators $\left\{\widetilde{H}^{ss},\;\widetilde{Q}^\pm\right\}$, which  satisfy the ${\cal N}=2$ $su(1,1)$ superalgbra and have \emph{twice} the dimension of $\left\{H^{ss}_m,\;Q^\pm_m \right\}$, and then reduce such superalgebra to the 1-parameter ${\cal N}=2$ $su(1,1)$ superalgbra satisfied by $\left\{H^{ss}_m,\;Q^\pm_m \right\}$, by taking the average over the superstate $\smbr{\ket{m},\;\ket{m-1}}^T$. To this end, we have to make use of the \emph{higher-order spin matrix maps}, instead of the \emph{first-order} (plain) spin matrix maps in Eqs. \eqnm{spin_matrix_map_plus} and \eqnm{spin_matrix_map_minus}. In deed, soon we will see that, when combining the two plain spin matrix maps. \eqnm{spin_matrix_map_plus} and \eqnm{spin_matrix_map_minus} into a single average over the superstate $\smbr{\ket{m},\;\ket{m-1}}^T$, we actually encounter one of such higher-order spin matrix maps.

In Subsubsection \ref{su(11)_intuitive}, we have noticed that the derivation of the  three equations \eqnm{Qpm_nilpotent}, \eqnm{Hs_J_anticommutator}  and \eqnm{Qpm_conservation}, which define a ${\cal N}=2$ $su(1,1)$ superalgebra, does not invoke any concrete property of $J^\pm$. Instead, such superalgebra follows from the anti-diagonal structure of the supercharge $Q$ and the fermionic properties $\smbr{\sigma^\pm}^2=0$. We can construct a certain \emph{enlarged} supercharge $\widetilde{Q}$ that also comes with the same anti-diagonal structure as $Q$, and then the \emph{enlarged} super-Hamiltonian $\widetilde{H}^{ss}$ through $\widetilde{H}^{ss}=\widetilde{Q}^2$, a relation that is similar to  $H^{ss}_J=Q^2$. Since these key properties remain, we expect that the resultant operator set $\left\{\widetilde{H}^{ss},\;\widetilde{Q}^\pm\right\}$ also have the same $su(1,1)$ superalgebra structure.

The expected enlarged supercharge $\widetilde{Q}$ is obtained simply by performing the iteration $J^\pm\rightarrow Q^\pm$ and $Q\rightarrow \widetilde{Q}$  in the definition \eqnm{antidiagonal_matrix} for the supercharge $Q$, i.e.,
\begin{eqnarray}\label{wdtldQ}
 \widetilde{Q} \equiv \left(\begin{array}{cc}
                       0 & Q^+ \\
                      Q^- & 0
                   \end{array}\right)=\widetilde{Q}^+ + \widetilde{Q}^-,
\end{eqnarray}
where the two \emph{enlarged} supercharges $\widetilde{Q}^\pm$ are defined as
\begin{eqnarray}\label{supercharge_tildeQ_pm}
\begin{aligned}
 \widetilde{Q}^+\equiv \left(\begin{array}{cc}
                           0 & Q^+ \\
                           0 & 0
                        \end{array}\right) = \frac{1}{2}\sigma^+ \otimes Q^+, \\
\widetilde{Q}^- \equiv \left(\begin{array}{cc}
                           0 & 0\\
                           Q^-  & 0
                        \end{array}\right) =\frac{1}{2}\sigma^- \otimes Q^-.
\end{aligned}
\end{eqnarray}
That is, $\widetilde{Q}^\pm$ are defined as the \emph{second-order} (higher-order) perpendicular spin matrix maps of $J^\pm$ through
\begin{eqnarray}\label{supercharge_tildeQ_pm_second_order_spin_matrix_maps}
 \frac{1}{2}\sigma^\pm \otimes \frac{1}{2}\sigma^\pm: J^\pm \longmapsto \widetilde{Q}^\pm\equiv \frac{1}{2}\sigma^\pm \otimes \frac{1}{2} \sigma^\pm\cdot J^\pm.
\end{eqnarray}
These $\widetilde{Q}^\pm$ are the analogs of $Q^\pm$ in \eq{Sigmapm_maps}, where the latter are defined through the first-order perpendicular spin matrix maps of $J^\pm$. Indeed, these $\widetilde{Q}^\pm$ can also be obtained by performing the iterations $J^\pm\rightarrow Q^\pm$ and $Q^\pm \rightarrow \widetilde{Q}^\pm$ directly in \eq{supercharge_Q_pm}.

The nilpotent properties
\begin{eqnarray}\label{tildeQ_square}
 \left\{\widetilde{Q}^\pm,\; \widetilde{Q}^\pm\right\}=0
\end{eqnarray}
are then guaranteed by the properties $\smbr{\sigma^\pm}^2=0$.

The \emph{enlarged} super-Hamiltonian $\widetilde{H}^{ss}$ associated with $\widetilde{Q}$ can be defined as
\begin{eqnarray}\label{super_Hamiltonian_wtHs}
 \widetilde{H}^{ss} \equiv \widetilde{Q}^2 =\left\{\widetilde{Q}^+,\; \widetilde{Q}^-\right\}=\left(\begin{array}{cc}
                           Q^+ Q^- & 0 \\
                           0 & Q^- Q^+
                        \end{array}\right).
\end{eqnarray}
This $\widetilde{H}^{ss}$ is the parallel of $H^{ss}_J$ in \eq{superHamiltonian_HsJ}. Indeed, such $\widetilde{H}^{ss}$ can be obtained by performing the iterations $J^\pm\rightarrow Q^\pm$ and $H^{ss}_J \rightarrow \widetilde{H}^{ss}$ directly in \eq{superHamiltonian_HsJ}.

The conservation properties
\begin{eqnarray}\label{wtQpm_conservation}
 \left[\widetilde{Q}^\pm,\;\widetilde{H}^{ss}\right]=0
\end{eqnarray}
just follow from the nilpotent properties \eqnm{tildeQ_square}.

Eqs. \eqnm{tildeQ_square}, \eqnm{super_Hamiltonian_wtHs} and \eqnm{wtQpm_conservation} together manifest that the enlarged operator set $\left\{\widetilde{H}^{ss},\;\widetilde{Q}^\pm\right \}$ \emph{do} form an ${\cal N}=2$ $su(1,1)$ superalgebra, as the operator set $\left\{H^{ss}_J,\;Q^\pm\right \}$ do.

The above iteration can keep going \emph{ad infinitum}, and the resultant matrix operators $\left\{ {\widetilde H}^{ss}_n, \; \widetilde{Q}^\pm_n \right \}$ with dimension $2^n$ satisfy the common ${\cal N}=2$ $su(1,1)$ superalgebra for arbitrary positive integer $n$, where $\widetilde{Q}^\pm_n$ are the $n$-th order perpendicular spin matrix maps of $J^\pm$.

With $\widetilde{Q}$ introduced in \eq{wdtldQ}, $Q_m$ is defined as the average of $\widetilde{Q}$ over the superstate $(\ket{m},\ket{m-1})^T$,
\begin{eqnarray}\label{Qm_alternative}
 Q_m &\equiv& \smbr{\bra{m},\;\bra{m-1}}\widetilde{Q}\left(\begin{array}{c}
                            \ket{m} \\
                           \ket{m-1}
                        \end{array}\right).
\end{eqnarray}
$Q_m$ defined this way is the same that given in \eq{Qbarm}. It can also be viewed as the result of merging the two matrix elements between states $\ket{m}$ and $\ket{m-1}$ in  Eqs. \eqnm{spin_matrix_map_plus} and \eqnm{spin_matrix_map_minus} into a single matrix average over the superstate $(\ket{m},\ket{m-1})^T$.

$Q_m$ defined in \eq{Qm_alternative} $Q^+_m$ naturally splits into $Q_m =Q^+_m+Q^-_m$, with
\begin{eqnarray}\label{Qpmm}
\begin{aligned}
 Q^+_m &\equiv \smbr{\bra{m},\;\bra{m-1}}\widetilde{Q}^+\left(\begin{array}{c}
                            \ket{m} \\
                           \ket{m-1}
                        \end{array}\right),\\
 Q^-_m &\equiv \smbr{\bra{m},\;\bra{m-1}}\widetilde{Q}^-\left(\begin{array}{c}
                            \ket{m} \\
                           \ket{m-1}
                        \end{array}\right),
\end{aligned}
\end{eqnarray}
where the upper equation is just \eq{spin_matrix_map_plus} and the lower one \eq{spin_matrix_map_minus}, once the matrix elements $\bra{m}Q^+\ket{m-1}$ and $\bra{m-1}Q^-\ket{m}$ there are rewritten into the averages of $\widetilde{Q}^\pm$ over the superstate $\smbr{\ket{m},\,\ket{m-1}}^T$.

With Eqs. \eqnm{supercharge_tildeQ_pm}, \eqnm{raising_operator} and \eqnm{lowering_operator},  the two equations in \eq{Qpmm} can be reduced respectively to
\begin{eqnarray}
       Q^+_m&=& \bra{m} Q^+\ket{m-1},\nonumber \\
       &=&\frac{1}{2}\sigma^+\bra{m}J^+\ket{m-1},\nonumber \\
       &=&\frac{1}{2}\sigma^+\cdot\mfk{J}^+_m(\theta),\label{Qplusm}
\end{eqnarray}
and to
\begin{eqnarray}
  Q^-_m   &=&\bra{m-1} Q^-\ket{m},\nonumber \\
       &=&\frac{1}{2}\sigma^-\bra{m-1}J^-\ket{m},\nonumber \\
       &=&\frac{1}{2}\sigma^-\cdot\mfk{J}^-_m(\theta).\label{Qminusm}
\end{eqnarray}

Recall that, $\bra{m}J^+\ket{m-1}$ and $\bra{m-1}J^-\ket{m}$ are the only non-vanishing matrix elements of $J^\pm$ between any two azimuthal harmonics,  as shown in \eq{J_plus_minus_matrices}. So Eqs. \eqnm{Qplusm} and \eqnm{Qminusm} show that $\smbr{\ket{m},\,\ket{m-1}}^T$ is the only superstate over which the average of $\widetilde{Q}^\pm$ are non-vanishing, and these averages are defined as $Q^\pm_m$.

The trivial splitting
\begin{eqnarray}
 &&\sum_{m',\,m''\in \mathbb{Z}}\left(\begin{array}{c}
                            \ket{m'} \\
                           \ket{m''}
                        \end{array}\right)\otimes\smbr{\bra{m'},\;\bra{m''}},\nonumber\\
                        &=&\left(\begin{array}{cc}
                            \sum_{m'\in \mathbb{Z}}\ket{m'}\bra{m'} &0\\
                           0&\sum_{m''\in \mathbb{Z}}\ket{m''}\bra{m''}
                        \end{array}\right)\\
                        &&+\frac{1}{2}\sigma^+\sum_{m',\,m''\in \mathbb{Z}}\ket{m'}\bra{m''} + \frac{1}{2}\sigma^-\sum_{m',\,m''\in \mathbb{Z}}\ket{m''}\bra{m'}\nonumber
\end{eqnarray}
can be rewritten as the completeness relation \text{--} a resolution of the $2\times 2$ unit matrix $I_2$ \text{--} in the form
\begin{eqnarray}\label{I2_resolution}
 I_2 &=&\left(\begin{array}{cc}
                            \sum_{m'\in \mathbb{Z}}\ket{m'}\bra{m'} &0\\
                           0&\sum_{m''\in \mathbb{Z}}\ket{m''}\bra{m''}
                        \end{array}\right),\nonumber \\
 &=&\sum_{m',\,m''\in \mathbb{Z}}\left(\begin{array}{c}
                            \ket{m'} \\
                           \ket{m''}
                        \end{array}\right)\otimes\smbr{\bra{m'},\;\bra{m''}},\\
    &&-\frac{1}{2}\sigma^+\sum_{m',\,m''\in \mathbb{Z}}\ket{m'}\bra{m''} - \frac{1}{2}\sigma^-\sum_{m',\, m''\in \mathbb{Z}}\ket{m''}\bra{m'}. \nonumber
\end{eqnarray}

Now we compute the average of the anti-commutator $\left\{\widetilde{Q}^+,\; \widetilde{Q}^-\right\}$ over the superstate $\smbr{\ket{m},\;\ket{m-1}}^T$. To calculate the average of $\widetilde{Q}^+ \widetilde{Q}^-$ or $\widetilde{Q}^- \widetilde{Q}^+$, we insert the completeness relation \eqnm{I2_resolution} between $\widetilde{Q}^\pm$ and $\widetilde{Q}^\mp$. Note that, as shown in \eq{supercharge_tildeQ_pm_second_order_spin_matrix_maps}, $\widetilde{Q}^\pm$ respectively carries two folds of matrix $\sigma^\pm$ as their factors,  because of the nilpotent properties $\smbr{\sigma^\pm}^2=0$, the second and third terms in the completeness relation \eqnm{I2_resolution} do not contribute when they encounter $\widetilde{Q}^\pm$. Hence, we have
\begin{eqnarray}
&& \smbr{\bra{m},\;\bra{m-1}}\widetilde{Q}^+\widetilde{Q}^- \left(\begin{array}{c}
                            \ket{m} \\
                           \ket{m-1}
                        \end{array}\right),\nonumber \\
&=&\sum_{m',\,m''\in \mathbb{Z}}\smbr{\bra{m},\;\bra{m-1}}\widetilde{Q}^+\left(\begin{array}{c}
                            \ket{m'} \\
                           \ket{m''}
                        \end{array}\right)\nonumber \\
                        &&\cdot \smbr{\bra{m'},\;\bra{m''}}     \widetilde{Q}^- \left(\begin{array}{c}
                            \ket{m} \\
                           \ket{m-1}
                        \end{array}\right), \nonumber \\
&=& \smbr{\bra{m},\;\bra{m-1}}\widetilde{Q}^+\left(\begin{array}{c}
                            \ket{m} \\
                           \ket{m-1}
                        \end{array}\right) \nonumber \\
                        &&\cdot \smbr{\bra{m},\;\bra{m-1}}\widetilde{Q}^- \left(\begin{array}{c}
                            \ket{m} \\
                           \ket{m-1}
                        \end{array}\right), \nonumber \\
&=& Q^+_m Q^-_m, \label{tildeQplustildeQ_average}
\end{eqnarray}
where the sum over $m'$ and $m''$ collapses due to the fact that the only non-vanishing matrix elements of $\widetilde{Q}^\pm$ are $Q^\pm_m$ defined in Eqs. \eqnm{Qplusm} and \eqnm{Qminusm}. Similarly,
\begin{eqnarray}\label{tildeQtildeQplus_average}
 \smbr{\bra{m},\;\bra{m-1}}\widetilde{Q}^-\widetilde{Q}^+ \left(\begin{array}{c}
                            \ket{m} \\
                           \ket{m-1}
                        \end{array}\right) = Q^-_m Q^+_m.
\end{eqnarray}

\eq{tildeQplustildeQ_average} and \eq{tildeQtildeQplus_average} together give us
\begin{eqnarray}\label{wdtldQpm_average}
 \smbr{\bra{m},\;\bra{m-1}}\left\{\widetilde{Q}^+,\;\widetilde{Q}^-\right\} \left(\begin{array}{c}
                            \ket{m} \\
                           \ket{m-1}
                           \end{array}\right) =\left\{ Q^+_m,\; Q^-_m\right\}.
\end{eqnarray}

Eqs. \eqnm{Hs_m_anticommutator} and \eqnm{wdtldQpm_average} together remind us to define $H^{ss}_m$ consistently as
\begin{eqnarray}\label{tildeHs_average}
 H^{ss}_m &\equiv& \smbr{\bra{m},\;\bra{m-1}}\widetilde{H}^{ss}\left(\begin{array}{c}
                            \ket{m} \\
                           \ket{m-1}
                        \end{array}\right).
\end{eqnarray}

With Eqs.\eqnm{tildeHs_average} and \eqnm{wdtldQpm_average}, the average of \eq{super_Hamiltonian_wtHs} over the superstate $\smbr{\ket{m},\,\ket{m-1}}^T$ gives rise to
\begin{eqnarray}\label{Hssm_decomposition}
 H^{ss}_m=\left\{Q^+_m,\;Q^-_m\right\},
\end{eqnarray}
which is just \eq{Hs_m_anticommutator}.

In a similar fashion, the average of \eq{wtQpm_conservation} over the superstate $\smbr{\ket{m},\;\ket{m-1}}^T$ gives rise to
\begin{eqnarray}
 \left[Q^\pm_m,\;H^{ss}_m\right]=0,
\end{eqnarray}
and the average of \eq{tildeQ_square} to
\begin{eqnarray}\label{Qpm_nilpotent2}
 \left\{Q^\pm_m,\;Q^\pm_m\right\}=0.
\end{eqnarray}

That is, the average of the \emph{2-parameter} ${\cal N}=2$ $su(1,1)$ superalgebra satisfied by the \emph{enlarged} operator set $\left\{\widetilde{H}^{ss},\;\widetilde{Q}^\pm \right\}$ over the super-state $\smbr{\ket{m},\,\ket{m-1}}^T$ gives rise to the same yet \emph{1-parameter} superalgegra satisfied by the operator set $\left\{H^{ss}_m,\;Q^\pm_m \right\}$.

\eq{Hssm_decomposition} can also be derived in the following component-wise way. Inserting the last expression for $\widetilde{H}^{ss}$ in \eq{super_Hamiltonian_wtHs}, \eq{tildeHs_average} then becomes
\begin{eqnarray}\label{tildeHs_average1}
 H^{ss}_m  = \bra{m} Q^+Q^-\ket{m}+ \bra{m-1} Q^+Q^-\ket{m-1}.
\end{eqnarray}

Notice that $\bra{m} Q^+\ket{m'}=\delta_{m',m-1}\bra{m} Q^+\ket{m-1}$ and $\bra{m'} Q^-\ket{m}=\delta_{m',m-1}\bra{m-1} Q^-\ket{m}$. By inserting the completeness relation $\sum_{m'}\ket{m'}\bra{m'}=1$, the first term on the right hand side of \eq{tildeHs_average1} becomes
\begin{eqnarray}
 \bra{m} Q^+Q^-\ket{m}&=&\bra{m} Q^+\sum_{m'}\ket{m'}\bra{m'}Q^-\ket{m}, \nonumber \\
                      &=&\bra{m} Q^+\ket{m-1}\bra{m-1} Q^-\ket{m},\nonumber \\
                      &=&Q^+_m Q^-_m,
\end{eqnarray}
and the second term becomes
\begin{eqnarray}
 \bra{m-1} Q^-Q^+\ket{m-1}&=&\bra{m-1} Q^-\sum_{m'}\ket{m'}\bra{m'}Q^+\ket{m-1}, \nonumber \\
                      &=&\bra{m-1} Q^+\ket{m}\bra{m} Q^-\ket{m-1},\nonumber \\
                      &=&Q^-_m Q^+_m,
\end{eqnarray}
where $Q^\pm_m$ are defined in Eqs. \eqnm{Qplusm} and \eqnm{Qminusm}, respectively. Therefore, \eq{tildeHs_average1} reduces to \eq{Hssm_decomposition}.

In order to look for the representation of the reduced 1-parameter ${\cal N}=2$ $su(1,1)$superalgegra satisfied by the operator set $\left\{H^{ss}_m,\;Q^\pm_m \right\}$, we have to impose the super eigenequation \eq{matrix_eigen_equ_1D} for $H^{ss}_m$, which is equivalent to the two component eigenequations \eqnm{Hermitian_factorization_theta_coord} and \eqnm{superpartner_eigenequ_theta_coord}.

The super eigenequation \eqnm{matrix_eigen_equ_1D} is consistent with the following raising/lowering recurrence relations
\begin{eqnarray}\label{Qpm_recurrence_relations}
 \begin{aligned}
  Q^+_m \left(\begin{array}{c}
                            0 \\
                           \Theta_{jm-1}
                        \end{array}\right)=c^-_{jm}\left(\begin{array}{c}
                           \Theta_{jm} \\
                            0
                        \end{array}\right), \\
Q^-_m \left(\begin{array}{c}
                           \Theta_{jm}\\
                            0
                        \end{array}\right)=c^-_{jm}\left(\begin{array}{c}
                             0\\
                           \Theta_{jm-1}
                        \end{array}\right).
 \end{aligned}
\end{eqnarray}
In fact, by acting the upper equation with $Q^-_m$ and the lower one with $Q^+_m$ and make use of the properties in \eq{Qpm_nilpotent2}, the super eigenequation \eqnm{matrix_eigen_equ_1D} is recovered.

The first recurrence relation in \eq{Qpm_recurrence_relations} is equivalent to \eq{J_tilde_raising} and the second one in \eq{Qpm_recurrence_relations} to \eq{J_tilde_lowering}.

The rest procedure of constructing the SUSYQM algorithm for solving these eigenequations is the same as that demonstrated in Subsubsection \ref{eigen_solutions_associated_equ_theta_coord}.

\section{Conclusions and Remarks}\label{conclusion}

In this paper, we have built quite comprehensive SUSYQM algorithms for resolving the eigenspectra of the base as well as the associated Legendre differential operators. The two symmetries \text{--} supersymmetry and shape invariance symmetry \text{--} hidden in these operators are brought to the fore and promoted to the two governing principles that guide us to build these algorithms. Key steps in this type of algorithms, that also apply to many other types of linear differential operators of second order, are summarized here. 1) Based on the eigen differential equation, prepare the negative of a square of a certain linear differential operator of first order \text{--} a symmetric double layers \text{--} to be identified with the quantum kinetic energy operator in the original coordinates. In many cases, this can be achieved by a simple multiplier and an additive constant to the eigen differential operator. 2) If the eigen differential operator still does not come out as a pure negative square form plus plus extra non-differential functions, but as a certain form of product of two \emph{asymmetric } first-order differential operators, that already contains a negative square form as its one term, plus extra non-differential functions, we then search for the supersymmetric factorization of that asymmetric factorization (product), by making use of the active supersymmetrization transformation, that is made up of the integration factors coming out of the two factors in the asymmetric factorization, and the accompanying momentum operator map that maps the base of the square form \text{--} a linear differential operator of first order \text{--} to the standard quantum momentum operator in new coordinate. 4) Absorb the non-differential functions to the previous partial supersymmetric factorization and modify the initial superpotential into the overall superpotential. Up to this stage, a standard Schr\"odinger eigenequation in its supersymmetric factorization is achieved, in which the Hamiltonian is expressed as the product of a Hermite conjugate pair. The latter will work as the ladder operators for constructing the eigenfunction of generic level in bottom-up or top-down order. 5) Obtain the superparnter eigenequation from this supersymmetrically factorized raw eigenequation, simply by performing the \emph{intertwining (leftward braiding)} action. 6) Design an upward/downward connected symmetry relation, called shape/pattern invariance relation,  between the superpartner Hamiltonian and the raw  Hamiltonian of the corresponding \emph{nearest neighboring} level. 7) Make use of this shape/pattern invariance relation to \emph{renormalize} the superpartner eigenequation into the raw eigenequation yet of that nearest neighboring level. At this point, the raising and lowering roles played by the two ladder operators become clear, and the supersymmetrically factorized eigenequation is equivalently split into a pair of two-term raising/lowering recurrence relations. Meanwhile, simple algebraic recurrence relations for the parameters of such supersymmetric factorization and for the eigenvalues are also constructed from the requirement of the supersymmetric factorization itself and the shape/pattern invariance relation. 8) An annihilation condition is obtained as a simple differential equation of first order, simply by setting the eigenvalue to be vanishing in the supersymmetrically factorized eigenequation of the lowest or top level. The initial eigenfunction is then obtained by integrating such annihilation condition. 9) With the vanishing initial eigenvalue, all factorization parameters are then fixed. Iteratively build the eigenfunction of generic level in a bottom-up/top-down approach, by applying the raising/lowering recurrence relation consecutively. If the procedure goes in a bottom-up way and the supersymmetrically factorized the eigen differential operator comes out in such a way that the lowering operator is situated on the right side of the raising operator, the renormalization of this supersymmetric factorization has to be performed first to move the raising operator to the right of the lowering one, and the intertwining action can be performed next. Since for each supersymmetrically factorizable eigen differential operator, the active supersymmetrization transformation plus the momentum operator map establish  1-to-1 correspondences between the raw coordinate representation of eigen operator, ladder operators, ladder relations, shape/pattern invariance relation, annihilation condition, etc.,  and the standard coordinate representation of these entities (in which the Hamiltonian takes the form of \emph{standard} supersymmetric factorizattion), the SUSYQM algorithm can also be constructed/represented \emph{directly in the raw coordinate} with almost the same amount of work, without passing to the standard coordinate representation.

Traditionally, on constructing the ladder operators and reaching a supersymemtric factorization of a linear eigen differential operator of second order, two popular approaches are followed. The first approach is to combine some three-term recurrence relations to derive the two-term ladder relations. These recurrence relations either come out of the already-known analytical eigensolutions or of the generating functions of these solutions as certain special functions. The second approach consists of \emph{two} distinct stages: 1) passing from a linear second-order eigen differential equation to the corresponding Schr\"odinger equation, through the analytical method that aims at eliminating the first-order derivative in the eigen-equation via deriving and solving another second-order differential equation for a certain coordinate transformation; 2) \emph{passively} looking for the supersymmetric factorization of the resultant Schr\"odinger equation. In contrast, our SUSYQM algorithms developed in this paper are completely detached from these popular traditional approaches and even do not receive any inputs from these methods. Based on the basic construct of kinetic energy operator in QM and SUSYQM themselves, from the eigen differential operator in question, a negative of the square of a certain differential operator of first order is firstly prepared as the quantum kinetic energy operator, by multiplying the eigenequation with some simple function for symmetrizing the asymmetric product of two first-order differential operators inside the eigen differential operator. This is then followed by two simple algebraic operations \text{--} \emph{momentum operator map} and \emph{active supersymmetrization transformation}. In so doing one directly passes from the \emph{asymmetric} factorization of an eigen differential operator in various coordinates to its (initial) \emph{supersymmetric} factorization \text{--} the product of a pair of ladder operators \text{--} \emph{in one run} in a purely algebraic way. Such algebraic method is full of SUSYQM flavor.

Take the base hypergeometric-like operators $h(x)$ as a concrete example for demonstrating how to prepare the negative square of a certain first-order differential operator as the quantum kinetic energy operator and pin down the momentum operator map, reach the initial asymmetric factorization, construct the active supersymmetrization transformation to supersymmetrize the asymmetric factorization (see the details in Subsection \ref{AST_hypergeometric_operator}). From the scratch, we start with the self-adjoint form \eqnm{hypergeometric-like_operator} of $h(x)$, an asymmetrically factorized product of two distinct first-order linear differential operators, and \emph{partially} symmetrize such asymmetric product into the form \eqnm{asymmetric_form_ph} by multiplying $p(x)$, so that it contains $-\smbr{p(x)d/dx}^2$  \text{--} the negative of a square of the operator $p(x)d/dx$, and such negative square form is identified with a standard quantum kinetic energy operator $-d^2/dy^2$ in $y$ coordinate, once we introduce the corresponding momentum operator map $p(x)d/dx =d/dy$. The next step is to evenly partition the non-differential term $2W_0(x)$ inside the first factor of the asymmetric product over the two factors, by separating a common copy of $W_0(x)$ as the superpotential and lumping the residual in each factor into a certain integration factor and its inverse. Such integration factors then make up the active supersymmetrization transformation ${\cal S}_b(-)$, that works to supersymmtrize the asymmetric factorization of $p(x)h(x)$. The Legendre operator $L(x)$ is a special hypergeometric-like operator $h(x)$ in that $W_0(x)=0$ and ${\cal S}_b(-)=1\cdot(-)\cdot 1$.

Highly similar procedures for conducting the initial supersymmetric factorization of the associated Legendre differential operators in $x$ and $\theta$ coordinates have been demonstrated in Subsections \ref{second_momentum_map} and \ref{Legendre_operator_theta_coordinate}, respectively. The active supersymmetrization transformations for the base as well as the associated Legendre operators and the accompanying momentum operator maps are summarized in Tables \ref{tab:table1}, \ref{tab:table2} and \ref{tab:table3}. As shown in the first row in Table \ref{tab:table1} for the multiplied Legendre operator $p(x) L(x)$, the first row in Table \ref{tab:table2} for the multiplied Legendre operator $\sin^2\theta J_0(\theta)$ and the third row in Table \ref{tab:table3} for the multiplied associated Legendre operator $\sin^2\theta \widehat{J}_j(\theta)$, these three operator, with the help of the multiplier $p(x)=\sin^2\theta$, come naturally in their supersymmetrically factorized forms, so these are supersymmetrized by the same trivial active supersymmetrization transformation $1\cdot \smbr{-}\cdot 1$. We also note that, the same non-trivial active supersymmetrization transformation ${\cal S}_x(-)={\cal S}_\theta(-)$ that supersymmetrizes the base Legendre operators $L(x)=J_0(\theta)$ (see the second row in Tables \ref{tab:table1} and \ref{tab:table2}) also supersymmetrizes the associated Legendre operators $L_m(x)$ and $J_m(\theta)$ (see the first and second rows in Table \ref{tab:table3}). The reason for this lies in that, $L_m(x)$ is just $L(x)$ plus a certain function of $x$ and $J_m(\theta)$ is just $J_0(\theta)$ plus another function of $\theta$, and the actions of the active supersymmetrization transformations ${\cal S}_x(-)={\cal S}_\theta(-)$ keep these ordinary functions intact, and these two functions, because of their particular forms, when absorbed into the supersymmetric factorizations of $L(x)$ and $J_0(\theta)$, happen to be capable of completing the two supersymmetric factorizations of $L_m(x)$ and $J_m(\theta)$.

Furthermore, in building all these SUSYQM algorithms, the eigenvalues of the base and associated Legendre operators are not needed to be input from any traditional method. Instead, they also come out of these SUSYQM algorithms themselves in purely algebraic ways.

The algebraic method of reaching the supersymmetric factorization of a linear differential operator of second order proposed in this paper and also in \cite{Zhou_SUSY_hypergeometric_operator_2024} is not only so elementary that it is accessible even to the one who has only preliminary  calculus background, but is also much more efficient than the traditional analytical methods.
The methods of constructing the SUSYQM algorithms that are summarized in the first paragraph of this section are fully general and are applicable to building many other classical special functions, especially those that are related to hypergeometric-type functions. Because of the complete independency of traditional methods, the aesthetic appeal of the two guiding symmetry principles that are brought to the fore, the simplicity of the algebraic operations involved, and the wide applicability as well as the high efficiency, these algorithms are promising candidates for supplanting the traditional analytical power series method of building special functions that still predominates the present text books and monographs.

\begin{table*}[t]
\caption{The two active supersymmization transformations on Legendre operator parametrized in $\bf x$ coordinate and the accompanying momentum operator maps.
}\label{tab:table1}
\scalebox{0.8}{
\begin{tblr}{width=1.25\linewidth, colspec={X[l,2] X[l,2.6] X[l,2.0] X[l,1.5] X[l,1.5] X[l,1.5]}}
\hline \hline
{Initial (a)symmetric \\ factorization} & {Initial supersymmetric \\ factorization} & {Active supersymmetrization \\ transformation} & Superpotential & {Momentum \\ operator map} & {Kinetic energy \\ operator map} \\
$pL(x)=-\left(p\od{}{x}\right)^2$ & $pL(x)=-\left(p\od{}{x}\right)^2$ & $1\cdot(-)\cdot1$ & $W_0(x)=0$ & $-ip\od{}{x}=-i\od{}{y}$ & $-\smbr{p\od{}{x}}^2=-\frac{d^2}{d y^2}$ \\
{$L(x)=$ \\ $\sqbr{-\sqrt{p}\od{}{x}+2 W^a_0(x)}\sqbr{\sqrt{p}\od{}{x}}$} & {${\cal S}_x\left(L(x)\right)=$ \\ $\left[-\sqrt{p}\od{}{x}+ W^a_0(x)\right]\left[\sqrt{p}\od{}{x}+ W^a_0(x)\right]$} & ${\cal S}_x(-)=p^{1/4}(-)p^{-1/4}$ & $W^a_0(x)=-\frac{\sqrt{p}^{(1)}}{2}$ & $-i\sqrt{p}\od{}{x}=-i\od{}{\theta}$ & $-\smbr{\sqrt{p}\od{}{x}}^2=-\frac{d^2}{d\theta^2}$ \\
\hline \hline
\end{tblr}
}
\end{table*}

\begin{table*}[t]
\caption{The two active supersymmization transformations on the Legendre operator parametrized in $\bf \theta$ coordinate and the accompanying momentum operator maps.
}\label{tab:table2}
\scalebox{0.8}{
\begin{tblr}{width=1.25\linewidth, colspec={X[l,2] X[l,2.5] X[l,2.4] X[l,1.5] X[l,1.5] X[l,1.7]}}
\hline \hline
{Initial (a)symmetric \\ factorization} & {Initial supersymmetric \\ factorization} & {Active supersymmetrization \\ transformation} & Superpotential & {Momentum \\ operator map} & {Kinetic energy \\ operator map} \\
$\sin^2\theta J_0(\theta)=-\left(\sin\theta \od{}{\theta}\right)^2$ & $\sin^2\theta J_0(\theta)=-\left(\sin\theta \od{}{\theta}\right)^2$ & $1\cdot(-)\cdot1$ & ${\cal W}_0(\theta)=0$ & $-i\sin\theta\od{}{\theta}=-i\od{}{y}$ & $-\smbr{\sin\theta\od{}{\theta}}^2=-\frac{d^2}{d y^2}$ \\
{$J_0(\theta)=-\frac{1}{\sin\theta}\od{}{\theta}\sin\theta\od{}{\theta}$ \\ $=\sqbr{-\od{}{\theta}+2 {\cal W}^a_0(\theta)}\sqbr{\od{}{\theta}}$} & {${\cal S}_\theta\left(J_0(\theta)\right)=$ \\ $\left[- \od{}{\theta}+ {\cal W}^a_0(\theta)\right]\left[\od{}{\theta}+ {\cal W}^a_0(\theta)\right]$} & ${\cal S}_\theta(-)=\sin^{1/2}\theta(-)\sin^{-1/2}\theta$ & ${\cal W}^a_0(\theta)=\frac{1}{2}\cot\theta$ & $-i\od{}{\theta} = -i\od{}{\theta}$ & $-\frac{d^2}{d\theta^2}=-\frac{d^2}{d\theta^2}$ \\
\hline \hline
\end{tblr}
}
\end{table*}
\begin{table*}[t]
\caption{The three active supersymmization transformations on the \emph{associated} Legendre operators parametrized in $\bf x$ as well as in $\bf \theta$ coordinates.
}\label{tab:table3}
\scalebox{0.8}{
\begin{tblr}{width=1.25\linewidth, colspec={X[l,2.3] X[l,2.2] X[l,2.4] X[l,1.8] X[l,1.5] X[l,1.7]}}
\hline \hline
{(A)symmetric \\ factorization} & {Supersymmetric \\ factorization} & {Active supersymmetrization \\ transformation} & Superpotential & {Momentum \\ operator map} & {Kinetic energy \\ operator map} \\
{$L_m(x)$\\$=\sqbr{\sqrt{p}\od{}{x}+(m+1)\sqrt{p}^{(1)}}$  \\
$\cdot \sqbr{-\sqrt{p}\od{}{x} +m\sqrt{p}^{(1)}}$} & {${\cal S}_x(L_m(x))$\\$=\sqbr{\sqrt{p}\od{}{x}+W^a_m(x)}$ \\ $\;\;\; \cdot\sqbr{-\sqrt{p}\od{}{x}+W^a_m(x)}$}& ${\cal S}_x(-)=p^{1/4}(-)p^{-1/4}$ & $W^a_m(x)=(m-1/2)\sqrt{p}^{(1)}$ & $-i\sqrt{p}\od{}{x}=-i\od{}{\theta}$ & $-\smbr{\sqrt{p}\od{}{x}}^2=-\frac{d^2}{d\theta^2}$ \\
{$J_m(\theta)$ \\ $=\sqbr{\od{}{\theta}+(1-m)\cot\theta}$ \\ $\quad\cdot\sqbr{-\od{}{\theta}-m\cot\theta}$} & {${\cal S}_\theta\left(J_m(\theta)\right)$ \\ $=\left[- \od{}{\theta}+ {\cal W}^a_m(\theta)\right]$ \\$\;\;\cdot \left[\od{}{\theta}+ {\cal W}^a_m(\theta)\right]$} & ${\cal S}_\theta(-)=\sin^{1/2}\theta(-)\sin^{-1/2}\theta$ & ${\cal W}^a_m(\theta)=\smbr{1/2-m}\cot\theta$ & $-i\od{}{\theta}\equiv -i\od{}{\theta}$ & $-\frac{d^2}{d\theta^2}=-\frac{d^2}{d\theta^2}$ \\
{$\sin^2\theta \widehat{J}_j(\theta)$ \\ $=\sqbr{-\sin\theta \od{}{\theta}-j\cos\theta}$\\ $\;\;\;\cdot\sqbr{\sin\theta \od{}{\theta}-j\cos\theta}$} & {$\widehat{H}_j(y)=\sin^2\theta \widehat{J}_j(\theta)$ \\ $=\left[- \od{}{y}+ {\cal W}^a_j(y)\right]$ \\$\;\;\,\cdot \left[\od{}{y}+ {\cal W}^a_j(y)\right]$} & $1\cdot(-)\cdot1$ & {${\cal W}^a_j(y)=j\tanh y$\\$\qquad\;\; =-j\cos\theta$} & $-i\sin\theta\od{}{\theta}\equiv -i\od{}{y}$ & $-\smbr{\sin\theta\od{}{\theta}}^2=-\frac{d^2}{dy^2}$ \\
\hline \hline
\end{tblr}
}
\end{table*}

We have also provided a SUSYQM answer to the innocent question on why there exist the base as well as the associated functions for the \emph{same} Legendre differential operator.
Side-by-side comparing the bottom-up iterative procedure of simultaneously building the base Legendre differential equation and its eigenfunctions in Subsection \ref{inductive_iterative_base_Legendre_function} with the highly similar procedure of simultaneously building the associated Legendre differential equation and its eigenfunctions in Subsection \ref{Positive_Association_Levels}, the very SUSYQM reason for the co-existence of two types of special functions that are associated with the same Legendre operator becomes perspicuous. First of all, from the same Legendre operator $L(x)$, two different types of kinetic energy operators $-\smbr{p(x)d/dx}^2$ and $-\smbr{\sqrt{p(x)}d/dx}^2$ can be constructed, hence there exist two different types of momentum operator maps $-p(x)d/dx= d/dy$ and $\sqrt{p(x)}d/dx=d/d\theta$, that are listed in the fifth column of Table \ref{tab:table1}. Secondly, corresponding to these two types of momentum operator maps, the \emph{same} $L(x)$ admits two distinct initial supersymmetric factorizations, that are listed in the second column of Table \ref{tab:table1} (parametrized in $x$ coordinate) and in the second column of Table \ref{tab:table2} (parametrized in $\theta$ coordinate), hence there exist two distinct initial supersymmetrically factorized eigen differential equations \eqnm{0th_eigenequ} and \eqnm{Legendre_equ_right_dagger} associated with the same $L(x)$. Thirdly,  each of these two initial supersymmetric factorizations proliferates into a hierarchy of higher-level supersymmetric factorizations in a self-similar (pattern invariant) way, by repeating the same two algebraic operations \text{--} intertwining action plus renormalization, hence two distinct hierarchies of eigenequations and eigensolutions are resulted in. In fact, just as demonstrated by the side-by-side comparison of the two flow charts of such proliferation shown in Diagram 1 in \cite{Zhou_SUSY_hypergeometric_operator_2024}, this is a general result for all hypergeometric-like differential operators.

In Section \ref{direct_factorization_of_associated_Legendre}, the SUSYQM algorithms are constructed for resolving the eigen-spectrum of the associated Legendre differential operator, starting with directly searching for the supersymmetric factorizations of  associated Legendre equation, that is assumed to be known to us beforehand, as it comes out as the $\theta$-component eigenequation after separating variables of a partial differential eigenequation involving Laplacian. Once again, because two distinct types of negative square forms $-\smbr{\sin\theta d/d\theta}^2$ and $-\smbr{d/d\theta}^2$, as listed in the last column of Table \ref{tab:table3}, can be constructed from the same associated Legendre differential operator and identified as two distinct kinetic energy operators, two different SUSYQM algorithms have been constructed for the same eigenspectrum issue. A pair of ladder operators in one SUSYQM algorithm depend on the quantum number $j$ only through the $j$-dependent superpotential, and these operators change $j$ by one unit each step with fixed $m$ quantum number when acting on the eigenfunctions, while a pair of ladder operators in the other SUSYQM algorithm depend on the quantum number $m$ only through the $m$-dependent superpotential, and these operators change $m$ by one unit each step with $j$ fixed when acting on the eigenfunctions. Become of such distinction of the roles played by the ladder operators, the iterative procedures of building the eigenfunctions of generic levels are different in these two SUSYQM algorithms.

In contrast to starting with the broken form \text{--}  also ``ill-organized" form \text{--} of the associated Legendre differential equation and looking for its supersymmetric factorizations (as demonstrated in Section \ref{direct_factorization_of_associated_Legendre}), a much more elegant approach is deriving one of such supersymmetric factorizations from the \emph{more-than-well-organized} $so(3)$ Lie algebra and its representation, given the fact that the issue of the eigenspectrum of associated Legendre operator is intimately related to the representation of the angular momentum algebra $so(3)$. In this approach, which is implemented in Section \ref{angular_momentum_algebra}, all the ingredients for building the SUSYQM algorithm of the associated Legendre operator are \emph{inherited} in an elegant way from the representation of the angular momentum algebra $so(3)$ or from the equivalent ${\cal N}=2$ $su(1,1)$ superalgebra behind this $so(3)$ Lie algebra, simply by peeling off the azimuthal angle harmonics, and the broken form of associated Legendre equation does not even appear explicitly, instead, the associated Legendre equation comes out of this reduction procedure automatically in its supersymmetric factorized form. Of course, such elegance comes out as the reward of sticking to the general principle \text{--} symmetry comes first, where the symmetry is the $SO(3)$ transformation invariance.

More concretely, this Lie algebra reduction approach developed in Section \ref{angular_momentum_algebra} can be summarized as the following. Firstly, at the very formal level, in terms of supercharge formulation we have revealed that there always exists a 2-parameter ${\cal N}=2$ $su(1,1)$ superalgebra behind an $so(3)$ Lie algebra, where the superalgebra elements can be constructed both through falling into the representation of this $so(3)$ Lie algebra and through the more straightforward approach \text{--} plain (first-order) \emph{spin matrix maps}. Secondly, such 2-parameter ${\cal N}=2$ $su(1,1)$ superalgebra is then reduced to a 1-parameter (the $\theta$ coordinate) counterpart, either through entering the representation of the $so(3)$ Lie algebra or through taking the average of another \emph{enlarged} ${\cal N}=2$ $su(1,1)$ superalgebra over a specific two-component superstate. In the latter case, the supercharge matrices of such enlarged superalgebra are constructed iteratively from the plain (first-order) spin matrix maps by \emph{second-order} spin matrix maps.  The resultant “well-organized'' spectrum representation of such reduced 1-parameter superalgebra then seamlessly makes contact with the rest of the SUSYQM algorithm of resolving the spectrum of the associated Legendre operator parametrized in $\theta$ coordinate, that has already been detailed in Subsection \ref{direct_SUSYQM_AL_theta_coord}, without making use of the part responsible for finding the supersymmetric factorization of the ``ill-organized'' associated Legendre equation somewhat by brutal force.

The methods of plain and higher-order spin matrix map developed in this paper for revealing the $su(1,1)$ superalgebraic structure behind the Lie algebra $so(3)$ are also applicable to unraveling the dynamic Lie algebraic structures of the Hamiltonian systems that own other Lie algebraic symmetries.

Finally, it would be worthwhile to embed the two symmetry algebras \text{--} the ${\cal N}=2$ $su(1,1)$ superalgebra and the shape invariance algebra \text{--} into a one bigger symmetry algebra.

\appendix

\section{Searching for Coordinate Transformations by Focusing on Eliminating First Order Derivatives}\label{traditional_analytical_method}
Formally from \eq{Legendre_equ_x_coord} represented in $x$-coordinate to the Schr\"odinger equation \eqnm{type1_S_Equ_standard} represented in $y$-coordinate, the major change is the disappearance of the first order derivative. Traditionally, this has been achieved in a way that is full of mathematical flavor (see, e.g. \cite{Courant_and_Hilbert_1989} and \cite{Gangopadhyaya_SUSY_QM_2017} and references therein), and is conceptually and methodologically different from the momentum operator map method used in this paper.

Firstly, in the traditional approach, $\Phi_j(x)$ is conceived of being the composite function $\widehat{\Phi}_j(y(x))$, where the intermediate variable $y=y(x)$ is as yet to be determined. The $x$-derivatives are then turned into $y$-derivatives via the chain rule of differentiation, i.e., $d/dx =y'(x) d/dy$ and $d^2/d x^2 = \sqbr{y'(x)}^2 d^2/dy^2+ y''(x)d/d y$, where $y'\equiv d y/dx$. In consequence, the $x$-differential equation \eqnm{Legendre_equ_x_coord} is turned into the $y$-differential equation
\begin{eqnarray}\label{eigenequ_z_coord}
 &&-\sqbr{1-x^2}\sqbr{y'(x)}^2 \frac{d^2}{d y^2}\widehat{\Phi}_j\smbr{y} \nonumber \\
 &&+ \sqbr{2x y'(x)-\smbr{1-x^2} y''(x)} \od{}{y}\widehat{\Phi}_j\smbr{y}\nonumber \\
&&=\lambda_j \widehat{\Phi}_j\smbr{y}.
\end{eqnarray}

Secondly, the coefficient of the first order $y$-derivative is forced to be vanishing, i.e.,
\begin{eqnarray}\label{ODE_2nd_order}
 2x y'(x)-\smbr{1-x^2} y''(x)=0,
\end{eqnarray}
This is a quite special second-order differential equation of $y(x)$ in \eq{eigenequ_z_coord} \text{--} a \emph{first-order} differential equation of $y'(x)$. Directly integrating once, we obtain
\begin{eqnarray}\label{first_integral}
 y'= a (1-x^2)^{-1},
\end{eqnarray}
where $a$ is an integration constant. Integrating \eq{first_integral} once more, we obtain the coordinate transformation
\begin{eqnarray}\label{second_integral}
 y(x)= \frac{a}{2}\ln\frac{1+x}{1-x}=a \arctanh(x).
\end{eqnarray}

Thirdly, $y(x)$ in \eq{second_integral} is inverted into $x=\tanh y$ (where we take $a=1$) and the coefficient in front of the second order derivative in \eq{eigenequ_z_coord} becomes $\sqbr{1-x^2(y)}\sqbr{y'(x)}^2=(1-x^2)^{-1}= \cosh^2 y$. Consequently, \eq{eigenequ_z_coord} is then turned into the following Schr\"odinger-type equation
\begin{eqnarray}
 -\frac{d^2}{d y^2}\widehat{\Phi}_j\smbr{y}
-\lambda_j \sech^{-2} y \widehat{\Phi}_j\smbr{y}=0
\end{eqnarray}
that can be tweaked into the same form of Schr\"odinger equation \eqnm{type1_S_Equ_standard}. One then seeks for a supersymmetric factorization of this equation in the same procedure illustrated at the end of Subsubsection \ref{MO_map_SUSY_factorization} (after \eq{type1_S_Equ_standard}), and then continues to construct the SUSYQM algorithm as done in Subsections \ref{SIS_Vjy} and \ref{detailed_factorization_scheme_y_coordinate}.

Note that, when taking $a=1$, \eq{first_integral} can be rearranged as
\begin{eqnarray}
 (1-x^2) \od{}{x} = \od{}{y}.
\end{eqnarray}
This is just the momentum operator map \eqnm{momentum_operator_map1}, and the coordinate transformation \eqnm{second_integral} is the same as \eqnm{x_to_y_trans1}.

To turn \eq{associated_Legendre_equ_theta_coord} into a Schr\"odinger-type equation for the purpose of further performing  supersymmetric factorization, the traditional approach  goes as the following, which has been demonstrated in, e.g.,  \cite{Dutt_spherical_harmonics_SUSY_shape_invariance_1997, Gangopadhyaya_SUSY_QM_2017}. Firstly, one starts  with the {\it expanded} form
\begin{eqnarray}
J_0(\theta)=-\frac{d^2}{d\theta^2}-\cot\theta \od{}{\theta}
\end{eqnarray}
in \eq{associated_Legendre_equ_theta_coord}.

Secondly, one aims at constructing a Schr\"odinger equation for the function $\widehat{\Phi}_{jm}(y)$, which is the composition of $\Theta_{jm}(\theta)$ and the intermediate variable $\theta=\theta(y)$.
By the chain rule of differentiation, we have $d/d\theta=(\theta')^{-1}d/dy$ and $d^2/d\theta^2=(\theta')^{-1}d/dy[(\theta')^{-1}d/dy]=\smbr{\theta'}^{-2} d^2/dy^2-\theta''\smbr{\theta'}^{-2} d/dy$, where $\theta'\equiv d\theta/dy$ and $\theta''\equiv d^2\theta/dy^2$. Then $J_0(\theta)$ becomes
\begin{eqnarray}
J_0(\theta)=-\frac{1}{\theta'^2}\frac{d^2}{dy^2}+ \frac{1}{\theta'}\smbr{\frac{\theta''}{\theta'}-\cot\theta} \od{}{y}.
\end{eqnarray}
Substituting for $J_0(\theta)$ in this form,  the original differential equation \eqnm{associated_Legendre_equ_theta_coord} for $\Theta_{jm}(\theta)$ with respect to the coordinate $\theta$ is then turned into the following differential equation for  $\widehat{\Phi}_{jm}(y)$ with respect to the coordinate $y$,
\begin{eqnarray}\label{Schr_equ_widehatTheta}
 &&-\frac{d^2\widehat{\Phi}_{jm}(y)}{dy^2}+ \theta'\smbr{\frac{\theta''}{\theta'}-\cot\theta}\od{\widehat{\Phi}_{jm}(y)}{y}\nonumber \\
 &&+\theta'^2\smbr{\frac{m^2}{\sin^2\theta}-\lambda_j} \widehat{\Phi}_{jm}(y)=0.
\end{eqnarray}

Thirdly, in order that the above equation becomes a Schr\"odinger equation, one views the last term on the left hand side as the term of potential subtracted by the eigenvalue and forces the coefficient of the first-order derivative to be vanishing. In consequence, the following ordinary differential equation for $\theta(y)$ is obtained,
\begin{eqnarray}
 \frac{\theta''}{\theta'}=\cot\theta.
\end{eqnarray}

Directly integrating the above equation with respect to $\theta$ once and choosing the integration constant to be zero, one obtains
\begin{eqnarray}\label{1st_order_equ_theta_y}
 \theta'=\sin\theta.
\end{eqnarray}
This can be rearranged as
\begin{eqnarray}
 \sin\theta \od{}{\theta} =\od{}{y},
\end{eqnarray}
which is just the momentum operator map given in \eq{momentum_operator_map_1A}.

Integrating \eq{1st_order_equ_theta_y} one more time, we obtain the coordinate transformation
\begin{eqnarray}\label{y_theta_coord_relation}
 y(\theta)=\ln \tan \frac{\theta}{2}=-\arctanh (\cos\theta),
\end{eqnarray}
which is just the one given \eq{theta_y_coord_transformation}.

The combination of Eqs. \eqnm{1st_order_equ_theta_y} and \eqnm{y_theta_coord_relation} yields
\begin{eqnarray}
 \theta'^2=\sin^2\theta=\sech^2 y.
\end{eqnarray}

With the above substitutions, \eq{Schr_equ_widehatTheta} then reduces to the Schr\"odinger-type equation
\begin{eqnarray}
 -\frac{d^2\widehat{\Phi}_{jm}(y)}{dy^2} -\lambda_j \sech^2 y \widehat{\Phi}_{jm}(y)=-m^2\widehat{\Phi}_{jm}(y),
\end{eqnarray}
which is \eq{symmetrized_AL_equ_theta_coord_y_coord} that is obtained simply by momentum operator map.

This traditional method of turning a Sturm-Liouville type of eigenequation into a Schr\"odinger type eigenequation by solving any other second-order differential equation of the coordinate transformation, as demonstrated by the two above examples, probably date back to Courant and Hilbert (see this, for instance, on page 292 in \cite{Courant_and_Hilbert_1989}). This analytical method is not as efficient as our purely algebraic method of momentum operator map.

\section{The Supercharge Formulation of the Supersymmetrically Factorized Two Partner Eigenequations -- the ${\cal N}=2$ $su(1,1)$ Superalgebra}\label{super_charge_formulation}
Taking the $y$-coordinate representation of the SUSYQM of the base Legendre operator as an example, the supersymmetrically factorized eigenequation \eqnm{type1_S_Equ_standard} and its superpartner eigenequation \eqnm{sp_eigen_equ_before_rnmlz_y_coord} can be packaged into a 2-D matrix eigenequation as
\begin{eqnarray}\label{super_Shr_equ}
\widehat{H}^{ss}_j \widehat{\Phi}_j^{s}=j^2 \widehat{\Phi}_j^{ss},
\end{eqnarray}
where the super Hamiltonian $\widehat{H}^{ss}_j$ is the diagonal matrix consisting of $\widehat{H}_j$ and $\widehat{H}^\dagger_j$, i.e.,
\begin{eqnarray}
\widehat{H}^{ss}_j\equiv\left(\begin{array}{cc}
                      \widehat{A}^\dagger_j \widehat{A}_j & 0 \\
                      0 & \widehat{A}_j \widehat{A}^\dagger_j
                   \end{array}\right),
\end{eqnarray}
and the two-component eigenspinor $\widehat{\Phi}_j^{ss}$ is given by
\begin{eqnarray}
\widehat{\Phi}^{ss}_j \equiv \left(\begin{array}{c}
                                    \widehat{\Phi}_j \\
                                    \widehat{\Phi}_{j-1}
                                   \end{array}\right).
\end{eqnarray}

On the other hand, the $2\times 2$ unity matrix  has the natural decomposition $I_2 =P_u + P_d$, with
\begin{eqnarray}
P_u &\equiv& \left(\begin{array}{cc}
         1 & 0 \\
         0 & 0
       \end{array}\right), \quad  P_d \equiv \left(\begin{array}{cc}
         0 & 0 \\
         0 & 1
       \end{array}\right).
\end{eqnarray}
It is easy to check that $P_u$ and $P_d$ can be factorized as
\begin{eqnarray}
P_u =\frac{1}{4} \sigma^+\sigma^-, \;  P_d =\frac{1}{4}\sigma^-\sigma^+,
\end{eqnarray}
where
\begin{eqnarray}
\begin{aligned}
 \sigma^+ \equiv \sigma_1 + i \sigma_2 = \left(\begin{array}{cc}
                           0 & 2 \\
                           0 & 0
                        \end{array}\right),\\
\sigma^- \equiv \sigma_1 - i \sigma_2= \left(\begin{array}{cc}
                                                                         0 & 0 \\
                                                                         2 & 0
                                                                       \end{array}\right),
\end{aligned}
\end{eqnarray}
and $\sigma_{1,2}$ are the first two Pauli matrices.

$\widehat{H}^{ss}_j$ is then naturally decomposed as
\begin{eqnarray}\label{super_Hamiltonian_decomposition}
 \widehat{H}^{ss}_j&=&\widehat{A}^\dagger_j \widehat{A}_j P_u + \widehat{A}_j\widehat{A}^\dagger_j  P_d, \nonumber \\
                   &=& \smbr{\frac{1}{2}\widehat{A}^\dagger_j \sigma^+} \smbr{\frac{1}{2}\widehat{A}_j \sigma^-}+\smbr{\frac{1}{2}\widehat{A}_j \sigma^-}\smbr{\frac{1}{2}\widehat{A}^\dagger_j \sigma^+},\nonumber \\
                   &=&\cubr{\widehat{Q}^+_j, \widehat{Q}^-_j},
\end{eqnarray}
where the pair of curly brackets denote the anticommutator, and
\begin{eqnarray}\label{super_charges_spin_matrix_maps}
\widehat{Q}^+_j \equiv \frac{1}{2}\sigma^+ \widehat{A}^\dagger_j, \quad \widehat{Q}^-_j \equiv \frac{1}{2} \sigma^-\widehat{A}_j
\end{eqnarray}
are the pair of supercharges that are Hermitian conjugate to each other.

Furthermore,  $\widehat{Q}^\pm_j$ inherit the nilpotent (fermionic) properties of $\sigma^\pm$ ($\smbr{\sigma^\pm}^2=0$), i.e.,
\begin{eqnarray}\label{super_charge_nilpotent}
 \cubr{\widehat{Q}^+_j, \widehat{Q}^+_j}=\cubr{\widehat{Q}^-_j, \widehat{Q}^-_j}=0.
\end{eqnarray}
With these properties and the decomposition \eqnm{super_Hamiltonian_decomposition} of $\widehat{H}^{ss}_j$, it is easy to check that $\widehat{Q}^\pm_j$ are conserved in the sense that
\begin{eqnarray}\label{supercharge_commuting}
 \sqbr{\widehat{Q}^\pm_j, \widehat{H}^{ss}_j}=0.
\end{eqnarray}
This conservation implies the eigenvalue degeneracy between the state $\widehat{\Phi}_j^{ss}$ and the states $\widehat{Q}^\pm_j \widehat{\Phi}^{s}_j$. In particular,
\begin{eqnarray}\label{recurrence_relations_supercharge}
 \left(\begin{array}{c}
   \widehat{\Phi}_j \\
      0
     \end{array}\right) \quad  \xrightleftharpoons[j^{-1} \widehat{Q}^+_j]{j^{-1}\widehat{Q}^-_j}  \left(\begin{array}{c}
     0 \\
    \widehat{\Phi}_{j-1}
     \end{array}\right),
\end{eqnarray}
which are the rephrasing of the lowering and raising recursion relations \eqnm{lowering_recurrence_relation_y_coord} and \eqnm{base_raising_recurrence_y_coord}, and tell us that the supercharges $\widehat{Q}^\pm_j$ switch the bosonic and fermionic states (if one is viewed as a bosonic state then the other as fermionic one), while the eigenvalue remains degenerate.

The super-eigenequation \eqnm{super_Shr_equ} is recovered by the combination of the two recurrence relations in \eq{recurrence_relations_supercharge}, \eq{super_Hamiltonian_decomposition} and \eq{super_charge_nilpotent}.

In terms the two supercharges $\widehat{Q}^\pm_j$, Eqs. \eqnm{super_Hamiltonian_decomposition}, \eqnm{super_charge_nilpotent} and \eqnm{supercharge_commuting} together define the ${\cal N}=2$ \emph{su(1,1)} superalgebra \cite{Witten_SUSY_breaking_1981, Witten_1982, Cooper_SUSY_and_QM_1995} that underlies the SUSYQM. This is the very deep reason for why the corresponding symmetry is termed supersymmetry. This superalgebra, the matrix equation \eqnm{super_Shr_equ} and the two component eigenequations \eqnm{type1_S_Equ_standard} and \eqnm{sp_eigen_equ_before_rnmlz_y_coord} are the three equivalent definitions of the same SUSYQM.

In the above, we have derived the ${\cal N}=2$ su(1,1) superalgebra in a heuristic way, starting with a pair of eigenequations that are superparnters of each other. We can also view  $\widehat{Q}^\dagger_j$ and $\widehat{Q}_j$ in \eq{super_charges_spin_matrix_maps} as the definitions of these two supercharges, through the \emph{spin matrix maps} of the two ladder operator $\widehat{A}^\dagger_j$ and $\widehat{A}_j$. The ferminonic properties satisfied by these two supercharges are stil given by \eq{super_charge_nilpotent}, which follow from the nilpotent properties $\smbr{\sigma^\pm}^2=0$. The super-Hamiltonian is defined as $\widehat{H}^{ss}_j\equiv \widehat{Q}^2_j$ for $\widehat{Q}_j\equiv \widehat{Q}^\dagger_j+\widehat{Q}_j$. With \eq{super_charge_nilpotent} in hand, both \eq{super_Hamiltonian_decomposition} for $\widehat{H}^{ss}_j$ and \eq{supercharge_commuting} follow straightforwardly. This latter procedure can also be treated as a standard one of deriving the ${\cal N}=2$ $su(1,1)$ superalgebra.

\section{$\theta$-coordinate Representation of the SUSYQM Algorithm of $J_0(\theta)$}\label{theta_coord_J0}

In $\theta$ coordinate, the Legendre differential equation reads
\begin{eqnarray}\label{Legendre_equ_theta_coord}
 J_0(\theta) \Theta_j(\theta)=\lambda_j\Theta_j(\theta),
\end{eqnarray}
where the Legendre operator $J_0(\theta)$ in its lumped form is given in \eq{J0_theta}.

Multiplying with $\sin^2\theta$ for preparing the square $-\smbr{\sin\theta d/d\theta}^2$ as the kinetic energy operator and adding $E_j\Theta_j(\theta)$ on both sides for completing the supersymmetric factorization, \eq{Legendre_equ_theta_coord} is turned into
\begin{eqnarray}\label{Legendre_equ_theta_coord_symmetrized}
 \sqbr{-\smbr{\sin\theta \od{}{\theta}}^2-\lambda_j \sin^2\theta +E_j}\Theta_j(\theta)= E_j\Theta_j(\theta).
\end{eqnarray}

The momentum operator map
\begin{eqnarray}
 \sin\theta \od{}{\theta} = \od{}{y},
\end{eqnarray}
which is the momentum operator map \eqnm{momentum_operator_map_1A}, that maps \eq{Legendre_equ_theta_coord_symmetrized} into the standard Schr\"odinger equation \eqnm{Sch_equ_y_coord}.

The operator in \eq{Legendre_equ_theta_coord_symmetrized} has the formal supersymmetric factorization
\begin{eqnarray}
 -\smbr{\sin\theta \od{}{\theta}}^2-\lambda_j \sin^2\theta +E_j = {\cal B}_j(\theta){\cal A}_j(\theta),
\end{eqnarray}
where
\begin{eqnarray}\label{Bj_Aj_theta_coord}
 {\cal B}_j(\theta)\equiv-{\cal A}_0(\theta)+ {\cal W}_j(\theta),\quad {\cal A}_j(\theta)\equiv {\cal A}_0(\theta)+ {\cal W}_j(\theta),\nonumber \\
\end{eqnarray}
\begin{eqnarray}
 {\cal A}_0(\theta)\equiv \sin\theta \od{}{\theta},\quad {\cal W}_j(\theta)\equiv -e_j \cos\theta,
\end{eqnarray}
\begin{eqnarray}\label{lambda_E_e_relations}
 \lambda_j=e_j(e_j+1),\quad E_j = e^2_j.
\end{eqnarray}
That is, \eq{Legendre_equ_theta_coord_symmetrized} is supersymmetrically factorized as
\begin{eqnarray}\label{Legendre_equ_theta_coord_supersymmetrized}
 {\cal B}_j(\theta){\cal A}_j(\theta) \Theta_j(\theta)= E_j\Theta_j(\theta).
\end{eqnarray}

$E_j$ in \eq{Legendre_equ_theta_coord_supersymmetrized} can be vanishing, such solution is labeled by $j=0$ and $E_0=0$. Consequently, \eq{lambda_E_e_relations} gives $e_0=0$ and $\lambda_0=0$. In this case, \eq{Legendre_equ_theta_coord_supersymmetrized} reduces to
\begin{eqnarray}
 {\cal A}_0(\theta) \Theta_0(\theta)=\od{\Theta_0(\theta)}{\theta}=0,
\end{eqnarray}
which means that $\Theta_0(\theta)$ is a constant.

The intertwining action on \eq{Legendre_equ_theta_coord_supersymmetrized} with ${\cal A}_j(\theta)$ yields its superpartner eigenequation
\begin{eqnarray}\label{Legendre_equ_theta_coord_superparnter_equ}
 {\cal A}_j(\theta){\cal B}_j(\theta) \Theta^s_j(\theta)= E_j\Theta^s_j(\theta),
\end{eqnarray}
where $\Theta^s_j(\theta)\propto {\cal A}_j(\theta) \Theta_j(\theta)$.

The downward-connected pattern invariance condition
\begin{eqnarray}\label{pattern_invariance_condition_theta_coord}
 {\cal A}_j(\theta){\cal B}_j(\theta)-{\cal B}_{j-1}(\theta){\cal A}_{j-1}(\theta)=\Delta_j
\end{eqnarray}
can be imposed for a certain constant $\Delta_j$ as yet to be determined. Such condition allows us to renormalize \eq{Legendre_equ_theta_coord_superparnter_equ} into
\begin{eqnarray}\label{renormalized_superpartner_equ}
 {\cal B}_{j-1}(\theta){\cal A}_{j-1}(\theta)\Theta^s_j(\theta)=E_{j-1}\Theta^s_j(\theta),
\end{eqnarray}
where $E_{j-1}\equiv E_j-\Delta_j$.

\eq{renormalized_superpartner_equ} has the same pattern as \eq{Legendre_equ_theta_coord_supersymmetrized}. This allows us to make the identification $\Theta^s_j(\theta)\propto {\cal A}_j(\theta) \Theta_j(\theta)\propto \Theta_{j-1}(\theta)$. That is, ${\cal A}_j(\theta)$ works as the lowering operator that lowers the index $j$ by one unit, and we can choose a normalization such that the lowering recurrence relation reads
\begin{eqnarray}\label{lowering_recurrence_theta_coord_base_L}
 {\cal A}_j(\theta) \Theta_j(\theta)= E_j\Theta_{j-1}(\theta).
\end{eqnarray}

The combination of Eqs. \eqnm{Legendre_equ_theta_coord_supersymmetrized} and \eqnm{lowering_recurrence_theta_coord_base_L} then yields the raising recurrence relation
\begin{eqnarray}\label{raising_recurrence_theta_coord_base_L}
 \Theta_j(\theta)={\cal B}_j(\theta)\Theta_{j-1}(\theta).
\end{eqnarray}

Substituting for ${\cal A}_j(\theta)$, ${\cal B}_j(\theta)$, ${\cal B}_{j-1}(\theta)$ and ${\cal A}_{j-1}(\theta)$ via \eq{Bj_Aj_theta_coord}, \eq{pattern_invariance_condition_theta_coord} then reduces to
\begin{eqnarray}
 \smbr{e_j+e_{j-1}}\sqbr{\smbr{1-e_j+e_{j-1}}\sin^2\theta+ e_j-e_{j-1}} =\Delta_j.
\end{eqnarray}
It then follows that
\begin{eqnarray}\label{ej_recurrence_Legendre}
 e_j=e_{j-1}+1,\quad e^2_j-e^2_{j-1}=\Delta_j.
\end{eqnarray}

With $e_0=0$, the recurrence series in \eq{ej_recurrence_Legendre} and $\lambda_j$ and $E_j$ in \eq{lambda_E_e_relations} are then solved as
\begin{eqnarray}
 e_j=j,\quad \Delta_j=2j-1,\quad \lambda_j=j(j+1),\quad E_j=j^2.
\end{eqnarray}

Consecutively applying \eq{raising_recurrence_theta_coord_base_L} for $j\rightarrow j,\;j-1,\;\cdots,\;2\;\text{and}\;1$, we obtain
\begin{eqnarray}\label{Theta_j_theta}
 \Theta_j(\theta)={\cal B}_j(\theta){\cal B}_{j-1}(\theta)\cdots {\cal B}_2(\theta) {\cal B}_1(\theta) \Theta_0(\theta).
\end{eqnarray}

${\cal B}_j(\theta)$ in \eq{Bj_Aj_theta_coord} can be rewritten as
\begin{eqnarray}
 {\cal B}_j(\theta)=-\sin\theta \od{}{\theta}-j\cos\theta=-\sin^{j-1}\theta \od{}{\theta} \sin^j\theta.
\end{eqnarray}
Using this lumped form of ${\cal B}_j(\theta)$ to shorten the operator product, \eq{Theta_j_theta} then becomes
\begin{eqnarray}
 \Theta_j(\theta)=\smbr{-1}^j \sin^{-j}\theta\smbr{\sin\theta \od{}{\theta}\sin\theta}^j \Theta_0(\theta).
\end{eqnarray}
This gives us the usual Legendre polynomial $P_j(\cos\theta)$ if we choose $\Theta_0(\theta)=1$.

\acknowledgments

This work is supported in part by National Natural Science Foundation of China under Grant No. 12075070.

\bibliographystyle{unsrt}
\bibliography{SUSY_factorization}

\end{document}